\documentclass[twocolumn]{aastex7}

\usepackage{subfigure}
\usepackage{amsmath}
\usepackage{svg}
\usepackage{xspace}
\usepackage{appendix}
\usepackage{hyperref} 

\newcommand{\ha}{H$\alpha$\xspace}
\newcommand{\hb}{H$\beta$\xspace}
\newcommand{\hii}{\ion{H}{2}\xspace}

\newcommand{\sii}{[\ion{S}{2}]\xspace}
\newcommand{\nii}{[\ion{N}{2}]\xspace}
\newcommand{\oi}{[\ion{O}{1}]\xspace}

\newcommand{\oiii}{[\ion{O}{3}]\xspace}

\newcommand{\niiha}{[\ion{N}{2}]/H$\alpha$\xspace}

\newcommand{\oiiihb}{[\ion{O}{3}]/H$\beta$\xspace}

\newcommand{\angstrom}{\mbox{\normalfont\AA} \xspace}

\newcommand{\fin}{$f_{\text{inner}}$}
\newcommand{\fout}{$f_{\text{outer}}$}

\begin{document}
\shortauthors{Casey et al.}
\shorttitle{MaNGA Fading AGN}

\title{AGN Archaeology: A Systematic Search for Fading AGN in MaNGA}

\author[0009-0005-9002-4800]{Quinn O. Casey}
\affiliation{Department of Physics and Astronomy, Dartmouth College, Hanover, NH 03755, USA}
\email[show]{quinn.o.casey.gr@dartmouth.edu}

\author[0000-0003-1468-9526]{Ryan C. Hickox}
\affiliation{Department of Physics and Astronomy, Dartmouth College, Hanover, NH 03755, USA}
\email{Ryan.C.Hickox@dartmouth.edu}

\author[0000-0003-1420-6037]{Jonathan H. Cohn}
%\affiliation{Department of Physics and Astronomy, Dartmouth College, Hanover, NH 03755, USA}
%\email{Jonathan.Cohn@dartmouth.edu}
\affiliation{Department of Physics, Virginia Commonwealth University, Richmond, VA 23284, USA}
\email{cohnj3@vcu.edu}

\author[0000-0001-7151-009X]{Nikko J. Cleri}
\affiliation{Department of Astronomy and Astrophysics, The Pennsylvania State University, University Park, PA 16802, USA}
\affiliation{Institute for Computational and Data Sciences, The Pennsylvania State University, University Park, PA 16802, USA}
\affiliation{Institute for Gravitation and the Cosmos, The Pennsylvania State University, University Park, PA 16802, USA}
\email{cleri@psu.edu}

\author[]{Julia M. Comerford}
\affiliation{Department of Astrophysical and Planetary Sciences, University of Colorado, Boulder, CO 80309, USA}
\email{julie.comerford@colorado.edu}

\author[0000-0002-4235-7337]{K. Decker French}
\affiliation{Department of Astronomy, University of Illinois, 1002 W. Green St., Urbana, IL 61801, USA}
\email{deckerkf@illinois.edu}

\author[0009-0006-7814-2334]{Sandra Jaison}
\affiliation{Departamento de Astronomia, Instituto de Física, Universidade Federal do Rio Grande do Sul, CP 15051, 91501-970, Porto Alegre, RS, Brazil}
\email{sandrajaison77@gmail.com}

\author[0000-0003-4440-259X]{Mar Mezcua}
\affiliation{Institute of Space Sciences, ICE-CSIC, Campus UAB, Carrer de Can Magrans s/n, 08913 Bellaterra, Barcelona, Spain}
\affiliation{Institut d'Estudis Espacials de Catalunya (IEEC), c/ Esteve Terradas 1, Edifici RDIT, Campus PMT-UPC, 08860 Castelldefels, Spain}
\email{marmezcua.astro@gmail.com}

\author[0000-0002-1321-1320]{Rogério Riffel}
\affiliation{Departamento de Astronomia, Instituto de Física, Universidade Federal do Rio Grande do Sul, CP 15051, 91501-970, Porto Alegre, RS, Brazil}
\email{riffel@ufrgs.br}

\author[0000-0003-2212-6045]{Dominika Wylezalek}
\affiliation{Astronomisches Rechen-Institut, Zentrum für Astronomie der Universität Heidelberg, Mönchhofstr. 12-14, D-69120 Heidelberg,
Germany}
\email{dominika.wylezalek@uni-heidelberg.de}

\author[0000-0002-1333-147X]{Peixin Zhu}
\affiliation{Research School of Astronomy and Astrophysics, Australian National University, Australia}
\affiliation{ARC Centre of Excellence for All Sky Astrophysics in 3 Dimensions (ASTRO 3D), Australia}
\affiliation{Center for Astrophysics | Harvard \& Smithsonian, 60 Garden Street, Cambridge, MA 02138, USA}
\email{peixin.zhu@cfa.harvard.edu}

\correspondingauthor{Quinn Casey}

\begin{abstract}
Active galactic nuclei (AGN) vary on broad timescales, but are primarily thought to ``flicker" on and off during episodes of high and low accretion.
Supermassive black holes (SMBHs) which are actively accreting material often show optical spectra dominated by AGN emission in their nuclei and beyond. 
In this work we develop an empirical model to quantify the fractional AGN contribution to optical spectra using the Baldwin-Phillips-Terlevich (BPT) emission line diagnostic diagram.
We leverage this model against integral field unit (IFU) data from the Mapping Nearby Galaxies at Apache Point Observatory (MaNGA) survey to select 34 galaxies with weak nuclear AGN activity relative to their AGN signatures at large radii; this signifies past AGN activity which is fading with time (i.e., fading-AGN).
%The majority of our fading-AGN reside in edge-on systems not undergoing major-mergers which suggests we are probing secular galaxy evolution.
The majority of our fading-AGN reside in edge-on systems undergoing minor mergers.
%, which suggests that such interactions may play an important role in AGN triggering.}
We construct a simple model consisting of a disk galaxy with ionization cones and find that the orientation of such a system will impact its BPT classification in both single-fiber and IFU observations.
The mean distance to the extended AGN emission in our sample corresponds to a light-travel time of $1.1\times10^4$ years. 
We use this to set a lower-limit on the typical lifetime of the AGN phase ($t_{\text{AGN}}>1.3\times10^5$ years) which is consistent with SMBHs growing via brief accretion bursts. 
%and suggests the instantaneous accretion rate of a system is not necessarily representative of its average accretion rate.
These fading-AGN allow us to probe the elusive period of post-AGN galaxy evolution, and they give a $\sim10^4$ year baseline for understanding the accretion history in these systems.
\end{abstract}


\section{Introduction}\label{sec:intro}

AGN have been shown to vary on timescales ranging from hours to millions of years; this variability occurs across the entire electromagnetic spectrum \citep[e.g.,][]{Peterson93, Ulrich97, Peterson98, Vaughan03, Hickox14, Shappee14, Schawinski15, Padovani17}. 
There are numerous causes for the short-term variability including dense gas eclipsing the AGN along the observer's line of sight \citep[e.g.,][]{Ricci23}, thermal fluctuations in the accretion disk \citep{Kelly09}, or tidal disruption events \citep[e.g.,][]{Gezari21} to name a few. 
SMBHs need a relatively stable gas supply to remain classified as an AGN using traditional techniques (see \citealt{Hickox18} for a review of these techniques), and this nuclear fueling can occur through numerous physical processes \citep[e.g.,][]{Jogee06, Storchi19, Riffel24, Alexander25}.
The AGN phase is a finite event meaning that AGN will turn on and off over cosmic time analogous to a ``flickering" light bulb \citep{Hickox14, Schawinski15}, and this variability occurs on much longer timescales relative to the short-term variability previously discussed.
However, we can observationally identify AGN which have recently turned off or on based on the spatial extent and location of the AGN emission \citep[e.g.,][]{Schawinski15, Esparza20, French23}.

Fading-AGN are AGN where the SMBH has halted (or substantially reduced) its accretion of material and therefore the entire system will begin to appear orders of magnitude dimmer. 
However, once the SMBH has exited its accretion phase, signatures of the past AGN will still be detectable at locations outside the central region of the galaxy. 
These signatures, which are referred to as AGN light echoes (LEs) or extended emission line regions (EELRs), persist due to gas cooling timescales and (often most significantly) light travel time (i.e., the time it takes light to travel some distance). 
Fading-AGN can also lend insight into the host galaxy's large-scale gas environment and trace the history of accretion onto the SMBH.
LEs and EELRs can inform us about the geometry of AGN narrow-line regions (NLRs) which allow for tests of the unified model \citep[e.g.,][]{Antonucci93, Moiseev23}. 
EELRs, in particular, are the longest tracer of AGN activity that account for $\sim10^{5}$ years in some cases \citep[e.g.,][]{Lintott09, Knese20}. 
A blind search for fading-AGN allows us to place empirical limits on the typical lifetime of the AGN phase \citep{French23}.
Such searches are challenging as spectroscopic data are frequently needed to determine the primary source of ionization, and the majority of (non-IFU) spectroscopic surveys focus their observations on galaxy nuclei where you would either not detect the EELR (due to the size on the sky) or the EELR is diluted by the host galaxy.

Perhaps the most well known example of a fading-AGN is Hanny's Voorwerp near IC 2497 \citep{Lintott09} which was discovered through the Galaxy Zoo project \citep{Lintott08}. 
Hanny's Voorwerp mimics a NLR with its high-ionization ($>35.12$ eV, \citealt{Berg21}) emission line ratios, despite being located $\sim15-35$ kpc from the galaxy's nucleus \citep{Schawinski10, Keel12}. 
Similarly, past searches for fading-AGN have focused on identifying extended \oiii~$\lambda5007$ emission from photometric surveys; these are regions which are unusually luminous in the Sloan Digital Sky Survey (SDSS) $g-$band in the local Universe \citep[e.g.][]{Keel12, Keel13, Sun17}.
Furthermore, merging systems are the ideal location to search for extended emission, as galaxy scale interactions can drive gas towards the SMBH and produce extended gas clouds to be illuminated by the central AGN \citep{Keel17, Keel22, Keel24, Esparza20}.
However, the reliance on merging systems may skew our understanding of fading-AGN regarding their ubiquity.

More recently, IFU spectroscopy has been utilized to find and characterize these systems to a higher degree of accuracy \citep[e.g.][]{Menezes16, Comerford17, Xu23, French23, Mo24, Finlez25}. 
%IFU spectroscopy is the preferable mode of study when searching for extended AGN emission because of the power and accuracy of spatially resolved emission line diagnostic diagrams. 
Extended AGN emission is relatively straightforward to identify in IFU data due to spatially resolved emission line diagnostic diagrams and spectroscopic redshift determinations.
Furthermore, the spatial component of IFU data can be more informative when differentiating between AGN and shock ionization because one can probe the radius and local velocity dispersion of high ionization emission line ratios \citep[e.g.,][]{Riffel13, Dagostino19, Dahmer19, Kewley19, Riffel21, Zhu25}.
%grants us the ability to differentiate between AGN ionization versus shock ionization.
This is of particular importance because shock ionization can masquerade as AGN activity \citep[e.g.][]{Dopita95, Moy02, Rich10, Rich11, Kewley13, Kewley19}.

In this paper we utilize the IFU spectroscopy from the MaNGA survey \citep{Bundy15} in order to discover and characterize a sample of 34 fading-AGN. 
In Section \ref{sec:data} we discuss the MaNGA survey and the data model that we adopt. 
Section \ref{sec:selection} discusses our AGN decomposition model, the selection algorithm used to find candidate fading-AGN, and the removal of interlopers.
We investigate properties of the fading-AGN such as their luminosity and morphology in Section \ref{sec:properties}. In Section \ref{sec:discussion} we place a lower limit on the AGN phase lifetime in the local Universe, perform 3D modeling to test the effects of orientation, and compare our results to previous work. 
In Section \ref{sec:conclusions} we summarize our findings and their importance. 
When necessary we utilize the following cosmology: $H_{0}$=70 km s$^{-1}$ Mpc$^{-1}$, $\Omega_{M}$=0.3, and $\Omega_{\Lambda}$=0.7.

%\begin{table}
%\renewcommand{\arraystretch}{1.75}
%\caption{The system of equations needed to determine the fractional AGN contribution in BPT space where S is emission due to star formation and A is emission due to the AGN. Subscripts $x$ and $y$ denote the \niiha and \oiiihb starting/ending locations in the SFG/AGN loci. The free parameter $f_{\text{AGN}}$ determines the fractional AGN contribution for a given location.}
%    \centering
%    \begin{tabular}{|c|c|}
%    \hline 
%    $\text{S}_{\text{H}\alpha}=1$ & $\text{A}_{\text{H}\alpha}=2.86\text{A}_{\text{H}\beta}$ \\ 
    
%    \hline
    
    %$\text{S}_{\text{H}\beta}=\text{S}_{\text{H}\alpha}/2.86$ & $\text{A}_{\text{H}\beta}=\text{A}_{\text{O}3} / 10^{\text{A}_{y}}$ \\ 
    
    %\hline
    
    %$\text{S}_{\text{N}2} = \text{S}_{\text{H}\alpha}10^{\text{S}_{x}}$ & $\text{A}_{\text{N}2} = \text{A}_{\text{H}\alpha}10^{\text{A}_{x}}$ \\ 
    
    %\hline
    
    %$\text{S}_{\text{O}3} = \text{S}_{\text{H}\beta}10^{\text{S}_{y}}$ & $\text{A}_{\text{O}3} = (-f_{\text{AGN}}\text{S}_{\text{O}3})/(f_{\text{AGN}}-1)$ \\ 
    
    %\hline
    %\hline
    %\end{tabular}
    %\label{tab:equations}
%\end{table}

\begin{table}
\renewcommand{\arraystretch}{1.75}
\caption{The system of equations needed to determine the fractional AGN contribution in BPT space where $F_{\text{SF}}$ is flux due to star formation and $F_{\text{AGN}}$ is flux due to the AGN. 
The other subscripts refer to the particular emission line (i.e., $\text{N2}=\text{\nii}\lambda6584$, $\text{O3}=\text{\oiii}\lambda5007$).
$x$ and $y$ denote the \niiha and \oiiihb BPT locations in the SFG/AGN loci (i.e., the starting/ending points for a given evolutionary track). 
The free parameter $f_{\text{AGN}}$ determines the fractional AGN contribution for a given location.}
    \centering
    \begin{tabular}{|c|c|}
    \hline 
    $\text{F}_{\text{SF,H}\alpha}=1$ & $\text{F}_{\text{AGN,H}\alpha}=3.1\text{F}_{\text{AGN,H}\beta}$ \\ 
    
    \hline
    
    $\text{F}_{\text{SF,H}\beta}=\text{F}_{\text{SF,H}\alpha}/2.86$ & $\text{F}_{\text{AGN,H}\beta}=\text{F}_{\text{AGN,O}3} / 10^{y_{\text{AGN}}}$ \\ 
    
    \hline
    
    $\text{F}_{\text{SF,N}2} = \text{F}_{\text{SF,H}\alpha}10^{x_{\text{SF}}}$ & 
    $\text{F}_{\text{AGN,N}2} = \text{F}_{\text{AGN,H}\alpha}10^{x_{\text{AGN}}}$ \\ 
    
    \hline
    
    $\text{F}_{\text{SF,O}3} = \text{F}_{\text{SF,H}\beta}10^{y_{\text{SF}}}$ & 
    $\text{F}_{\text{AGN,O}3} = \frac{f_{\text{AGN}}\text{F}_{\text{SF,O}3}}{1-f_{\text{AGN}}}$ \\ 
    
    \hline
    %\hline
    \end{tabular}
    \label{tab:equations}
\end{table}

\section{Data}\label{sec:data}
This study uses data from the MaNGA survey which was a core program of SDSS-IV \citep[][]{Blanton17}.
MaNGA is a spectroscopic IFU survey that observed $\sim$10,000 unique galaxies at $0.01 < z < 0.15$ \citep{Bundy15, Drory15, Law15}; the instrument is mounted on the SDSS 2.5m telescope at Apache Point Observatory \citep{Gunn06}. 
The observations employ 2\arcsec\xspace fibers arranged in hexagonal bundles; the fiber bundles vary in size from 19 fibers to 127 fibers which corresponds to a 12.5\arcsec\xspace to 32.5\arcsec\xspace diameter on the sky.
The center of the IFU bundles are positioned on the targets' photometric centroid as determined through modeling the light profile of SDSS imaging \citep[e.g.,][]{Pier03}.
Sources in the primary MaNGA sample have IFU coverage out to 1.5 effective radii ($R_e$, which is defined as the radius containing half of a systems' total luminosity; e.g., \citealt{Binney08}) as measured in the $r$-band, whereas sources in the secondary sample extend to $2.5R_e$.
Approximately two-thirds and one-third of the MaNGA sample are observed out to $1.5R_e$ and $2.5R_e$, respectively; this split provides a stellar mass distribution that is roughly flat \citep[e.g.,][]{Bundy15, Wake17}.
MaNGA has an average footprint of $\sim500$ arcsec$^{2}$ per IFU. 
The fiber bundles are fed into the dual-channel Baryon Oscillation Spectroscopic Survey (BOSS) spectrographs, which provide spectral coverage over $3600-10300$\xspace$\angstrom$ with a median resolution of $R\sim2,000$ \citep{Smee13}.

In this study we use MaNGA Data Release 17 \citep[DR17,][]{Abdurrouf22} which is the final data release containing all MaNGA observations. 
The sky-subtraction and spectrophotometric calibrations are performed with the Data Reduction Pipeline \citep[{\fontfamily{cmtt}\selectfont DRP},][]{Law16, Yan16}. 
The {\fontfamily{cmtt}\selectfont DRP} smooths the spectra to a spatial resolution of 2.5\arcsec\xspace and the spectra are calibrated to percent level accuracy.
The MaNGA Data Analysis Pipeline \citep[{\fontfamily{cmtt}\selectfont DAP},][]{Westfall19, Belfiore19, Law21a} measures stellar kinematics, emission-line fluxes, and stellar spectral indices for individual ``spaxels" (spectral-pixels). 
The flux measurements are derived using Penalized Pixel Fitting \citep[pPXF,][]{Cappellari04, Cappellari17, Cappellari23}.
The stellar continua are fit using various simple-stellar populations from the MILES-HC library \citep{Falcon11}. 
%Specifically, our work uses the {\fontfamily{cmtt}\selectfont HYB10-MILESHC-MASTARHC2} data model, which utilizes a Voronoi algorithm \citep{Cappellari03} to bin spaxels and achieve a signal-to-noise (S/N) ratio $\gtrsim10$ for the stellar kinematic measurements; emission-line measurements are performed on a spaxel-by-spaxel basis. 
Specifically, our work uses the {\fontfamily{cmtt}\selectfont HYB10-MILESHC-MASTARHC2} data model which performs emission-line measurements on a spaxel-by-spaxel basis. 
This is the optimal model as our work seeks to investigate emission as a function of location; the treatment of the stellar kinematics makes no impact on our study.
%and the stellar kinematics are not crucial to this study.

\section{Selecting Fading-AGN}\label{sec:selection}

\begin{figure}
    \centering
    \includegraphics[width=\linewidth]{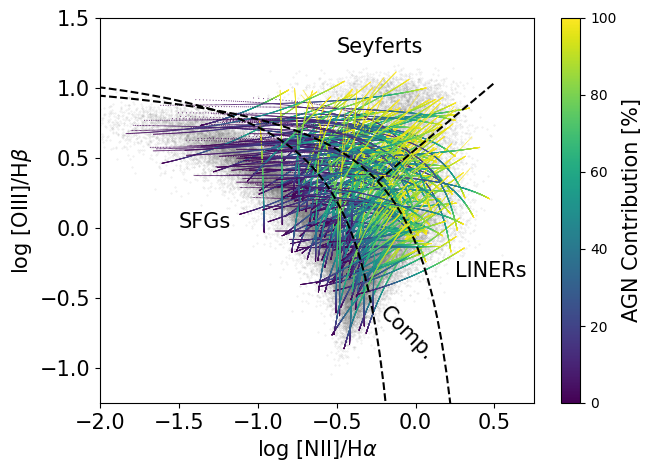}
    \caption{The fractional AGN contribution model grid on the BPT diagram where the colored lines represent the AGN contribution (i.e., 100\% is a pure AGN and 0\% is a pure star-forming galaxy). 
    The underlying gray points are eBOSS galaxies from the \citet{MatthewsAcuna25} catalog used to define starting and ending points of each track.
    This figure only shows a subset ($N=250$) of the total number of tracks in this model ($N=25,000$). Black dashed lines are the demarcation lines of \citet{Kewley01}, \citet{Kauffmann03}, and \citet{Law21b}. The relevant classification regions are labeled in black.}
    \label{fig:frac_agn}
\end{figure}

\begin{figure*}
    \centering
    \includegraphics[width=\linewidth]{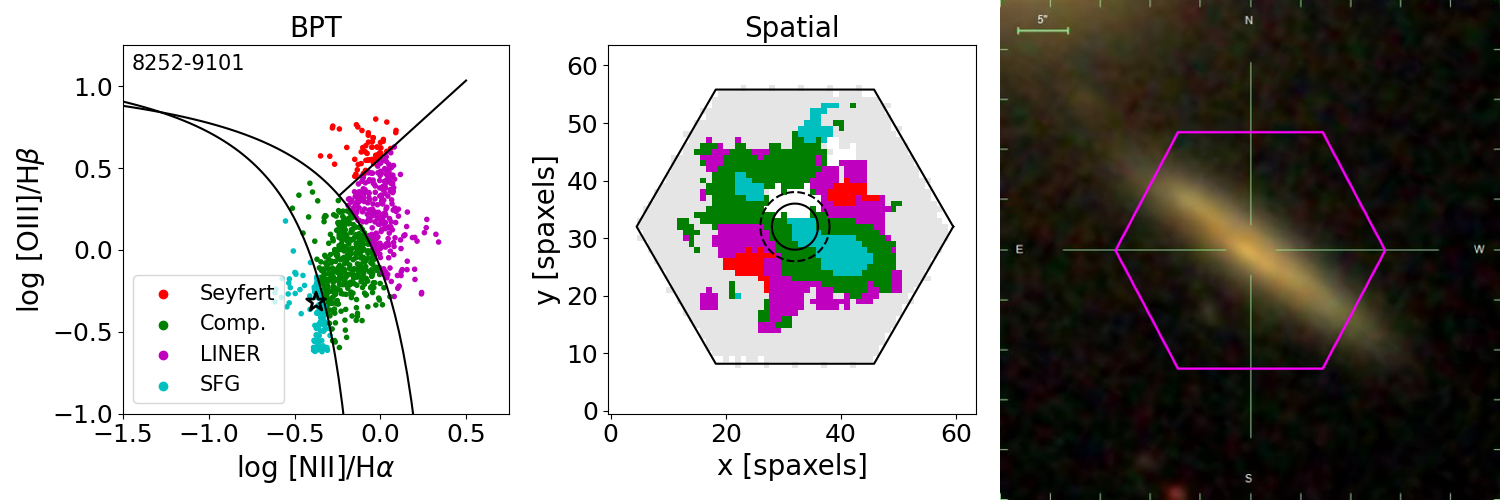}
    \caption{Example fading-AGN for MaNGA galaxy 8252-9101. 
    \textit{Left panel:} spaxels plotted on the BPT diagram and color-coded by their emission type. 
    The black star represents the BPT location of this system as determined by the 3\arcsec\ SDSS single-fiber observation (see Section \ref{subsec:sdss}), and the black lines are the \citet{Kewley01}, \citet{Kauffmann03}, and \citet{Law21b} BPT demarcation lines.
    \textit{Middle panel:} The spatial distribution of the different types of emission as determined by the BPT diagram. 
    Gray spaxels have S/N$<3$ on any of their \nii-BPT emission lines, and white spaxels are masked by the {\fontfamily{cmtt}\selectfont DRP}.
    The black circles represent the inner ($r<1$ kpc, solid) and outer ($r>1.5$ kpc, dashed) radii used in the selection algorithm (see Section \ref{subsec:detection}).
    \textit{Right panel:} the corresponding SDSS $gri$ image. 
    In the left and middle panel, red points denote Seyfert emission, magenta points are classified as LINERs, green shows composite emission, and cyan shows star-formation. 
    The black and magenta hexagons represent the MaNGA IFU fiber bundle.}
    \label{fig:bpt-3panel}
\end{figure*}

\subsection{Fractional AGN Model} \label{subsec:fractional-agn}

The \nii Baldwin-Phillips-Terlevich (BPT; \citealt{Baldwin81}) diagram is commonly used in optical spectroscopic studies to classify a galaxy's main source of ionization based on their
\nii$\lambda6584$/\ha and \oiii$\lambda5007$/\hb emission line ratios. 
It is a powerful diagnostic to separate excitation mechanisms (i.e., star-formation, AGN), but there is clear evidence of a ``mixing sequence" within individual galaxies \citep{Kewley06, Davies14a, Davies14b, Dagostino18, Dagostino19, Sanchez20, Alban23, Riffel23}. 
The mixing sequence in IFU data is characterized by AGN spaxels spreading out towards the star-forming region of the BPT with an increasing radius from the center of the galaxy; this illustrates an ionization continuum within galaxies. 
Therefore, it is crucial to decompose the source of emission into its component parts: a fractional AGN contribution and a fractional star-formation contribution. 

To do this, we construct an empirical model similar to \citet{Kauffmann09}, \citet{Wild10}, \citet{Trump15}, and \citet{Jones16}, but vastly expanded to cover the majority of \nii-BPT space ($-2.25~<$~\niiha~$<~0.725$ and $-1.25~<$~\oiiihb~$<~1.325$).
We use the 2D distribution of integrated line flux star-forming galaxies (SFGs) and AGN on the \nii-BPT diagram to define the model parameter space. 
These data come from the SDSS extended Baryon Oscillation Spectroscopic Survey \citep[eBOSS;][]{Dawson16} DR17 \citep{Abdurrouf22} using the spectroscopic catalog of \citet{MatthewsAcuna25}; we select galaxies at $0.05<z<0.55$ to ensure coverage of \nii~$\lambda6584$ and \ha.
We select SFGs (i.e., below the \citealt{Kauffmann03} line) and AGN (i.e., above the \citealt{Kewley01} line) from the 2D distribution of galaxies and define them to be 0\% AGN and 100\% AGN, respectively. 
At a single SDSS SFG location on the BPT diagram we assume a Balmer ratio associated with case B recombination (H$\alpha$/H$\beta = 2.86$ for a $T=10^4 \text{ K}$ and $n_e=10^2 \text{ cm}^{-3}$ cloud; e.g., \citealt{Osterbrock06}) and continually add a fractional AGN component ($0<f_{\text{AGN}}<1$, in Table \ref{tab:equations}) until the evolutionary track matches that of an observed SDSS AGN. 
Note that we assume a Balmer ratio of 3.1 when determining the AGN line ratios to account for collisional excitation \citep{Osterbrock06}.
By solving the system of equations given in Table \ref{tab:equations} for each emission line on the BPT diagram, one can determine the fractional AGN contribution for any BPT location given a SFG and AGN sequence.
We repeat this process $\sim$25,000 times to create a grid covering the vast majority of BPT space (Figure \ref{fig:frac_agn}). 
The model is available on GitHub\footnote{\url{https://github.com/quinn-casey/f_agn}} and preserved on Zenodo at doi:\href{https://doi.org/10.5281/zenodo.23395801}{10.5281/zenodo.23395801}.
Note that the model does not distinguish between low-ionization nuclear emission-line regions (LINERs; \citealt{Heckman80}) and Seyfert emission; we assume they both originate from an AGN. 
This assumption is imperfect for certain use cases, but we address the potential contamination from LINER spectra in Section \ref{subsec:post-agb}.

\begin{figure*}
    \centering
    \includegraphics[width=\linewidth]{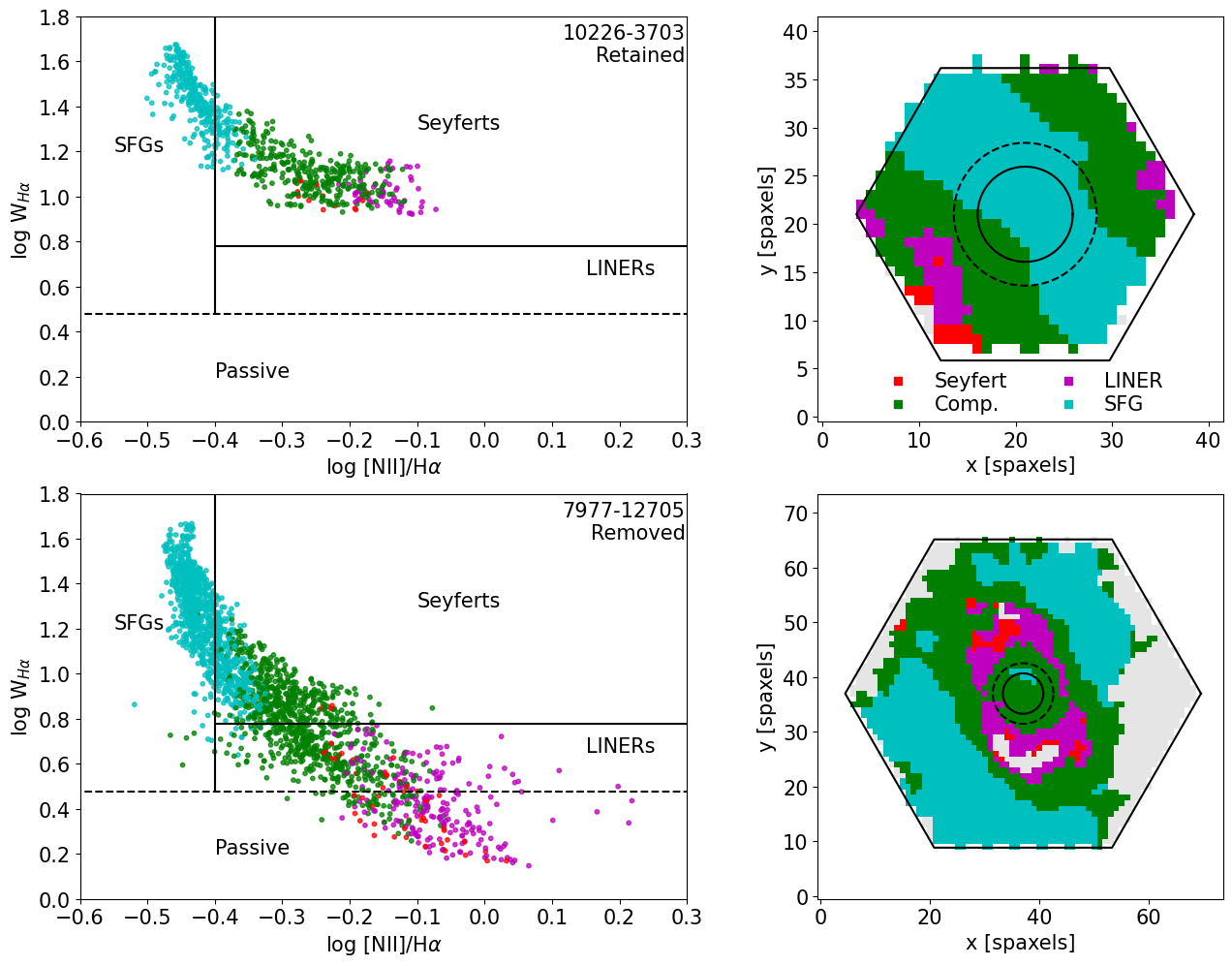}
    \caption{\textit{Left:} WHAN diagrams for candidate fading-AGN 10226-3703 (top) and 7977-12705 (bottom). 
    Relevant regions are labeled, and the passive region is dominated by emission from PAGB stars.
    \textit{Right:} the corresponding spatial \nii-BPT diagrams for the two candidates. 
    In each panel the spaxels are color-coded by their classification on the \nii-BPT diagram: cyan spaxels are star-forming, green are composite, magenta are LINERs, and red are Seyferts.
    Gray spaxels have S/N$<3$ on any of their \nii-BPT emission lines, and white spaxels are masked by the {\fontfamily{cmtt}\selectfont DRP}.
    The black circles represent the inner ($r=1$ kpc, solid) and outer ($r=1.5$ kpc, dashed) radii used in the selection algorithm.
    Galaxy 10226-3703 passes the cut described in Section \ref{subsec:post-agb} and is retained in our sample, but 7977-12705 does not pass the cut as only $\sim$36\% of the Seyfert/LINER spaxels reside in the Seyfert/LINER region of the WHAN diagram.}
    \label{fig:whan}
\end{figure*}

\subsection{Detecting Fading-AGN in MaNGA}\label{subsec:detection}
We use the model in Section \ref{subsec:fractional-agn} to calculate the fractional AGN component of each spaxel for all MaNGA galaxies.
%Many of the evolutionary tracks in Figure \ref{fig:frac_agn} overlap with each other, so a given spaxel's fractional AGN component is the mean $f_{\text{AGN}}$ value of all data points which lie within a radius of 0.075 dex.
Many of the evolutionary tracks in Figure \ref{fig:frac_agn} overlap with each other, so we calculate four fractional AGN contributions for each spaxel by taking the mean $f_{\text{AGN}}$ value of all data points which lie within a radius of $r<0.025, 0.05, 0.075,$ and $0.1$ dex on the \nii-BPT diagram.
With this method we can quantify the contribution of the AGN in various spatial regions. 
To select candidate fading-AGN, we calculate the fractional AGN contribution (using the four different radii on the \nii-BPT diagram) within 1 kpc (\fin) of the center of the IFU bundle (and thus the photometric center, see Section \ref{subsec:disc-location-assumptions}), and the fractional AGN contribution at $r>1.5$ kpc (\fout, using only Seyfert and LINER spaxels). 
This ensures that we select sources with a non-zero number of AGN spaxels, and reject purely star-forming galaxies. 
For any of the four radii, if $\text{\fin}+\sigma_{\text{\fin}} < \text{\fout}-\sigma_{\text{\fout}}$, where $\sigma_f$ is the standard deviation of fractional AGN contribution, then we consider the system to be a candidate fading-AGN. 
We adopt the {\fontfamily{cmtt}\selectfont DRP} masks, which exclude spaxels affected by broken fibers, foreground star contamination, detector artifacts, or those laying outside the IFU field of view \citep{Westfall19}.
%Additionally, we mask spaxels where the \nii, \ha, \oiii, and \hb flux measurements fall below $\text{S/N}>3$.
Additionally, we mask spaxels with $\text{S/N}<3$ on the \nii, \ha, \oiii, and \hb emission lines.
Out of the 10,735 MaNGA datacubes (10,010 unique sources), we find 108 ($r=0.025$ dex), 119 ($r=0.05$ dex), 118 ($r=0.075$ dex), and 122 ($r=0.1$ dex) candidate fading-AGN within each radius, with 131 unique sources. 
%We find 101 candidate fading-AGN out of the 10,735 MaNGA datacubes (10,010 unique sources) with this method. 
An example fading-AGN selected using this methodology is shown in Figure \ref{fig:bpt-3panel}.

We require $>15\%$ completeness in the IFU of the \nii BPT diagram \citep[e.g.,][]{Wylezalek18} to ensure that we can visually identify the different emission regions. 
We define completeness as the number of BPT spaxels with S/N $>3$ divided by the total number of available spaxels in the IFU after accounting for the {\fontfamily{cmtt}\selectfont DRP} masks. 
This cut removes 35 candidate fading-AGN. 
%; combined with the cut on AGN from \citet{Comerford24} we are left with 63 candidate fading-AGN. 

Fourteen galaxies in our sample of 96 are included in the \citet{Comerford24} catalog, which selects AGN using multi-wavelength observations from the final MaNGA sample.
Two of the candidates in our sample are selected as AGN in \citet{Comerford24} using observations of ultrahard X-rays from the Swift observatory's Burst Alert Telescope \citep{Barthelmy05}. 
Five are classified as type I AGN based on broad emission, which our selection algorithm does not account for.
Eleven are selected as AGN based on their \textit{Wide-field Infrared Survey Explorer} (\textit{WISE}; \citealt{Wright10}) mid-infrared (MIR) colors.
Two are classified as radio AGN 
%(mean $L_R\approx10^{45} \text{ erg s}^{-1}$) 
using the NRAO Very Large Array Sky Survey (NVSS; \citealt{Condon98}) and Faint Images of the Radio Sky at Twenty centimeters (FIRST; \citealt{Becker95}).
%; all of which are subclassified as low-excitation radio galaxies. 
Five galaxies are selected as AGN using two or more of the diagnostics from \citet{Comerford24}.
We match our sample to the \citet{Negus24} catalog of broad line sources in MaNGA and find nine matches.
We can remove candidate fading-AGN matched to the \citet{Comerford24} and \citet{Negus24} catalogs from our sample as these AGN indicators likely originate from the galaxies' nuclei and therefore indicate an AGN which is currently ``on." 
We are left with 73 candidates.

Upon visual inspection, three systems likely host off-center AGN merging with the central galaxy (Plate-IFUs: 11759-12705, 9504-9102, and 8157-3702). 
We identify these three systems by comparing their spatial BPT diagrams to the sources within the IFU.
The off-center AGN are within the IFU bundle and their spatial BPT diagrams show AGN classifications coincident with a stellar core (note that \citealt{Steffen23} identified 9504-9102 as a dual AGN system).
Therefore, these systems are incorrectly identified as fading-AGN and can be removed from our sample.
Additionally, we manually remove three interlopers, which have been designated as AGN in the literature: 11830-1902, 8626-12702, and 8274-9102.
11830-1902 is narrowly selected by our algorithm, but its central spaxels are squarely classified as Seyfert on the \nii, \sii, and \oi BPT diagrams; \citet{Pai24} classified this system and 8626-12702 as AGN based on their MIR variability. 
8274-9102\footnote{This galaxy was observed twice; the other Plate-IFU is 8256-9102} shows extended Seyfert emission which is expected for a fading-AGN, but this system was previously classified as a Wolf-Rayet galaxy by \citet{Liang20}. 
Photoionization models are known to underestimate the \oiiihb ratio for low-metallicity \hii regions \citep{Dopita06} where Wolf-Rayet stars reside and thus Wolf-Rayet emission can be misclassified as AGN \citep[e.g.,][]{Brinchmann08, Wylezalek18}.
To be conservative, we exclude this system from our analysis and are left with 67 sources. 
%Lastly, we exclude 8442-1901 which \citet{Mezcua20} identified as a dwarf AGN on account of broad \ha emission in the stacked spectra. 

\subsection{Post-AGB Emission} \label{subsec:post-agb}
The BPT diagram is a useful tool for selecting different sources of emission; however, it can fail in certain regimes. 
Nuclear LINER emission on the BPT diagram is often interpreted as being produced by low-luminosity AGN \citep{Ho93}; however, emission from post asymptotic giant branch (PAGB) stars can place non-AGN in the LINER region of the BPT \citep{Fernandes10}.
Diffuse gas ionized by these evolved stars can enhance BPT emission line ratios and move \hii regions towards the composite/LINER regions \citep{Zhang17, Lugo24}. 

The WHAN diagram \citep{Fernandes11} uses the \ha equivalent width ($W_{\text{H}\alpha}$) along with the ratio of \niiha to differentiate between AGN emission and emission from PAGB stars. 
We use the WHAN diagram (with $\text{S/N}>3$) to make an additional cut on our sample; we require that $>50\%$ of the BPT-selected Seyfert/LINER spaxels fall within the Seyfert/LINER region of the WHAN diagram ($\log \text{\nii}/\text{H}\alpha > -0.4$ and $W_{\text{H}\alpha}>3\text{\angstrom}$) for a system to remain in our sample -- 33 systems do not pass this cut and are removed.
Jaison et al. (submitted) show that high-metallicity regions of diffuse ionized gas (DIG) can mimic AGN emission on the WHAN diagram for IFU datasets and suggest increasing the boundary of \niiha.
If we apply their criteria ($\log \text{\nii}/\text{H}\alpha > -0.25$ and $W_{\text{H}\alpha}>3\text{\angstrom}$) to our candidate fading-AGN, only one additional source would be removed (8453-1901). 
This suggests DIG-regions do not heavily impact our sample; we choose to retain 8453-1901 in our sample as it is only narrowly selected by the Jaison et al. (submitted) criteria.
Figure \ref{fig:whan} shows two candidate fading-AGN; one which passes our cut when using the \citet{Fernandes11} criteria (10226-3703, top panel) and one which does not (7977-12705, bottom panel) and is excluded from our analysis.
After removing the 33 sources which do not pass our WHAN criteria, we are left with 34 candidate fading-AGN.

\begin{figure}
    \centering
    %\vspace{-0.1in}
    \includegraphics[width=\linewidth,trim={0 0 0 1in},clip]{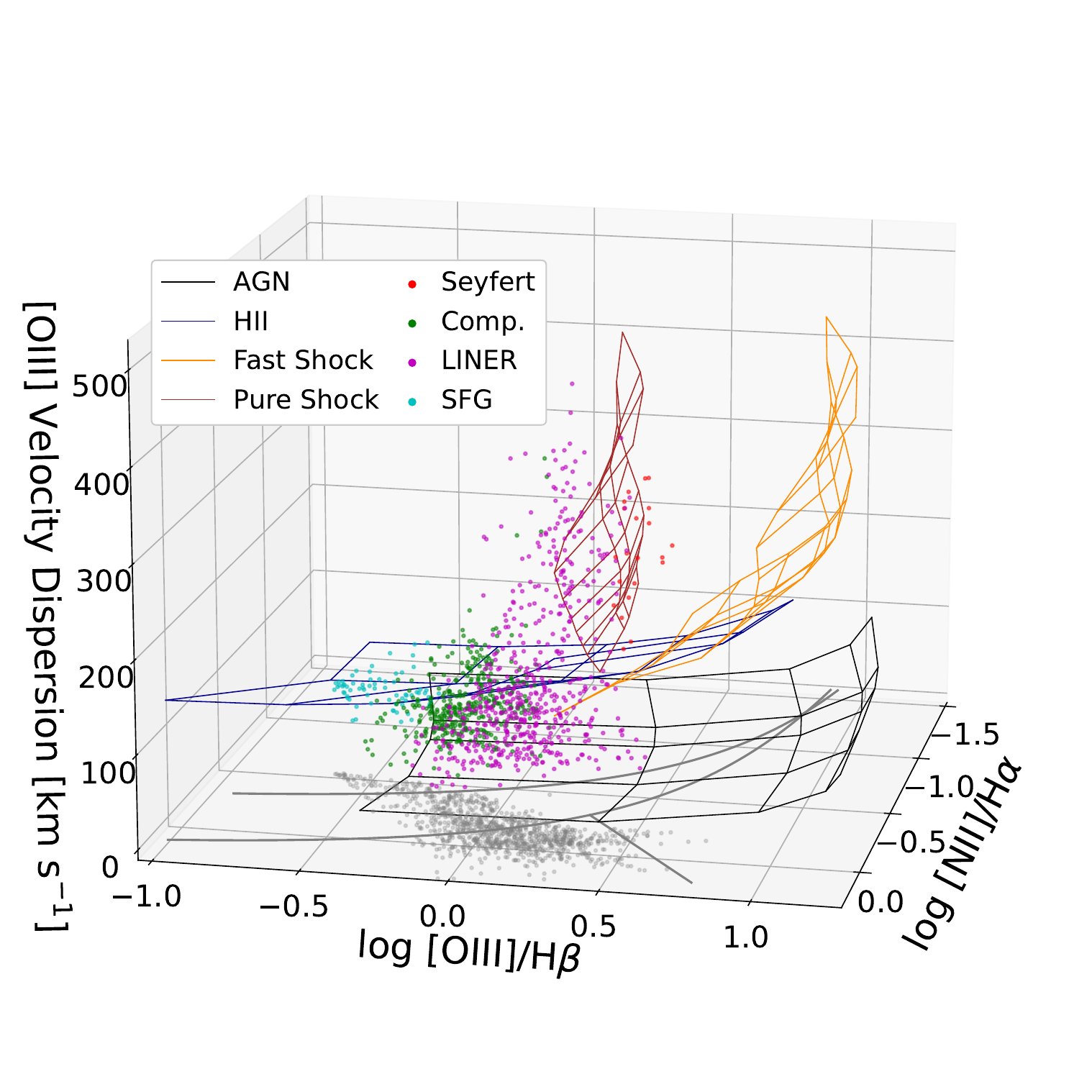}
    \vspace{-0.4in}
    \caption{3D shock diagnostic diagram from \citet{Zhu25}; the grids represent photoionization models varying metallicity, ionization parameter, and shock velocity. 
    The grids were tailored for NGC 5728, but can be used in a general case.
    We show MaNGA galaxy 9507-12704 as an example system not in our sample that exhibits ``pure shock" spaxels, where the pure shock grid traces post-shock emission.
    The spaxel color scheme is the same as Figures \ref{fig:bpt-3panel} and \ref{fig:whan}; we show a reflection of the spaxels and the BPT demarcation lines at the bottom in gray.}
    \label{fig:shock}
\end{figure}

\begin{table*}
\caption{Properties of the candidate fading-AGN. $z$ is the redshift and $\log M_*$ is the Sersic stellar mass, both of which are from the NASA-Sloan Atlas. $t_{\text{faded}}$ is the mean light travel time from the IFU center to the Seyfert/LINER spaxels. The 3\arcsec\ classification column represents how the fading-AGN are classified in the \nii, \sii, and \oi-BPT diagrams using the integrated line flux measurements from SDSS DR12. }
    %\centering
    \begin{tabular}{c c c c c c c}
    \hline \hline
    SDSS Name & MaNGA & $z$ & $\log M_*$ & Spaxel Scale  & $t_{\text{faded}}$ & \nii-, \sii-, and \oi-BPT \\
     & Plate-IFU & & ($M_{\odot}$) & (pc) & (lightyears) & 3\arcsec\ Classification  \\
    \hline
    J093846.69+483350.9 & 8252-9101  & 0.0248  & 9.77  & 249.6  & 1.0$\times10^{4}$  &  SFG/SFG/SFG  \\ 
    J141349.19+371608.8 & 8337-6104  & 0.0222  & 10.15  & 224.3  & 1.0$\times10^{4}$  &  SFG/SFG/SFG  \\ 
    J125855.94+275000.1 & 8950-12705  & 0.0251  & 10.25  & 252.8  & 6.5$\times10^{3}$  &  Comp/SFG/SFG  \\ 
    J131019.89+470732.0 & 8465-9101  & 0.0244  & 9.71  & 245.8  & 8.7$\times10^{3}$  &  Comp/SFG/LINER  \\ 
    J134143.70+554026.0 & 11020-6103  & 0.0253  & 10.26  & 254.6  & 1.0$\times10^{4}$  &  --  \\ 
    J080115.64+325056.0 & 9484-6101  & 0.0179  & 9.92  & 181.8  & 9.7$\times10^{3}$  &  SFG/SFG/SFG  \\ 
    J020748.72+133634.5 & 7993-9101  & 0.0650  & 10.67  & 623.9  & 1.8$\times10^{4}$  &  Comp/SFG/SFG  \\ 
    J162141.57+403527.9 & 12673-6101  & 0.0312  & 10.04  & 311.9  & 1.1$\times10^{4}$  &  Comp/SFG/SFG  \\ 
    J165726.74+404408.2 & 11945-6104  & 0.0290  & 10.48  & 290.5  & 1.3$\times10^{4}$  &  Comp/SFG/SFG  \\ 
    J082922.09+432129.0 & 11745-12702  & 0.0311  & 10.27  & 311.1  & 1.3$\times10^{4}$  &  SFG/SFG/SFG  \\ 
    J153207.84+490729.4 & 8485-6103  & 0.0250  & 9.85  & 251.7  & 1.2$\times10^{4}$  &  SFG/SFG/SFG  \\ 
    J115626.06+034600.4 & 12489-12705  & 0.0224  & 9.86  & 226.6  & 1.1$\times10^{4}$  &  Comp/SFG/SFG  \\ 
    J170408.72+312902.3 & 7959-3701  & 0.0330  & 10.47  & 329.1  & 1.1$\times10^{4}$  &  Comp/SFG/SFG  \\ 
    J160614.83+181058.3 & 11976-3701  & 0.0394  & 10.73  & 390.0  & 1.6$\times10^{4}$  &  Comp/SFG/SFG  \\ 
    J093818.39+040538.9 & 10844-1901  & 0.0230  & 10.12  & 232.1  & 6.3$\times10^{3}$  &  Comp/SFG/SFG  \\ 
    J162408.61+305708.9 & 9027-3701  & 0.0603  & 10.42  & 582.6  & 2.5$\times10^{4}$  &  Comp/SFG/SFG  \\ 
    J122015.71+333938.9 & 12483-12702  & 0.0216  & 10.14  & 218.3  & 1.4$\times10^{4}$  &  SFG/SFG/SFG  \\ 
    J214702.66-082628.1 & 12684-12701  & 0.0618  & 10.60  & 596.0  & 2.1$\times10^{4}$  &  Comp/SFG/SFG  \\ 
    J113859.53+494641.1 & 8990-12704  & 0.0518  & 10.34  & 505.6  & 1.5$\times10^{4}$  &  Comp/SFG/LINER  \\ 
    J023144.51-005643.8 & 10226-3703  & 0.0200  & 9.87  & 202.7  & 8.6$\times10^{3}$  &  SFG/SFG/SFG  \\ 
    J092138.71+434334.2 & 8250-6104  & 0.0399  & 10.38  & 394.5  & 1.8$\times10^{4}$  &  SFG/SFG/SFG  \\ 
    J130738.44+312913.0 & 12484-1901  & 0.0262  & 9.72  & 263.0  & 6.1$\times10^{3}$  &  Comp/SFG/SFG  \\ 
    J104101.22+003757.5 & 10837-1902  & 0.0222  & 9.49  & 224.3  & 6.6$\times10^{3}$  &  --  \\ 
    J093649.40+351855.7 & 8152-1901  & 0.0223  & 9.57  & 225.7  & 7.4$\times10^{3}$  &  SFG/SFG/SFG  \\ 
    J101327.72+473058.4 & 8453-1901  & 0.0253  & 9.68  & 255.1  & 7.2$\times10^{3}$  &  SFG/SFG/SFG  \\ 
    J083956.11+553402.2 & 8723-1902  & 0.0253  & 9.64  & 254.8  & 6.3$\times10^{3}$  &  Comp/SFG/SFG  \\ 
    J130107.37+544757.1 & 11863-9102  & 0.0314  & 10.48  & 313.6  & 1.2$\times10^{4}$  &  Comp/SFG/SFG  \\ 
    J130302.78+494042.9 & 11018-6101  & 0.0339  & 10.28  & 337.7  & 1.3$\times10^{4}$  &  Comp/SFG/--  \\ 
    J083724.90+245247.9 & 9502-12704  & 0.0295  & 10.11  & 296.0  & 9.7$\times10^{3}$  &  --  \\ 
    J031105.31-003247.4 & 8081-9101  & 0.0282  & 10.22  & 283.0  & 1.2$\times10^{4}$  &  Comp/SFG/SFG  \\ 
    J134743.83+035540.3 & 11958-3703  & 0.0244  & 9.39  & 245.7  & 7.7$\times10^{3}$  &  SFG/SFG/SFG  \\ 
    J170845.72+620236.1 & 8610-12703  & 0.0262  & 10.32  & 263.1  & 1.1$\times10^{4}$  &  Comp/--/SFG  \\ 
    J094545.59-005839.0 & 12512-3701  & 0.0300  & 9.75  & 300.0  & 9.1$\times10^{3}$  &  --  \\ 
    J162857.99+391908.7 & 9869-12702  & 0.0339  & 10.44  & 338.0  & 1.5$\times10^{4}$  &  Comp/SFG/SFG  \\
    \hline
    \end{tabular}
    \label{tab:properties}
\end{table*}

\subsection{Shock Ionization}\label{subsec:shocks}
Seyfert galaxies and shock-excited regions both reside at large \oiiihb ratios on the BPT diagram, and shocks can excite gas-rich non-Seyfert regions such that the BPT classification falsely identifies a Seyfert galaxy \citep[e.g.,][]{Rich11, Dagostino19, Benedetti24}. 
Furthermore, extended LINER emission ($\gtrsim1$ kpc) may be the result of fast shocks (via galaxy-scale outflows) in gas-poor environments \citep{Dopita95, Ho14, Kewley19}.
For these reasons, it is important to investigate the possibility of an increased fractional AGN component due to shocks in our candidate fading-AGN.

%We make use of the three-dimensional diagnostic presented in \citet{Dagostino19}. 
%This tool leverages the ionization information of the BPT diagram in the ``emission line ratio function" (ELR, Equation \ref{eq:elr}). 
%The ELR plotted against the velocity dispersion and the radius can be used to identify shock regions in IFU spectroscopy.
%At a low ELR and low velocity dispersion star-formation will dominate. 
%At low ($\sim0$ kpc) radii, high ELRs, and high velocity dispersions, AGN will dominate.
%At non-zero radii ($\sim$few kpc), high ELRs, and high velocity dispersions, shocks will dominate \citep{Kewley19}.
%Figure \ref{fig:shock} shows one of our candidate galaxies (panel \textit{a}) and a non-candidate galaxy (panel \textit{b}). 
%The non-candidate galaxy features a number of spaxels residing in the ``starburst-shock" region of this 3D space. 
%This indicates that shocks may play a major role in the BPT classification of those spaxels. 
%The fast majority of the Seyfert spaxels in the candidate fading-AGN reside at relatively low velocity dispersion and large ELR along the ``starburst-AGN" sequence.
%We find no conclusive evidence that any of our fading-AGN exhibit shock-like features.
%This leads us to believe that the Seyfert emission can only be explained by prior AGN activity.

We make use of the 3D diagnostic presented in \citet{Zhu25}, which leverages the standard BPT selection for determining the source of ionization, as well as the \oiii$\lambda5007$ velocity dispersion on the third axis to understand the kinematics of the gas. 
Narrow-line AGN will have low nuclear \oiii velocity dispersions in the absence of shocks.
Shock-ionized gas will look like Seyfert/LINER emission on the BPT and will reside at large velocity dispersions.
We use the \hii, AGN, fast shock, and pure shock model grids from \citet{Zhu25} and compare our candidate fading-AGN to them. 
We find no systems which are shock-dominated to remove from our sample. 
In Figure \ref{fig:shock} we show an example MaNGA galaxy (9507-12704) that exhibits shocks for reference, but this system is not in our sample.
Some systems in our sample host a handful of Seyfert/LINER spaxels which reside in the shock regions of Figure \ref{fig:shock}, but the majority of the Seyfert/LINER spaxels have low velocity dispersions for such systems.

%We can further investigate the physics of the interstellar medium (ISM) by looking at the ionization parameter, $U$.
%$U$ is a measure of the ratio between the local ionizing photon flux and the local density of hydrogen atoms. 
%This is known to be uniform across standard star-forming galaxies \citep{Moustakas10, Poetrodjojo18}, but the ionization parameter varies much more in AGN dominated galaxies \citep{Moy02}. 
%On average, AGN tend to produce a larger value of $U$ compared to typical star-forming galaxies because AGN have a harder ionizing continuum \citep{Ho93}. 
%$U$ is highly dependent on the metallicity of a system and IFU data is not sensitive enough to yield accurate spatially resolved metallicity estimates. 
%Therefore, we opt to investigate the ionization parameter by using the ratio of \oiii$\lambda5007$ to \oii$\lambda\lambda3727, 29$ as a proxy. 
%This ratio details the ionization state of a systems because of the difference in ionization potentials between the two atomic species. 
%We correct both emission lines for dust using the \citet{Salim18} reddening law and, due to the small number of Seyfert spaxels in some cases, group Seyfert/LINER spaxels together and compare their mean \oiiioii ratio to the star-forming/composite spaxels. 
%These values are presented in Table \ref{tab:properties}.
%With the exception of one galaxy (Plate-IFU 9514-1901), the \oiiioii ratio is higher on average for the Seyfert and LINER spaxels which is highly indicative of ionization dominated by AGN activity.

\section{Properties of the Fading-AGN Candidates} \label{sec:properties}

\begin{figure*}[btp]
    \centering
    \includegraphics[width=\linewidth]{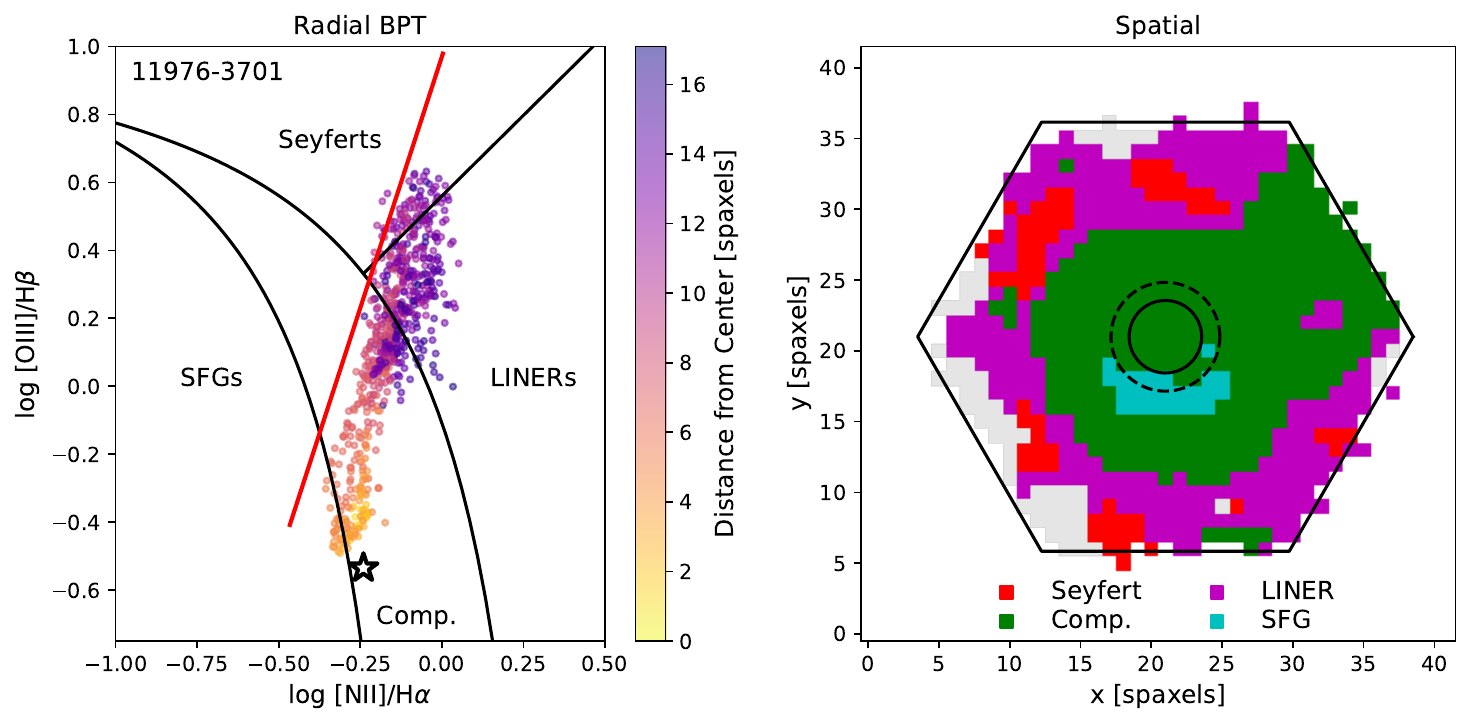}
    \caption{The candidate fading-AGN 11976-3701. 
    \textit{Left:} Radial BPT diagram where points are color-coded by their distance from the IFU center. 
    The black star represents the BPT location of this system as determined by the 3\arcsec\ SDSS single-fiber observation.
    The black lines represent the demarcation lines from \citet{Kewley01}, \citet{Kauffmann03}, and \citet{Law21b}. 
    The red line is the \hii-AGN mixing-sequence from \citet{Thomas18}. 
    \textit{Right:} The spatial distribution of the different types of emission as determined by the \nii-BPT diagram; cyan is classified as star-forming, green as composite, magenta as LINER, and red as Seyfert emission. 
    Gray spaxels have S/N$<3$ on any of their \nii-BPT emission lines, and white spaxels are masked by the {\fontfamily{cmtt}\selectfont DRP}.
    The black circles represent the inner ($r=1$ kpc, solid) and outer ($r=1.5$ kpc, dashed) radii used in the selection algorithm.
    The black hexagon represents the MaNGA fiber bundle. 
    The radial emission profiles of the fading-AGN go in the opposite direction of typical AGN \citep[c.f.,][]{Davies14a, Davies14b}.}
    \label{fig:radial-bpt}
\end{figure*}

\subsection{Comparing to Single Fiber Observations} \label{subsec:sdss}
Our final sample contains 34 candidate fading-AGN; some of their properties are listed in Table \ref{tab:properties}, and we show diagnostic figures for each source in the Appendix.
28 of the candidate fading-AGN were targeted in the primary MaNGA sample, three in the secondary secondary sample, and two were acillary targets. 
We are afforded a more complete picture of the ionization mechanisms with the MaNGA IFU data compared to the traditional single fiber observations employed in other SDSS surveys. 
However, it is useful to understand how the fading-AGN systems would be classified in the single fiber observations as aperture size plays an important role in classifying galaxies on BPT diagrams \citep[e.g.,][]{Alban23}. 
We cross-match the fading-AGN to the SDSS DR12 emission line catalog provided by the Portsmouth Group\footnote{\url{https://www.sdss4.org/dr12/spectro/galaxy_portsmouth/}} and find 28 matches. 
One source which wasn't matched to the DR12 catalog has measurements in the \citet{MatthewsAcuna25} eBOSS catalog which we adopt; the remaining four sources do not have SDSS single-fiber observations.
We report the \nii/\sii/\oi-BPT classifications (with $\text{S/N}>3$) as determined from the single-fiber observations in Table \ref{tab:properties}; we require a 3$\sigma$ detection on each emission line.
11/29 are classified as SFGs and the remaining are classified as composite on the \nii-BPT, 28/28 are SFGs on the \sii-BPT, and 26/28 are SFGs on the \oi-BPT.

%\citet{Alban23} showed that the aperture size plays an important role when classifying galaxies on BPT diagrams. 
%For aperture sizes up to $\sim6$ kpc, the number of galaxies classified as AGN decreases, but this trend reverses at apertures larger than $\sim6$ kpc. 
%The increase in the number of AGN selected when using larger apertures suggests that regions of diffuse ionized gas (DIG) can contaminate the AGN selection. 
%DIG-dominated regions can shift the BPT emission line ratios from the SFG region to the composite/LINER region of the BPT diagram \citep[e.g.,][]{Alban23, Lugo24}.
%This is likely not be an issue for our sample as DIG-dominated regions should be excluded through our use of the WHAN diagram in Section \ref{subsec:post-agb} \citep[e.g.,][]{Zhang17}.

On the \nii-BPT diagram the 3\arcsec\ classification and the central MaNGA spaxels of the fading-AGN primarily reside at low \oiiihb ratios relative to SFGs/composites in the full SDSS DR12 sample (e.g., Figure \ref{fig:radial-bpt}).
The fading-AGN share a common radial emission pattern where the central spaxels sit at low \oiiihb and low \niiha.
The spaxels then tend to spread out to larger \oiiihb and \niiha with increased radius, moving into the LINER/Seyfert region of the BPT. 
This radial sequence moves in the opposite direction of the mixing-sequence observed in typical AGN where nuclear AGN spaxels spread out into the star forming region of the BPT with increasing radius \citep{Davies14a, Davies14b}.
Figure \ref{fig:radial-bpt} shows a radial BPT diagram for the 11976-3701 system where points are colored by their distance from the center.
The red line in the left panel of Figure \ref{fig:radial-bpt} is the mixing sequence from \citet{Thomas18}, which was calibrated from SDSS DR7 observations. 
The majority of the fading-AGN follow this same SFG to AGN radial emission pattern where the sequence is slightly negatively offset from the \citet{Thomas18} mixing-sequence line.

%\subsection{Luminosity} \label{subsec:lumin}
\subsection{Tri-Color Images} \label{subsec:tri-color}
\begin{figure}
    \centering
    \includegraphics[width=\linewidth]{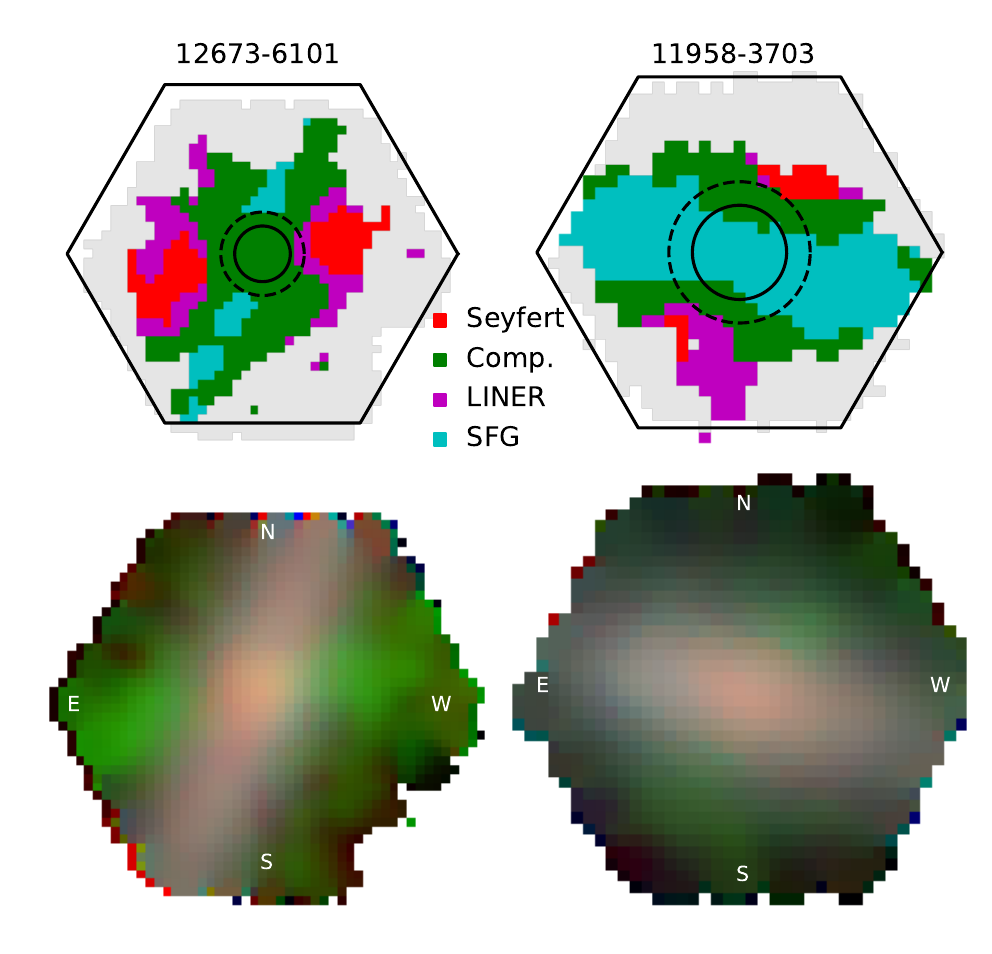}
    \caption{Spatial BPT diagrams and three-color images of two fading-AGN in our sample: 12673-6101 (left) and 11958-3703 (right). 
    The color scheme in the spatial BPT (top panel) is the same as Figure \ref{fig:bpt-3panel}, where red points denote Seyfert emission, magenta is LINER, green is composite, and cyan is star-formation.
    Gray spaxels have S/N$<3$ on any of their \nii-BPT emission lines, and white spaxels are masked by the {\fontfamily{cmtt}\selectfont DRP}.
    The black circles represent the inner ($r=1$ kpc, solid) and outer ($r=1.5$ kpc, dashed) radii used in the selection algorithm.
    In the RGB image (bottom panel) the stellar disks are clearly visible in red (which corresponds to star-formation/composite in the spatial BPT), and the EELRs/LEs are shown in green, which corresponds to emission from \oiii$\lambda5007$. 
    In both cases, the \oiii emission is perpendicular to the disk.
    %In the bottom left figure the EELR surrounds the majority of the disk, whereas in the bottom right figure the EELR is confined to a much smaller region NW and SE of the nucleus.
    }
    \label{fig:tricolor}
\end{figure}

We can visualize the signatures of the fading-AGN by constructing three-color images in the same fashion as \citet{French23}. 
We select two line-free continuum regions to represent the blue and red colors of the RGB image, where blue is $4500 < \lambda < 4700\text{\AA}$ and red is $7300 < \lambda < 7500\text{\AA}$. 
The spectral region directly surrounding \oiii$\lambda5007$ represents the green portion of the RGB image (i.e., $4997 < \lambda < 5017\text{\AA}$). 
This method allows us to view faint EELRs that would be otherwise undetectable in the photometry. 
EELRs are clearly visible in $\gtrsim60\%$ of our sample and appear more prominently in edge-on systems where the stellar continuum is low.
Two example fading-AGN and their three-color images are shown in Figure \ref{fig:tricolor}; green regions represent strong \oiii emission relative to the continuum (three-color images for each source are shown in the Appendix).
Very few of our systems show powerful \oiii signatures in the nuclei, which further suggests low levels of accretion onto the SMBH \citep[c.f.,][]{French23}. 

% (IF WE ARE SHOWING WISE LUMINOSITY-COLOR)
\subsection{Infrared Emission}
\begin{figure}
    \centering
    \includegraphics[width=\linewidth]{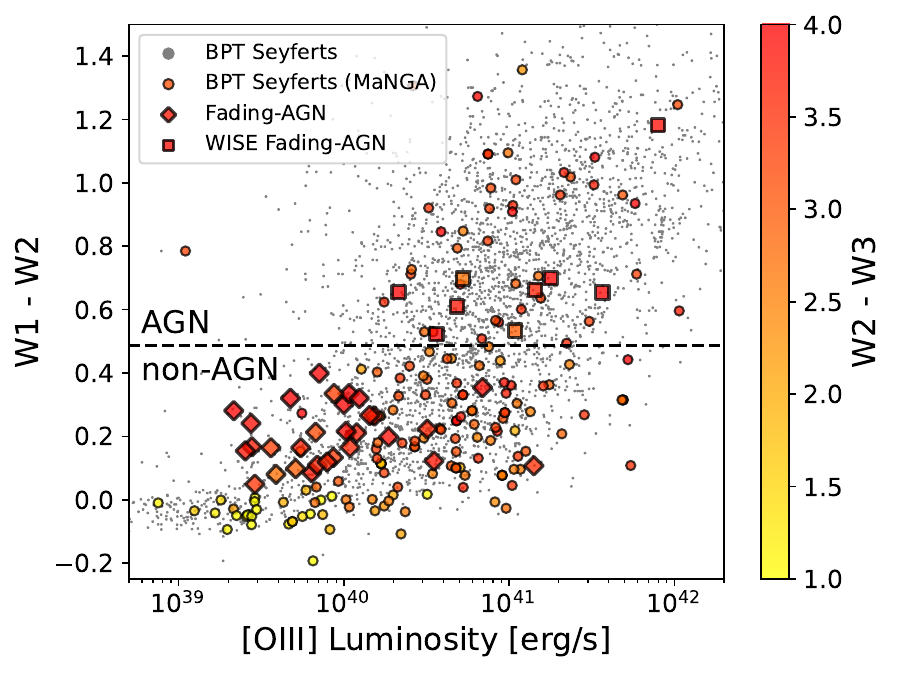}
    \caption{\oiii luminosity versus \textit{WISE} W1-W2 for a number of subsamples: gray points are BPT-selected Seyferts from SDSS DR12, circles are BPT-selected Seyferts that were observed in MaNGA, diamonds are our fading-AGN, and squares are fading-AGN that were originally selected in Section \ref{subsec:detection}, but excluded because they are \textit{WISE} AGN. 
    Circles, diamonds, and squares are colored by their W2-W3 color. 
    For plotting purposes, we only show sources brighter than 13.07 mag in W2 which excludes two fading-AGN (12484-1901 and 11958-3703), and two fading-AGN were not matched to \textit{WISE} (8453-1901 and 12512-3701). 
    The black dashed line is the 75\% reliability criteria from \citet{Assef18} used to select AGN.
    Our fading-AGN are primarily clustered at low \oiii luminosity and low W1-W2; their W2-W3 colors are more red compared to MaNGA AGN systems at the same location. }
    \label{fig:wise}
\end{figure}

\textit{WISE} IR color-color diagrams are frequently used to select luminous AGN that appear red in W1-W2 \citep[e.g.,][]{Stern12, Assef18}. 
We select BPT classified Seyferts from SDSS DR12 and cross-match these with the AllWISE catalog. 
We show \oiii$\lambda5007$ luminosity versus W1-W2 in Figure \ref{fig:wise}, where gray points are BPT selected Seyferts from DR12, circles are BPT selected Seyferts that were observed in MaNGA, our fading-AGN are diamonds, and \textit{WISE} AGN that were originally selected as candidate fading-AGN in Section \ref{subsec:detection} are shown as squares.
These are systems that we removed based on the \citet{Comerford24} AGN catalog; they were selected as AGN according to the \citet{Assef18} 75\% reliability criteria (i.e., the black dashed line in Figure \ref{fig:wise}). 
The \oiii luminosity is calculated by summing the \oiii flux of unmasked spaxels for systems with MaNGA observations; all other \oiii luminosities are calculated from the 3\arcsec\ fiber flux. 

Owing to our selection criteria, none of the fading-AGN are \textit{WISE} AGN; the \oiii luminosities of the fading-AGN are less than those of the \textit{WISE} fading-AGN by $\sim1$ dex. 
The fading-AGN are situated at low \oiii luminosities and blue W1-W2; their W2-W3 colors are more red compared to the other MaNGA Seyferts at the same location. 
The W3 and W4 luminosities of the fading-AGN are roughly equivalent, which is similar to what is observed in dust-obscured SFGs \citep[e.g.,][]{Hickox18}.
The IR serves as evidence that thermal emission from the torus in the fading-AGN is low, and therefore the nuclear activity is likely less than a typical AGN. 
However, we note that \textit{WISE} color cuts frequently miss dust obscured AGN and AGN with high levels of nuclear star-formation \citep[e.g.,][]{Hickox17, Goel26}.
This may explain the discrepancy between the \nii-BPT selected AGN and the \textit{WISE} selected AGN in Figure \ref{fig:wise}.

To further understand the torus emission of our sample, we quantify the fractional AGN contribution in the IR for our fading-AGN as well as the AGN from the \citet{Comerford24} catalog.
To do this, we perform simple multi-component spectral energy distribution (SED) fits to the \textit{WISE} W1 -- W4 data. 
We adopt the AGN template and host galaxy templates (elliptical, spiral, and irregular) from \citet{Assef10}, and fit for the fractional contribution of each component using {\fontfamily{cmtt}\selectfont scipy}'s NonNegative Least Squares ({\fontfamily{cmtt}\selectfont nnls}) routine.
The median reduced $\chi^2$ values for the SED fits are 1.1 and 0.54 for the fading-AGN and the \citet{Comerford24} AGN, respectively.

We find that the \citet{Comerford24} AGN have a larger AGN contribution in the IR ($f_{\rm{AGN-IR}}=0.27\pm0.02$, where the quoted uncertainty is the standard error on the mean) relative to the fading-AGN ($f_{\rm{AGN-IR}}=0.06\pm0.01$). 
Unsurprisingly, the \citet{Comerford24} sources that were selected as \textit{WISE} AGN show a much larger fractional AGN contribution ($N=123$, $f_{\rm{AGN-IR}}=0.59\pm0.03$) relative to the non-\textit{WISE} AGN ($N=249$, $f_{\rm{AGN-IR}}=0.12\pm0.01$).
However, if we compare the optical fractional AGN contribution as measured from the DR12 spectra (using the model presented in Section \ref{subsec:fractional-agn}), we find the \citet{Comerford24} AGN ($f_{\rm{AGN-[OIII]}}=0.58\pm0.01$) are much more similar to our fading-AGN ($f_{\rm{AGN-IR}}=0.45\pm0.01$).
On the timescales we are interested in, the torus luminosity traces the ``instantaneous" accretion rate, given that the torus cooling timescale is primarily governed by the light travel time \citep[e.g.,][]{Ichikawa17}.
The \oiii based fractional AGN contribution traces a much longer timescale given that the NLR is far more extended than the torus. 
The differences between the wavelength-dependent fractional AGN contribution support the idea that the fading-AGN are in a period of low accretion relative to their accretion rate in the past.

% (IF WE ARE SHOWING WISE COLOR-COLOR) 
%WISE mid-infrared color-color diagrams are frequently used to select luminous AGN which appear red in $W1-W2$ \citep[e.g.,][]{Stern12}. 
%We select BPT classified Seyferts from SDSS DR12 and cross-match these with the AllWISE catalog. 
%We show the WISE color-color diagram in Figure \ref{fig:wise} where gray points are BPT selected Seyferts, circles are BPT selected Seyferts which were observed in MaNGA, our fading-AGN are diamonds, and WISE AGN which were originally selected as candidate fading-AGN in Section \ref{subsec:detection} are shown as squares.
%The fading-AGN are clustered at high $W2-W3$ and low $W1-W2$.

%\begin{figure}
%    \centering
%    \includegraphics[width=\linewidth]{wise_color_color.pdf}
%    \caption{WISE color-color diagram for a number of subsamples: gray points are BPT-selected Seyferts from SDSS DR12, circles are BPT-selected Seyferts which were observed in MaNGA, diamonds are our fading-AGN, and squares are fading-AGN which were originally selected in Section \ref{sec:selection}, but excluded because they are WISE AGN. Circles, diamonds, and squares are colored by their \oiii luminosity which is calculated by summing the \oiii flux in the unmasked spaxels. Our fading-AGN are clustered at high $W2-W3$ and low $W1-W2$. }
%    \label{fig:wise}
%\end{figure}

%\subsection{Star Formation} \label{subsec:starformation}

\subsection{Morphology and Environment} \label{subsec:mergers}
%Approximately 21/33 of the fading-AGN are in edge-on galaxies, which is unsurprising given that NLRs (and EELRs) are more easily detectable in edge-on galaxies because stellar contamination is weaker. 
%This may influence our interpretation of these systems, see Section \ref{subsec:orientation} for more details. 
Visually, many of the fading-AGN candidates are in edge-on galaxies where dust obscuration and host galaxy contamination may play an important role; dust should not impact the BPT classifications (except in cases of extreme attenuation), as the emission-line ratios have small wavelength separations.
We cross-match the fading-AGN candidates with the MaNGA morphological catalog of \citet{Vazquez22} and find that $\sim 71\%$ (24/34) of sources are classified as edge-on, where edge-on is defined as an inclination $i>~70^{\circ}$. 
If we compare this to the AGN from \citet{Comerford24}, only $\sim6\%$ (24/387) are edge-on. 
Host galaxy inclination likely impacts the effectiveness of different AGN selection criteria \citep[e.g.,][]{Lagos11, Gkini21, Ji22}, so it is more reasonable to compare to \nii-BPT based classifications. 
%Using the single-fiber SDSS DR12 observations to classify galaxies on the \nii-BPT, we find that $\sim15\%$ 
We classify MaNGA sources on the \nii-BPT diagram using their single-fiber SDSS DR12 observations and find that $\sim15\%$ (41/266) of Seyferts are edge-on. 
This is compared to $28\%$, $24\%$, and $13\%$ of SFGs, composite galaxies, and LINERs which are classified as edge-on. 
Whether the excess of edge-on fading-AGN candidates in our sample is the result of heavy obscuration/stellar contamination, or the ease of finding such sources in edge-on systems is something we discuss in Section \ref{subsec:orientation}.

%We visually inspect the Legacy Survey \citep{Dey19} images of the fading-AGN for merger signatures and find at least six systems are interacting with a nearby massive system (8252-9101, 11020-6103, 11976-3701, 8250-6104, 8464-3702, 8081-9101). 
Given that most fading-AGN we know of reside in galaxies undergoing mergers \citep[e.g.,][]{Keel12, Keel17, Keel22, Keel24, Esparza20}, we seek to understand the environment of our sample.
We cross-match our fading-AGN candidates with the MaNGA Galaxy Merger Catalog of \citet{Comerford24} which is built from the \citet{Nevin23} SDSS merger catalog. 
%Out of the 33 candidate fading-AGN, four are classified as major mergers, seventeen as minor mergers, seven as non-mergers, and five have no classifications. 
There are 29 candidate fading-AGN matched to the merger catalog, and 22/29 ($\sim76\%$) are classified as mergers; 5/22 ($\sim23\%$) are designated as major mergers (where the mass ratio, $\mu$ is $\gtrsim$1:3) and 17/22 ($\sim77\%$) as minor mergers ($\mu \lesssim$1:5).
Seven sources are classified as non-mergers, and five have no classifications.
Of the 22 mergers, eleven are classified as post-coalescence (where the merger is defined to end 0.5 Gyr after coalescence), six are in the early merger stage, and five are in the late merger stage. 
We visually inspect the Legacy Survey \citep{Dey19} images of the sources which had no merger classification and find at least one source likely to be in a major merger (8252-9101, shown in Figure \ref{fig:bpt-3panel}), but do not speculate on the stage of the merger.
%Although we do not fit for signatures of previous mergers, 

We compare our merger rates to previous studies by compiling a sample of $N=61$ unique known fading-AGN from \citet{Keel12, Keel17, Keel22, Keel24}, \citet{Schirmer13}, \citet{Schweizer13}, \citet{Watkins18}, and \citet{Esparza20}. 
Out of this sample, 30 sources have measurements in the \citet{Nevin23} catalog; 80\% (24/30) are classified as mergers, where 62.5\% (15/24) are major mergers and 37.5\% (9/24) are minor mergers. 
While the overall merger fraction between our sample and the literature is similar, the majority
of our fading-AGN are not undergoing a major merger. 
This is in contrast with previous studies, but many previous searches intentionally targeted strongly interacting systems given that imaging signatures of previous AGN activity are more abundant \citep[e.g.,][]{Keel12, Keel17, Keel22, Keel24}.
%which is interesting since the majority of previously identified fading-AGN were discovered in merging systems given that imaging signatures of previous AGN activity are more abundant \citep[e.g.,][]{Keel12, Keel17, Keel22, Keel24}.
This suggests that we are either probing secular galaxy evolution, or minor mergers were effective in triggering AGN activity in many of our fading-AGN candidates.
Cosmological simulations have suggested that minor mergers (or mini mergers) have the ability to trigger AGN activity \citep[e.g.,][]{Capelo15, Byrne25} which may explain the abundance of low-luminosity AGN \citep{Yang19}. 
However, our sample may be subject to the same selection effects as previous studies (i.e., extended emission is more likely to be detected in interacting systems). 
%rather than a period of merger-induced accretion. 
%The majority of studies regarding the merger-AGN connection focus on major mergers \citep[e.g.,][]{DiMatteo05, Springel05, Hopkins08, Comerford24}, but the role of minor mergers in triggering AGN is less explored, despite the fact that minor mergers occur roughly three times more frequently than major mergers \citep[e.g.,][]{Jogee09}.
%Cosmological simulations have suggested that minor mergers (or mini mergers) have the ability to trigger AGN activity \citep[e.g.,][]{Capelo15, Byrne25}, and thus may explain the abundance of low-luminosity AGN \citep{Yang19}.
%The frequency of minor mergers in our fading-AGN sample seemingly suggests that such interactions can trigger AGN activity; however, our sample size is small and further work must be done to better understand these processes. 

Six fading-AGN show photometric signatures of their EELRs in their Legacy Survey images (11020-6103, 8485-6103, 8250-6104, 10844-1901, 9027-3701, 12483-12702). 
These signatures manifest as unusually bright and diffuse emission in the $g$-band where the \oiii$\lambda$5007 emission line is located at these redshifts; although we note that differentiating features arising from gravitational interactions from EELRs is a difficult task when the emission falls outside the MaNGA IFU.
Such systems would be ideal targets for follow-up observations to confirm their association and quantify the full extent of the EELR. 

\begin{figure}
    \centering
    \includegraphics[width=\linewidth]{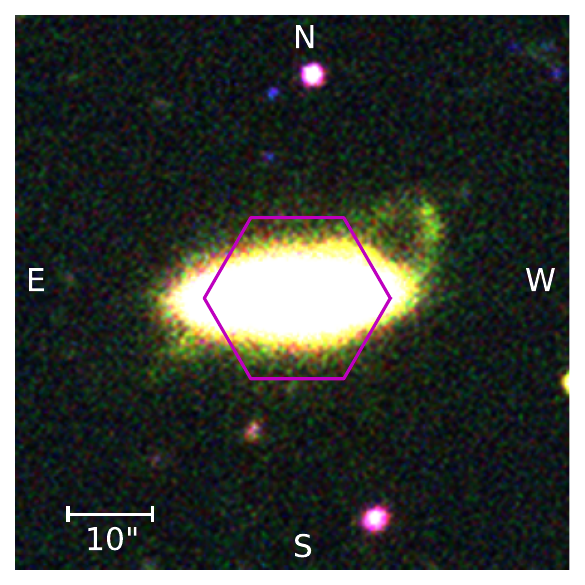}
    \caption{RGB image of MaNGA system 8485-6103 from the Legacy Survey, where the $r$-band is red, $g$-band is green, and the $z$-band is blue. The magenta hexagon is the MaNGA IFU FOV. A scale bar is shown for reference. The loop is $\sim$20\arcsec\ NW of the host galaxy and may be the result of an AGN outflow, a radio-jet, or a minor merger.} 
    \label{fig:loop}
\end{figure}

One system in our sample, 8485-6103, hosts a prominent ``loop" of emission $\sim~20$\arcsec\ (10.3 kpc) NW of the host galaxy (Figure \ref{fig:loop}). 
This loop is bright in the $g$-band and is analogous to the Teacup galaxy \citep{Gagne14, Keel17} or NGC 3393 \citep{Maksym17}; however the loop is located outside the MaNGA IFU and thus we cannot confirm it is at the same redshift.
These loops potentially result from powerful AGN outflows (see \citealt{Gabor13}, their Figure 6, and \citealt{Gabor14}, their Figure 1) or expanding radio-jet plasmas \citep[e.g.,][]{Gagne14} and are useful laboratories for studying the impact of AGN at large radii. 
The feature could also be a remnant of a previous merger as it was flagged as a minor merger in the \citet{Nevin23} catalog.
This source is a clear candidate for follow-up deep imaging and spectroscopy to inform the accretion history of the AGN.

\section{Discussion} 
\label{sec:discussion}
Within the MaNGA dataset we have identified a sample of 34 candidate fading-AGN based on their emission line properties. 
This search identifies the most extreme cases of fading-AGN wherein accretion onto the central SMBH has halted or been severely reduced such that stellar ionization plays a more pronounced role in the nuclear regions.
Our technique for finding fading-AGN candidates selects for a non-zero amount of star-formation in the galactic nuclei; these criteria may exclude cases where the nucleus is lacking both star-formation and AGN activity (i.e., the nuclear spectrum would be devoid of BPT emission lines); however, this scenario may be rare outside of post-starburst and quiescent galaxies and thus should minimally impact our sample.

\subsection{Assumptions Regarding IFU Center and SMBH Location} \label{subsec:disc-location-assumptions}
When selecting fading-AGN candidates (Section \ref{subsec:detection}), we assume that 1) the photometric center and the IFU center are aligned, and 2) the SMBH is located at the photometric center. 
For each fading-AGN candidate, we calculate the positional offset between the IFU center and the photometric center using the {\fontfamily{cmtt}\selectfont IFURA/IFUDEC} and {\fontfamily{cmtt}\selectfont OBJRA/OBJDEC} keys. 
The mean offset between the IFU center and photometric center of our sample is 0.006\arcsec (the maximum offset is 0.015\arcsec). 
This is well below the positional offsets incurred through fiber replugging ($\sim0.1$\arcsec; \citealt{Law16}), as well as the MaNGA spaxel scale (0.5\arcsec); therefore, it is reasonable to use the IFU center as the galaxy's photometric center.
Assuming the SMBH is located at the photometric center is a reasonable assumption as the two are aligned $\sim98\%$ of the time in AGN at $z\sim0.1$ \citep{Comerford14}.
In sources which do host offset AGN, fewer than $\sim1\%$ of these sources have SMBH-photometric center offsets larger than $0.6$\arcsec\xspace \citep{Skipper18}.
Thus, these assumptions should have little to no impact on our sample of fading-AGN candidates.

\subsection{Deriving an AGN Lifetime} \label{subsec{disc-agn-lifetime}}
We have carefully selected this sample in order to set a limit on the number of AGN turning off per year in the local Universe. 
We estimate the time since the AGN turned off by calculating the light travel time from the center of the IFU to the mean radius of the Seyfert/LINER spaxels (mean $t_{\text{faded}}=1.1\pm0.07\times10^{4}$ years for the entire sample, where the quoted uncertainty is the standard error). 
We calculate a turn-off rate in the sample of fading-AGN by dividing the number of fading-AGN ($N=34$) by the average fading timescale $t_{\text{faded}}$, finding $\sim3.0\times10^{-3}$ AGN turning off per year; equivalently, one AGN has turned off approximately every $\sim340$ years within the fading-AGN sample. 
Also, we estimate the fraction of \textit{all} AGN that turn off per year by dividing this turn-off rate by the total number of AGN in MaNGA ($N=387$; \citealt{Comerford24}). 
This corresponds to an AGN turnoff rate of $(7.6\pm1.4)\times10^{-6}$ per year.
Thus, for any individual AGN, the typical timescale to turn off is $t_{\text{AGN}}=(1.3\pm0.2)\times10^5$ years.

In reality, we do not quantify the full extent of the EELR/LE because the Seyfert/LINER spaxels are located at the edge of the MaNGA IFU, or the S/N drops below $3\sigma$ past these locations. 
Therefore, we interpret the fading timescale as a lower limit ($t_{\text{faded}} > 1.1\times10^{4}$ years).
Additionally, the number of MaNGA AGN in the \citet{Comerford24} catalog is a lower limit as it does not include Chandra selected AGN or AGN identified through IFU studies. 
For instance, \citet{Mezcua24} find 664 optically-selected dwarf AGN in MaNGA; therefore, the true number of MaNGA AGN may be in the thousands. 
Treating both the fading timescale and the number of MaNGA AGN as a lower limit corresponds to a lower limit on the lifetime of the AGN phase ($t_{\text{AGN}}>1.3\times10^5$ years).
%This is in good agreement with \citet{French23}, who calculated an AGN lifetime of $1.1\times10^5$ years using six EELRs in a sample of 93 post-starburst galaxies. 
%This also agrees well with \citet{Schawinski15}, who estimated that the AGN phase lasts $\sim10^5$ years using X-ray bright optically elusive galaxies (XBONGs).
This agrees well with observational estimates of the AGN lifetime; \citet{French23} calculated $t_{\text{AGN}}\sim1.1\times10^5$ years using EELRs in post-starburst galaxies, and \citet{Schawinski15} estimated $t_{\text{AGN}}\sim10^5$ years using X-ray bright optically elusive galaxies (XBONGs). 
Likewise, numerous simulations favor short episodes of accretion over long-term sustained accretion \citep[e.g.,][]{Novak11, Gabor13, Sijacki15, Alcazar21}.
Overall, our findings support the idea that black holes primarily grow through short bursts of accretion (i.e., AGN flicker on and off), and the instantaneous AGN luminosity does little to inform the average black hole accretion rate \citep[e.g.,][]{Hickox14}.

\subsection{Orientation Effects} \label{subsec:orientation}

\begin{figure*}
    \centering
    \includegraphics[width=\linewidth]{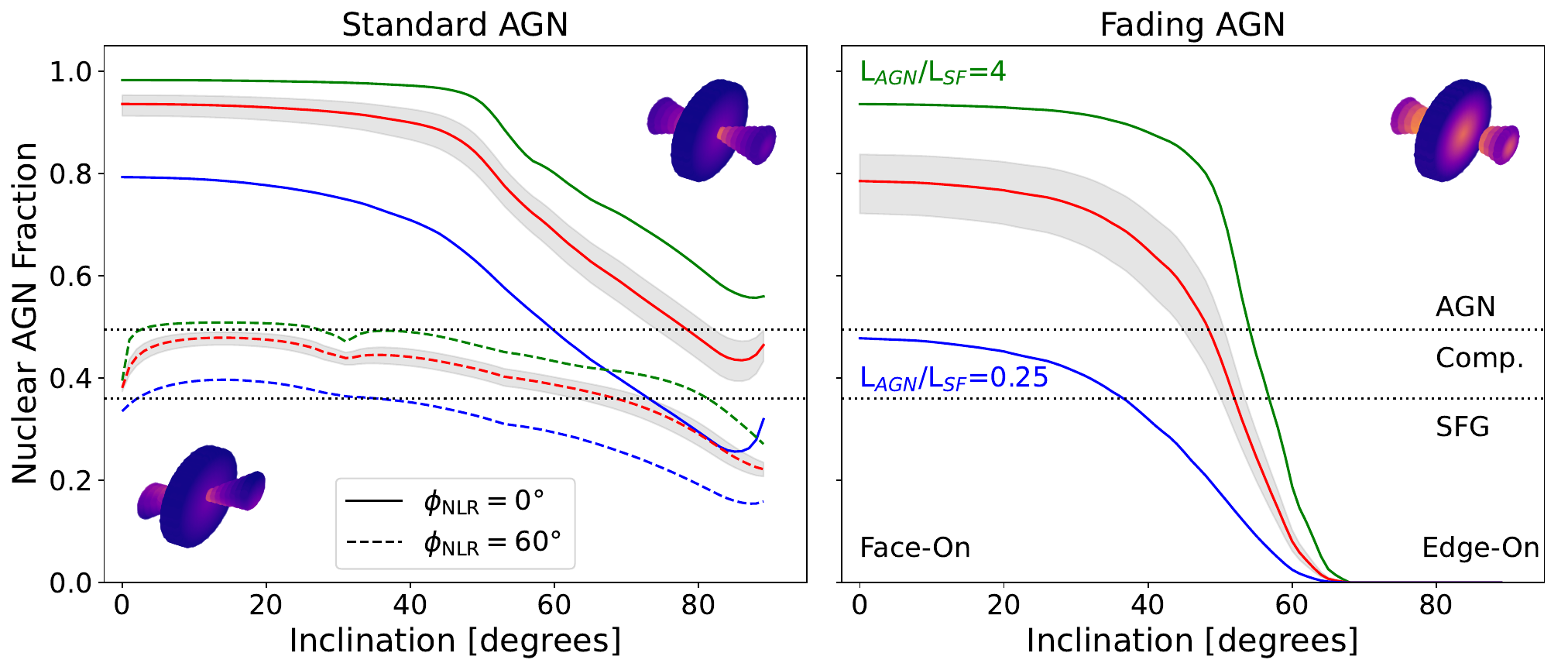}
    \caption{Results of simulating mock IFU observations for a regular AGN (left) and a fading-AGN (right) comprised of a disk component and ionization cones. The 3D models are shown at an arbitrary inclination, position angle, and location for visualization purposes only; the colors represent their spatial luminosities.
    %their location on the figure is arbitrary, and the colors represent the relative luminosity.
    The modeled systems are rotated between a fully face-on system ($i=0^{\circ}$) and a fully edge-on system ($i=90^{\circ}$), and we measure the fractional AGN contribution at the central kiloparsec for each angle.
    Green, red, and blue lines represent luminosity ratios of $L_{\text{AGN}}/L_{\text{SF}} = 4, 1, 0.25$, respectively. 
    Gray shaded regions represent the range of luminosity ratios in our MaNGA sample.
    The tilt of the NLRs are shown as solid ($\phi_{\text{NLR}}=0^{\circ}$) and dashed ($\phi_{\text{NLR}}=60^{\circ}$); note that the $\phi_{\text{NLR}}=60^{\circ}$ tilt yields a nuclear AGN fraction of 0 for the fading-AGN model at all inclinations.
    Dotted black lines represent BPT classification cutoffs determined from the model described in Section \ref{subsec:fractional-agn}.}
    \label{fig:3d-pa}
\end{figure*}

Our estimation of the AGN phase lifetime assumes that all 34 systems are indeed fading-AGN. 
We must consider that $\gtrsim70\%$ of the fading-AGN reside in edge-on systems (i.e., $i>70^{\circ}$, see Section \ref{subsec:mergers}), which may alter the physical interpretation. 
Fading-AGN may be more easily detectable in edge-on systems because the NLR is less susceptible to stellar contamination, \textit{or} edge-on systems produce false detections because the intervening material along the line of sight (i.e., dust and/or star-formation) is diluting a central AGN which can only be optically detected by observing the NLR. 
\textit{WISE} colors should inform the AGN nature of such edge-on systems (Figure \ref{fig:wise}), but the IR AGN selection is imperfect and host galaxy emission may still dilute the AGN signal \citep[e.g.,][]{Goulding12, Buchner17}.
In cases of such heavy obscuration, high resolution radio observations, the identification of high ionization MIR emission lines, or measuring the relative strengths of high-to-low excitation CO lines may be the only way to confidently identify the AGN \citep[e.g.,][]{Hickox18}.

Jaison et al. (submitted) studied off-nuclear AGN in MaNGA and found that off-nuclear AGN spaxels and nuclear AGN spaxels reside in face-on (edge-on) hosts $\sim50\%$ ($\sim16\%$) and $\sim72\%$ ($\sim9\%$) of the time, respectively. 
This suggests that while there may be a preference for finding off-nuclear AGN spaxels in edge-on systems, this preference is modest.
Additionally, they show that off-nuclear AGN spaxels have lower excitation ratios (as traced by \oiiihb) and \oiii luminosities relative to the nuclear spaxels regardless of the inclination angle. 
This implies that the off-nuclear AGN spaxels are in a distinct physical environment, rather than the result of projected nuclear emission.

To test whether the orientation of a system can significantly impact the observed nuclear AGN fraction, we construct a standard AGN model and a fading-AGN model in 3D. 
Our standard AGN model consists of a disk galaxy with proportions similar to that of the Milky Way, and a NLR modeled as two symmetric ionization cones with $r_{\text{NLR}}=2 \text{ kpc}$ \citep[e.g.,][]{Antonucci93, Lebowitz25} and an opening angle of $\sim95^{\circ}$ \citep[e.g.,][]{He18}.
We model the light in the disk (i.e., the SF luminosity) as an exponential profile, and the NLR (i.e., AGN luminosity) light profile as $1/r^2$.
Our fading-AGN model is the same as the standard AGN model, but we remove the inner 1 kpc of the ionization cones to produce a light echo.
%The schematic of this model is shown in the upper right hand side of Figure \ref{fig:3d-pa}.
We collapse the 3D model into a 2D projection to generate mock IFU observations by summing the luminosity components along the line of sight for a number of inclinations ($0^{\circ}\leq i \leq 90^{\circ}$),  luminosity ratios ($L_{\text{AGN}}/L_{\text{SF}}$), and NLR tilts ($\phi_{\text{NLR}}=0^{\circ}, 60^{\circ}$).
$\phi_{\text{NLR}}$ is defined as the angle between the NLR axis and the normal to the host galaxy's disk (e.g., $\phi_{\text{NLR}}=0^{\circ}$ is perpendicular to the disk, and $\phi_{\text{NLR}}=90^{\circ}$ would lie in the disk's plane).
These angles were chosen to illustrate the important effect of the NLR geometry on the observed properties of AGN, but we note that there is no correlation between a galaxy's disk and the angle of its NLR \citep[e.g.,][]{Fischer13}.
For each inclination and luminosity ratio we measure the fractional AGN contribution, $L_{\text{AGN}}/(L_{\text{AGN}} + L_{\text{SF}})$, in the central kiloparsec.
We define AGN, composite, and SFG cutoffs by calculating the mean $f_{\text{AGN}}$ value at the \citet{Kewley01} and \citet{Kauffmann03} dividing lines in our fractional AGN model (Figure \ref{fig:frac_agn}). 
These cutoffs are shown as dotted lines in Figure \ref{fig:3d-pa}.
We determine the observed luminosity ratios of our MaNGA fading-AGN sample by summing the \oiii luminosity multiplied by $f_{\text{AGN}}$ (or $1-f_{\text{AGN}}$ for $L_{\text{SF}}$) in each spaxel. 
Interestingly, most of the fading-AGN have luminosity ratios close to 1 (gray shaded regions, Figure \ref{fig:3d-pa}). 

\subsubsection{Standard AGN Model}\label{subsubsec:reg-agn}
This model shows that a standard AGN with $\phi_{\text{NLR}}=0^{\circ}$ will be classified as an AGN on the BPT diagram for inclinations above the AGN/composite dividing line in Figure \ref{fig:3d-pa}. 
For equal luminosity ratios this corresponds to a $i\lesssim80^{\circ}$.
A face-on AGN with $\phi_{\text{NLR}}=60^{\circ}$ will look like a composite system even at $L_{\text{AGN}}/L_{\text{SF}}\approx4$ (Figure \ref{fig:3d-pa}, green dashed line).
We can run the same selection algorithm described in Section \ref{subsec:detection} to determine whether a standard AGN would be incorrectly identified as a fading-AGN. 
Even though the AGN with $\phi_{\text{NLR}}=60^{\circ}$ shows a low ($\lesssim0.5$) nuclear AGN contribution across all inclinations, they are only falsely selected as fading-AGN for the highest inclinations ($\gtrsim80^{\circ}$). 
This is because they show extended AGN fractions similar to the nuclear AGN fraction. 
Thus, very edge-on systems that are classified as composite in their nuclear regions are more likely to host a buried AGN that is being diluted from star formation in the disk. 
Of the fading-AGN candidates that are classified as edge-on in the \citet{Vazquez22} catalog, twelve have nuclear spaxels classified as composite and thus are more likely to be contaminants.
%; if we assume these systems are contaminants and remove them from our sample then our estimate of the AGN phase lifetime increases to $t_{\text{AGN}}\sim1.8\times10^5$ years. 

\subsubsection{Fading-AGN Model} \label{subsubsec:fading-agn}
Likewise, we can test the effects of orientation for the modeled fading-AGN. 
%We construct this model similar to the standard AGN, but remove the inner 1 kpc of the ionization cones to produce a light echo. 
The right panel of Figure \ref{fig:3d-pa} shows that a face-on system looks like a standard AGN, but the nuclear AGN fraction quickly decreases between $30^{\circ}\lesssim i \lesssim 60^{\circ}$. 
The luminosity ratio also has a more pronounced effect on the nuclear AGN fraction compared to the standard AGN model. 
We run the selection algorithm on the $\phi_{\text{NLR}}=0^{\circ}$ model and find that fading-AGN are difficult to select in face-on systems because the light echo looks no different than a NLR; on the other hand, edge-on systems are easily detected. 
For the $\phi_{\text{NLR}}=60^{\circ}$ model, we would correctly detect a fading-AGN at every inclination; the nuclear AGN contribution is always 0 for this model which is why there are no dashed lines in the right panel of Figure \ref{fig:3d-pa}.

Across all models, systems oriented between $40^{\circ}\lesssim i \lesssim80^{\circ}$ degrees are most likely to be correctly recovered as fading-AGN.
Below this inclination range (i.e., face-on systems) we likely miss some fading-AGN because we cannot differentiate between the light echo and a typical NLR. 
Above this inclination range (i.e., edge-on systems) we likely have some ($\sim12$) false detections that host regular AGN, which are obscured and/or diluted by star formation in the disk. 
This, of course, is a simplified model that ignores some effects of NLR geometry (e.g., opening angle diversity, distance to the light echo) and differential reddening; however, we conclude that we detect a large fraction of edge-on fading-AGN because 1) the light echo is easier to detect, and 2) we likely have a handful of false detections. 
Regardless, the system orientation does not significantly impact our value of $t_{\text{AGN}}$; even a 50\% contamination rate only increases $t_{\text{AGN}}$ to $2.6\times10^5$ years. 

\subsection{Comparison to Other Work}
\citet{Comerford22} studied ten MaNGA galaxies that show off-nuclear Seyfert emission for systems not identified as AGN based on the SDSS 3\arcsec\xspace BPT classification.
They used Chandra to find that between seven to ten of these systems host (at least one) AGN depending on the strictness of the criteria.
They conclude these systems	were not selected as AGN in their optical spectra for a handful of reasons: the AGN are off-nuclear, low-luminosity, dust-obscured, and/or the AGN are flickering on and off.

We investigate how our selection algorithm classifies these ten off-nuclear Seyfert emitting galaxies from \citet{Comerford22}.
Nine of	these systems are not selected with our technique because their central kiloparsec fractional AGN contribution ($f_{inner}+\sigma_{f_{inner}}$) 
is greater than	the outer AGN contribution from the Seyfert and LINER spaxels ($f_{outer}-\sigma_{f_{outer}}$).
The single system that is selected by our algorithm (7443-12703, or Mrk 848) hosts an off-nuclear AGN that was removed when we cross-matched our candidates with the \citet{Comerford24} AGN catalog (note that this system was also identified as a binary AGN by \citealt{Fu18}).
This comparison shows that our selection algorithm, which treats emission line ratios as a mixture of star-formation and AGN emission, is more robust than using the BPT classification by itself.
The fractional AGN model is able to statistically identify star-forming/composite spaxels that host prominent AGN components and thus such systems are not selected by our algorithm.

\citet{Wylezalek18} searched for AGN in MaNGA DR14 and selected 308 using spatially resolved BPT diagrams. 
Seven of our fading-AGN candidates were included in DR14 and five were selected as AGN by \citet{Wylezalek18}.
We do not need to exclude these five from our sample as the \citet{Wylezalek18} selection includes systems with off-nuclear AGN signatures.
\citet{Wylezalek20} looked at AGN outflow signatures in this same sample using the $W_{80}$ measure of the \oiii$\lambda5007$ emission line.
Kinematic signatures may provide a longer timescale for detecting fading-AGN than ionization signatures \citep{Ishibashi15}. 
This is partially because \oiii$\lambda5007$ recombines $\sim70$ times faster than other BPT emission lines \citep{Binette87} and thus traditional AGN diagnostics will quickly fail to select fading-AGN. 

We look at the $W_{80}$ values in the fading-AGN and BPT selected SFGs from DR12, where $W_{80}~=~2.56\sigma_{\text{[OIII]}}$.
As described in \citet{Wylezalek20}, the $\sigma_{\text{[OIII]}}$ values may be unreasonable in the MaNGA {\fontfamily{cmtt}\selectfont DAP} as the {\fontfamily{cmtt}\selectfont DAP} only fits a single Gaussian to the emission line, but we report the values for comparison purposes. 
The fading-AGN mean $W_{80}$ value is $296\pm13 \text{ km s}^{-1}$ compared to $222\pm1.3 \text{ km s}^{-1}$ in the SFG sample as measured from the MaNGA {\fontfamily{cmtt}\selectfont DAP}, where the quoted uncertainty is the standard error. 
The fading-AGN have higher $W_{80}$ values relative to the SFGs, which is another way to determine prominent AGN signatures outside of the nucleus. 
This may suggest the EELRs/LEs are relic AGN outflows, but further analysis is needed to prove this. 

Jaison et al. (submitted) has followed up the \citet{Wylezalek18} analysis for the full MaNGA sample. 
They search for off-nuclear AGN by selecting sources with $>15\%$ of spaxels classified as AGN-like at $r>1$ kpc \citep{Alban23}; they find $\sim200$ off-nuclear AGN candidates. 
Compared to our sample of fading-AGN candidates, 23/34 are selected as off-nuclear AGN with their criteria. 
They classify one source (10837-1902) as a quiet (composite-dominated) AGN, and one source (11020-6103) is removed from their sample because it is a merger. 
The remaining sources (9/34) do not pass their $d_{\text{BPT}}>0.3$ criteria \citep{Wylezalek18} on the \sii-BPT diagram, and are thus excluded from their analysis. 
The overlap between these two samples which use distinct selection criteria is encouraging and highlights the importance of IFU spectroscopy when searching for extended AGN emission.

\section{Summary \& Conclusions}\label{sec:conclusions}
Fading-AGN remain elusive because the AGN signatures dissipate remarkably fast relative to the accretion timescales of AGN, and the orientation of the light echo relative to the disk impacts the number of systems that are detectable. 
The average fading-AGN in our sample is identified as the result of nuclear AGN flux emitted $\sim$11,000 years ago based on the light-travel time; this roughly corresponds to the end of the last glacial period on Earth \citep{Corrick20}. 
This paper, and others, have shown that IFU spectroscopy is by far the most accurate way to find fading-AGN \citep[e.g.,][]{Mezcua20, Mezcua24}.
The continued detection of fading-AGN will likely be driven by narrow-band photometry with follow-up targeted spectroscopy, or future large scale IFU surveys similar to MaNGA.

In this paper we introduce an empirical model to calculate the fractional AGN contribution to optical spectra relying on only four emission lines: \ha, \hb, \nii~$\lambda6584$, and \oiii~$\lambda5007$. 
We use this model to select MaNGA galaxies that have a statistically larger AGN contribution in their outskirts than in their nucleus. 
We cull the sample by removing AGN selected using multi-wavelength techniques and PAGB stars that mimick LINER emission. 
The finalized sample contains 34 fading-AGN, which greatly expands the number of known systems ($N\approx61$; \citealt{Keel12, Keel17, Keel22, Keel24, Schirmer13, Schweizer13, Watkins18, Esparza20}).
%\citep[c.f.,][]{Keel17, Esparza20, French23}.

The BPT classifications from SDSS 3\arcsec\ fiber observations suggest the fading-AGN are primarily SFGs or composites, but the MaNGA IFU data shows Seyfert and LINER emission at large radii in these systems. 
The fading-AGN follow a similar radial pattern in BPT space which is opposite the mixing-sequence seen in IFU data of typical AGN. 
We can visually identify the EELRs in $\gtrsim$ 60\% of the fading-AGN when comparing \oiii emission to the continuum. 
Compared to typical AGN, the fading-AGN have blue W1-W2 and red W2-W3 colors. 
Their W3 and W4 luminosities are roughly equivalent, which may indicate dust-obscured star-formation, and their IR fractional AGN contribution is lower than typical AGN which implies low levels of nuclear activity.

%We find more fading-AGN in edge-on galaxies because the NLR/EELR is less contaminated from host galaxy emission. 
More than $70\%$ of the fading-AGN candidates reside in edge-on galaxies. 
Through 3D modeling we show that very edge-on systems ($i>80^{\circ}$) identified as composite in their nucleus are more likely to be false detections of fading-AGN; star formation in the host galaxy can dilute the nuclear AGN signal making it appear less pronounced and impacting the observed BPT line ratios.
Furthermore, we likely do not detect a number of fading-AGN in face-on systems ($i < 40^{\circ}$) because we cannot differentiate between a LE and a standard NLR. 
Surprisingly, only five systems are undergoing major mergers whereas most previously identified fading-AGN were found in locations with extended gas freed up by gravitational interactions. 
Seventeen fading-AGN candidates are classified as minor mergers, and seven as non-mergers.
We find six systems with broadband photometric signatures of their fading-AGN nature, which would be ideal candidates to follow up in order to quantify the full extent of the EELR. 
One system, 8485-6103, shows a prominent loop of emission bright in the $g$-band, likely caused by a powerful relic AGN outflow. 

Using this sample of fading-AGN relative to the total number of AGN in MaNGA, we are able to place a lower limit on the lifetime of the AGN phase of $t_{\text{AGN}}>1.3\times10^5$ years. 
This agrees well with a similar study using a smaller sample size of post-starburst galaxies \citep{French23}, and the order-of-magnitude estimation derived from XBONGs \citep{Schawinski15}.
Overall, our study supports the idea that black holes grow through short bursts of accretion and the AGN phase tends to be short-lived ($\sim10^5$ years).
Quantifying the full extent of the EELRs/LEs may help to pin down this timescale with greater precision. 
Fading-AGN are unique tools that can be used to understand the role of AGN feedback and the connection between AGN and star-formation. 
Their relative scarceness had limited previous studies to small numbers of systems, but we have shown that fading-AGN may be more abundant than previously thought, which can garner statistically motivated studies of these systems.

\section{Acknowledgments}
We thank the anonymous reviewer, whose comments
and suggestions greatly improved the content of this paper.
We thank Christy Tremonti for helpful discussions.
Q.O.C. acknowledges support from the Dartmouth Fellowship.
R.C.H. acknowledges support from NASA ADAP grant numbers 80NSSC25K7569 and 80NSSC23K0485.
K.D.F. acknowledges support from NSF grant AST-2307375.
M.M. acknowledges support from the Spanish Ministry of Science and Innovation through the project PID2024-159201NB-C22. This work was also partly supported by the Spanish program Unidad de Excelencia Mar\'ia de Maeztu CEX2020-001058-M, financed by MCIN/AEI/10.13039/501100011033, and by the MaX-CSIC Excellence Award MaX4-SOMMA-ICE. 
R.R. acknowledges support from  Conselho Nacional de Desenvolvimento Cient\'{i}fico e Tecnol\'ogico  (CNPq, Proj. CNPq-445231/2024-6, 311223/2020-6, 404238/2021-1, and 310413/2025-7), Funda\c{c}\~ao de amparo \`{a} pesquisa do Rio Grande do Sul (FAPERGS, Proj. 19/1750-2 and 24/2551-0001282-6) and Coordena\c{c}\~ao de Aperfei\c{c}oamento de Pessoal de N\'{i}vel Superior (CAPES, 88881.109987/2025-01).

Funding for the Sloan Digital Sky Survey IV has been provided by the Alfred P. Sloan Foundation, the U.S. Department of Energy Office of Science, and the Participating Institutions. 

SDSS-IV acknowledges support and resources from the Center for High Performance Computing  at the University of Utah. The SDSS website is www.sdss4.org.

SDSS-IV is managed by the 
Astrophysical Research Consortium 
for the Participating Institutions 
of the SDSS Collaboration including 
the Brazilian Participation Group, 
the Carnegie Institution for Science, 
Carnegie Mellon University, Center for 
Astrophysics | Harvard \& 
Smithsonian, the Chilean Participation 
Group, the French Participation Group, 
Instituto de Astrof\'isica de 
Canarias, The Johns Hopkins 
University, Kavli Institute for the 
Physics and Mathematics of the 
Universe (IPMU) / University of 
Tokyo, the Korean Participation Group, 
Lawrence Berkeley National Laboratory, 
Leibniz Institut f\"ur Astrophysik 
Potsdam (AIP),  Max-Planck-Institut 
f\"ur Astronomie (MPIA Heidelberg), 
Max-Planck-Institut f\"ur 
Astrophysik (MPA Garching), 
Max-Planck-Institut f\"ur 
Extraterrestrische Physik (MPE), 
National Astronomical Observatories of 
China, New Mexico State University, 
New York University, University of 
Notre Dame, Observat\'ario 
Nacional / MCTI, The Ohio State 
University, Pennsylvania State 
University, Shanghai 
Astronomical Observatory, United 
Kingdom Participation Group, 
Universidad Nacional Aut\'onoma 
de M\'exico, University of Arizona, 
University of Colorado Boulder, 
University of Oxford, University of 
Portsmouth, University of Utah, 
University of Virginia, University 
of Washington, University of 
Wisconsin, Vanderbilt University, 
and Yale University.

\hfil 

\textit{Software:} astropy \citep{astropy2022}, matplotlib \citep{matplotlib}, numpy \citep{numpy2020}, pandas \citep{pandas2021}, scipy \citep{scipy}, Topcat \citep{Taylor2005}.

%\clearpage
%\input{bibliography}
%\appendix
%\label{appendix}

\begin{figure}[h]
    \centering
    \includegraphics[width=0.5\textwidth]{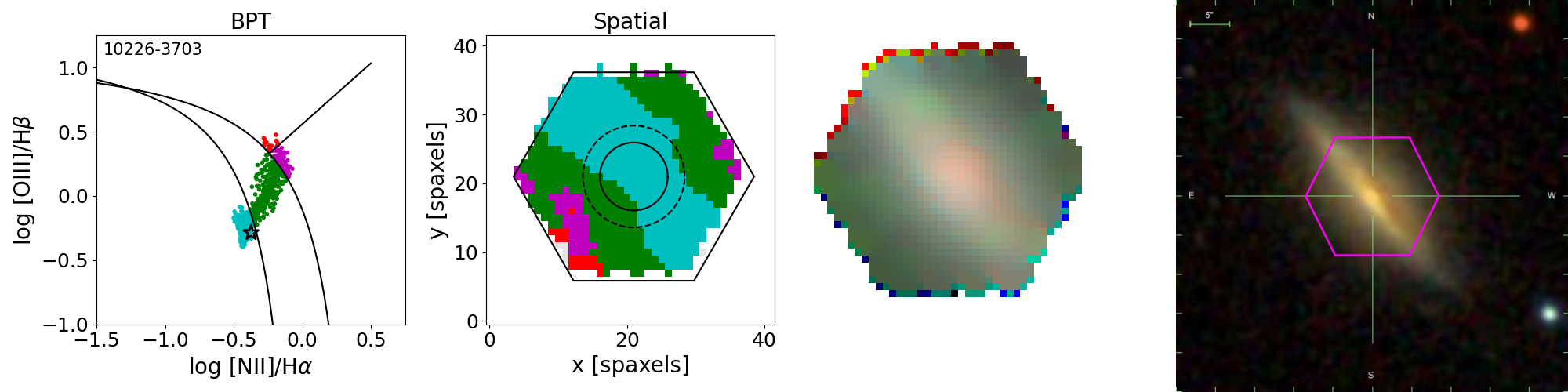}\hfill
    \includegraphics[width=0.5\textwidth]{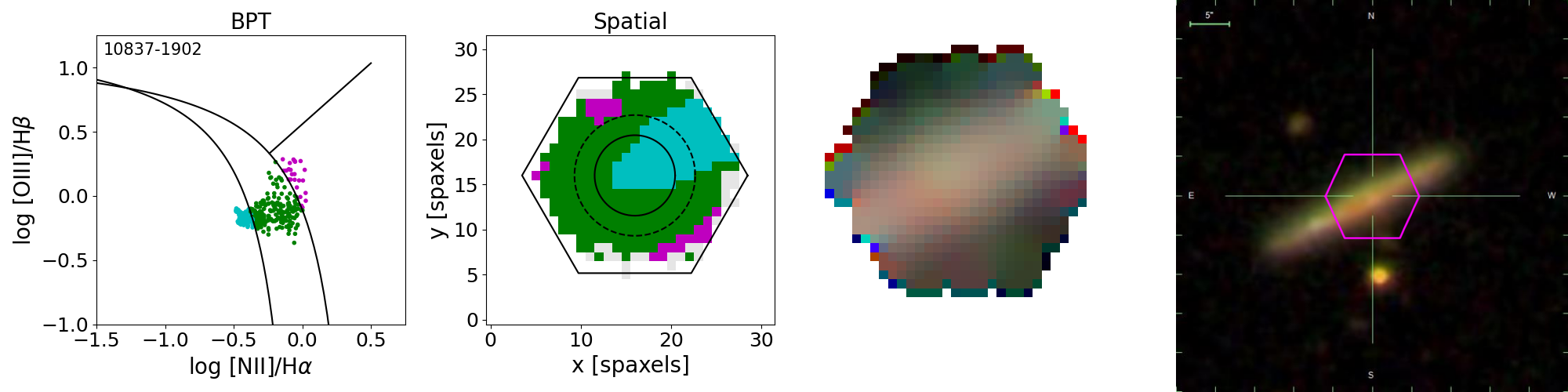}
    
    \includegraphics[width=0.5\textwidth]{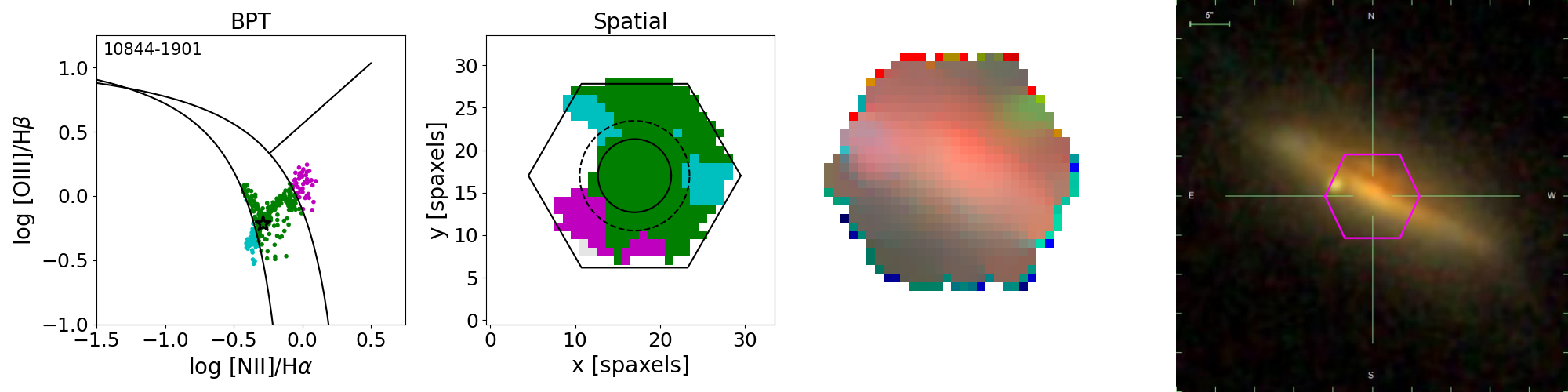}\hfill
    \includegraphics[width=0.5\textwidth]{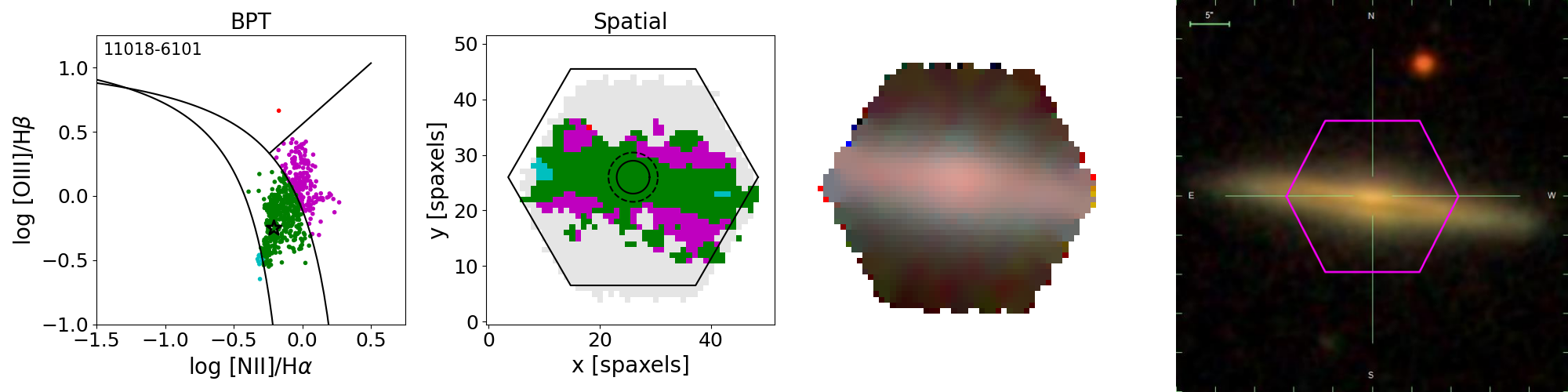}
    
    \includegraphics[width=0.5\textwidth]{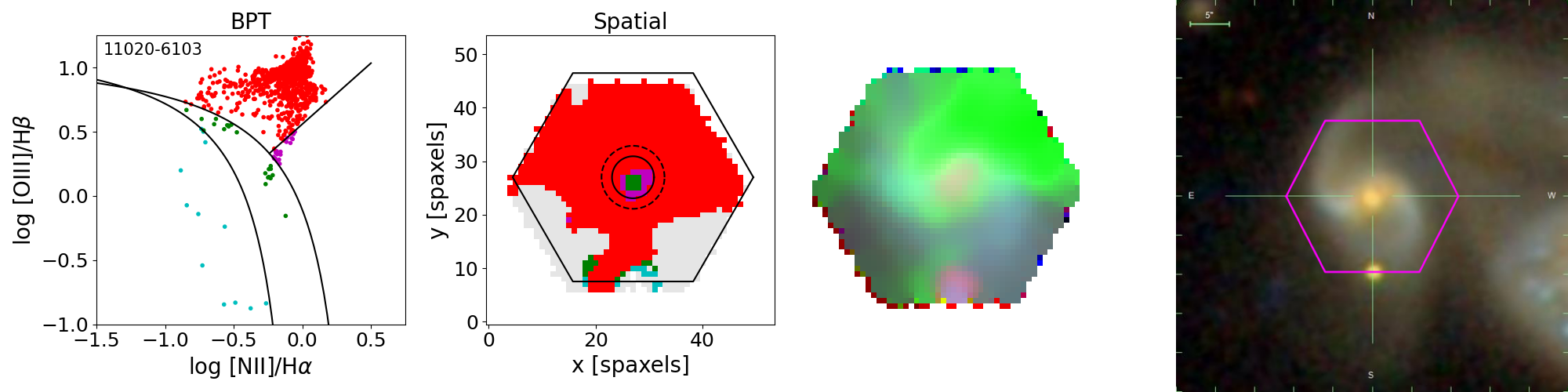}\hfill
    \includegraphics[width=0.5\textwidth]{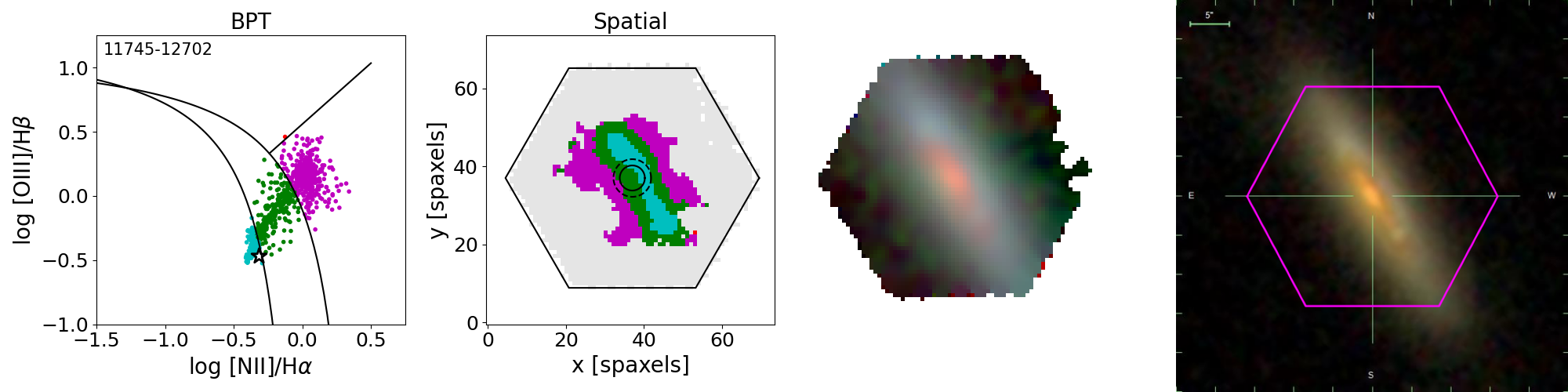}
    
    \includegraphics[width=0.5\textwidth]{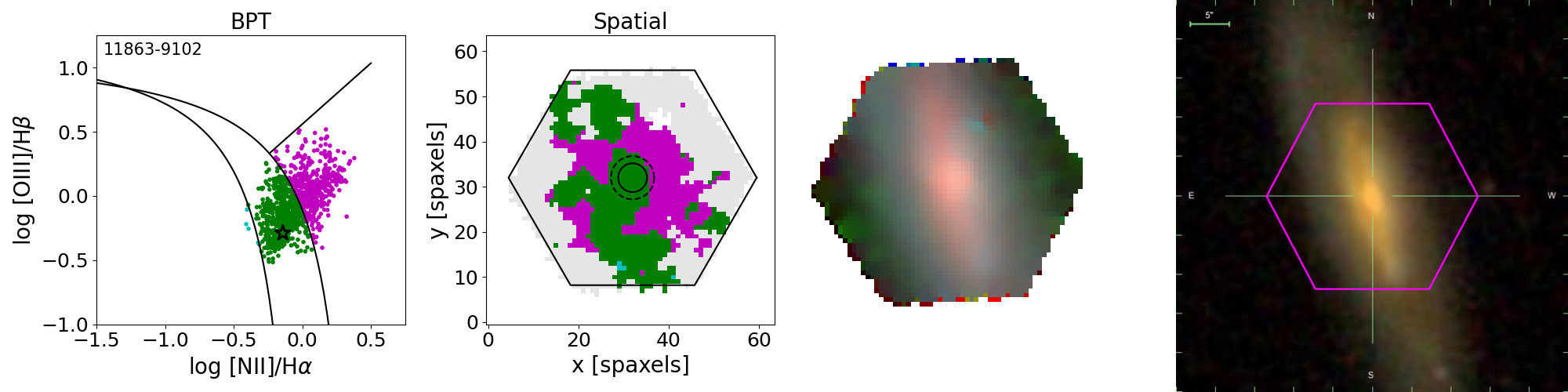}\hfill
    \includegraphics[width=0.5\textwidth]{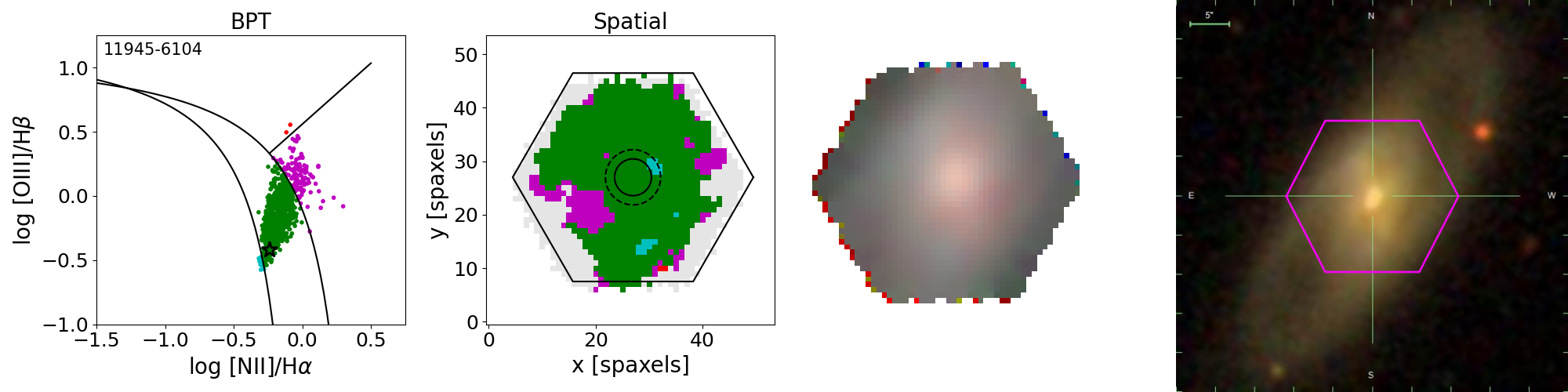}
    
    \includegraphics[width=0.5\textwidth]{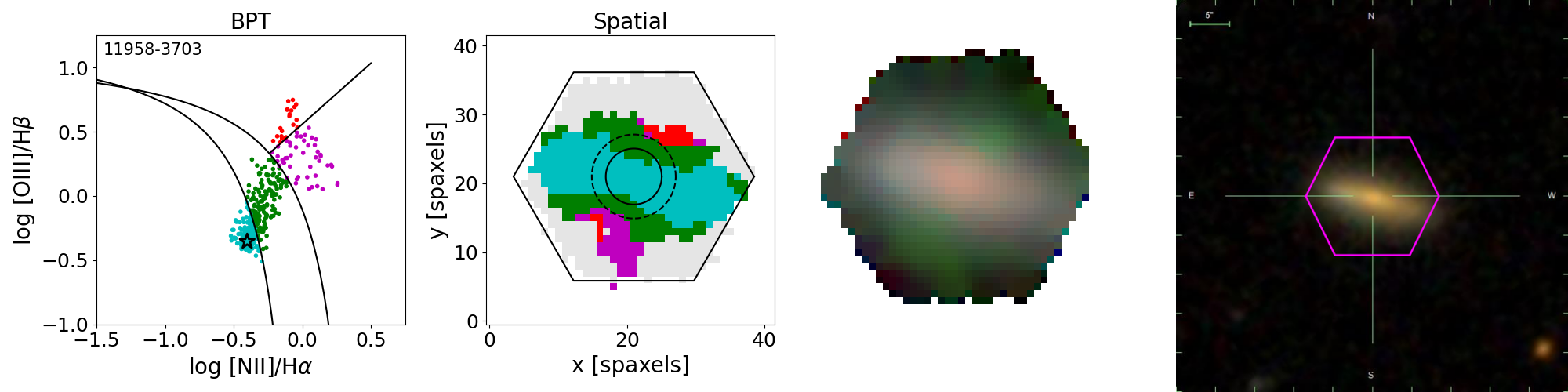}\hfill
    \includegraphics[width=0.5\textwidth]{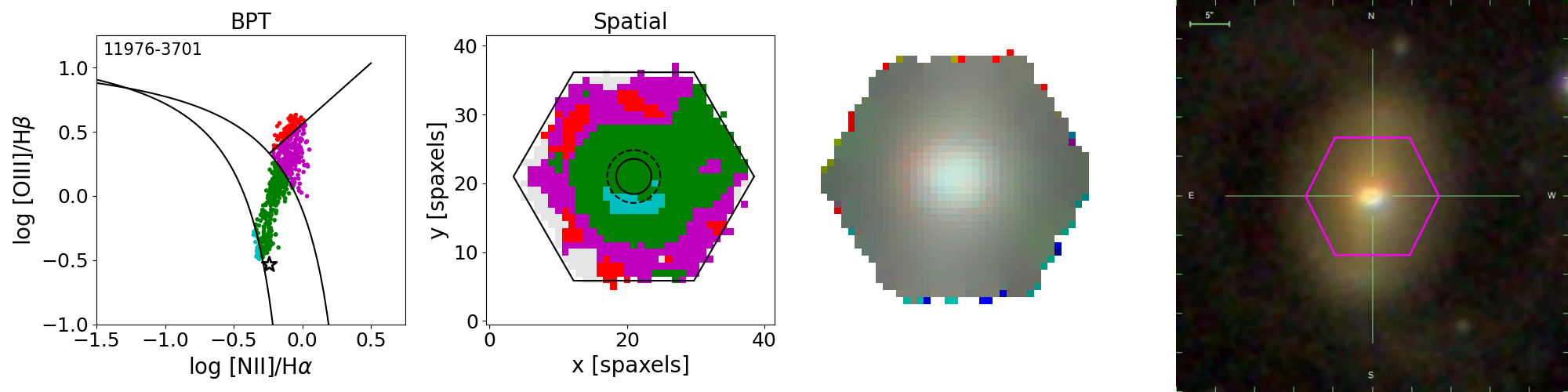}

    \includegraphics[width=0.5\textwidth]{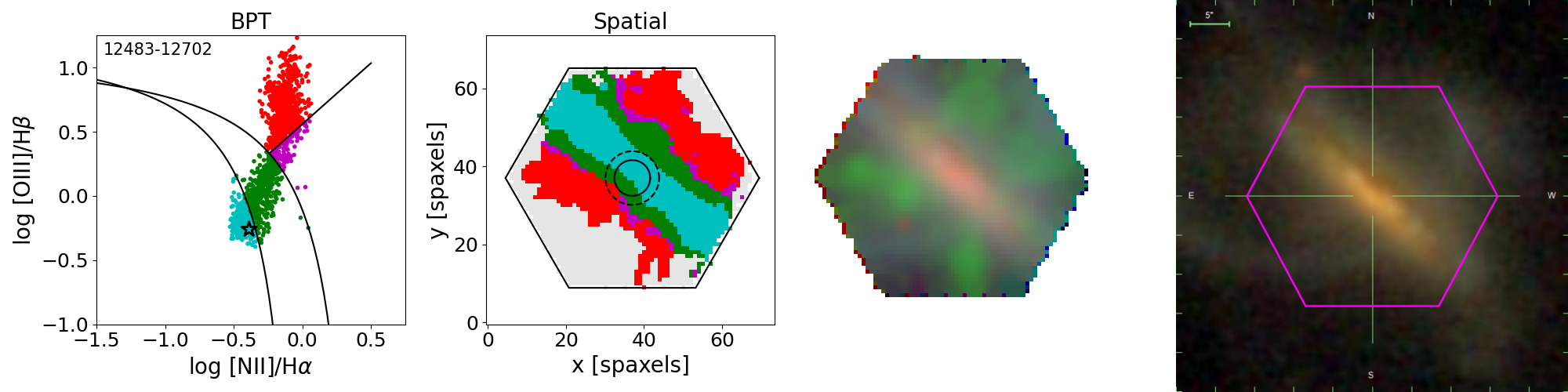}\hfill
    \includegraphics[width=0.5\textwidth]{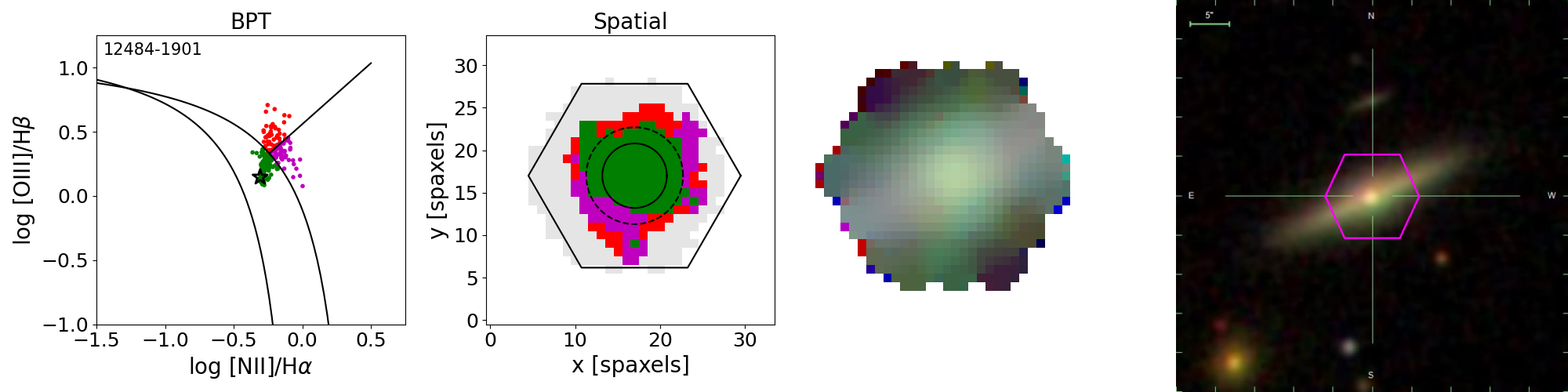}

    \includegraphics[width=0.5\textwidth]{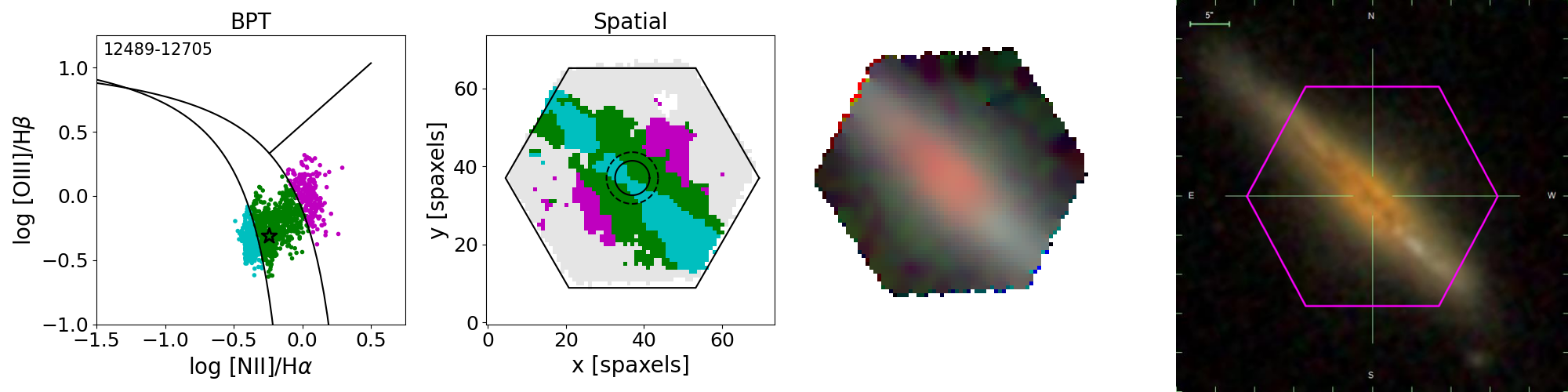}\hfill
    \includegraphics[width=0.5\textwidth]{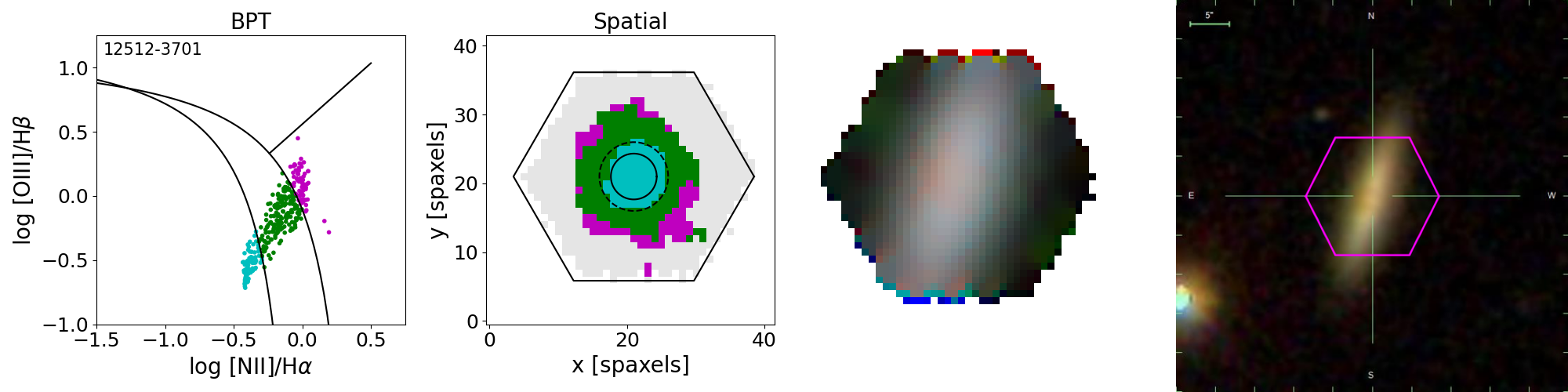}

    \includegraphics[width=0.5\textwidth]{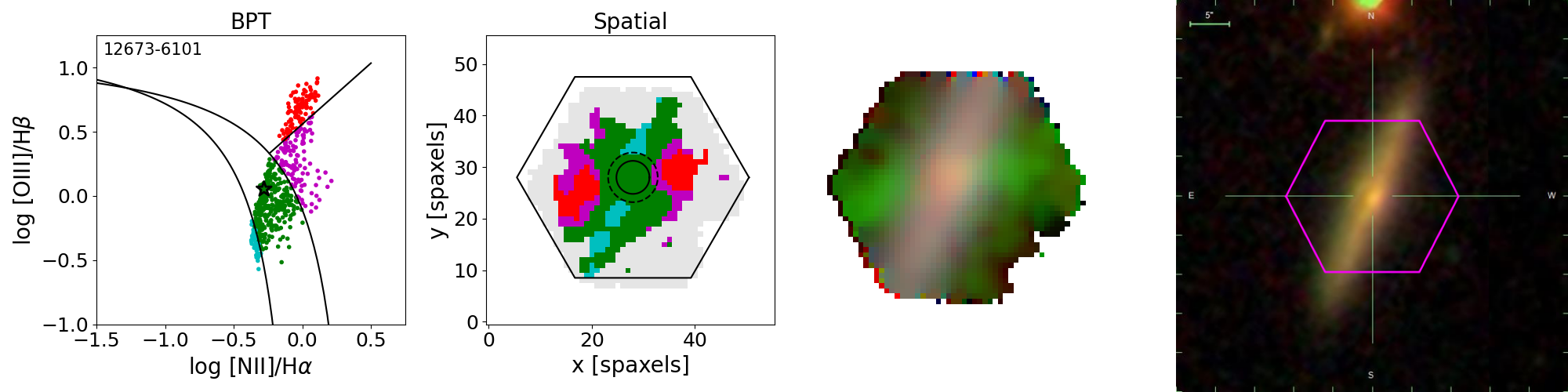}\hfill
    \includegraphics[width=0.5\textwidth]{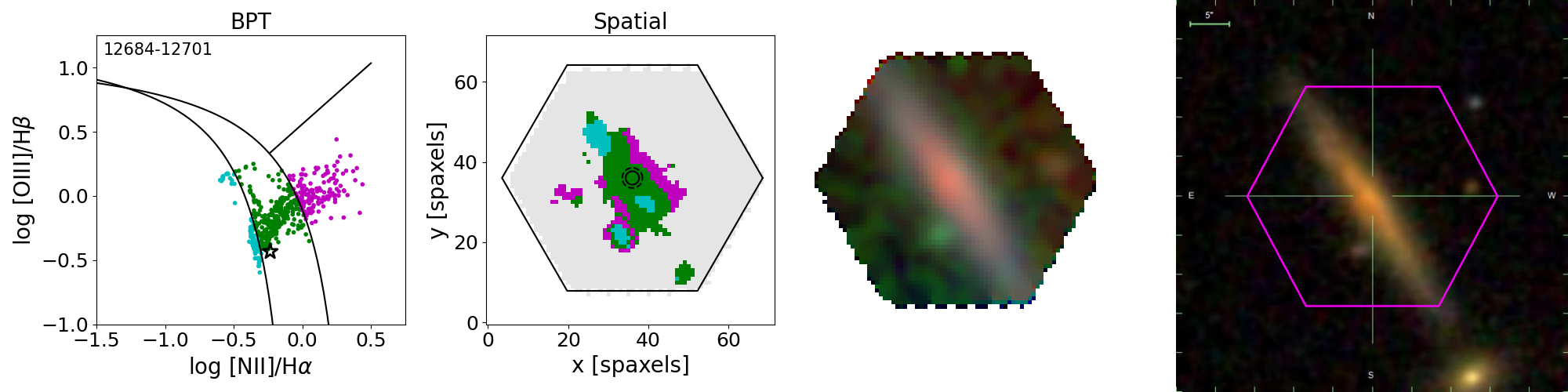}

    %\caption{Candidate fading-AGN in our sample. The left panel shows each spaxel on the \nii-BPT diagram. The black star represents the BPT location as determined by the 3\arcsec\xspace SDSS single-fiber observations (note that four sources do not have single-fiber measurements). The center left panel shows the spatial distribution of the emission types as determined by the \nii-BPT diagram. The black hexagon represents the MaNGA IFU fiber bundle; the solid and dashed circles denote radii of 1 and 1.5 kpc, respectively. These radii are used to determine the inner and outer fractional AGN contribution (see Section \ref{subsec:detection}). }
    \caption{The left and center left panels show the \nii-BPT for each spaxel and the spatial distribution of the emission types (see the caption of Figure \ref{fig:bpt-3panel} for more details). 
    The center right panel shows tri-color images, where green represents emission from \oiii $\lambda$5007 (see Section \ref{subsec:tri-color} and Figure \ref{fig:tricolor}). 
    The right panel shows the SDSS $gri$ image, where the magenta hexagon represents the MaNGA IFU fiber bundle.}
    \label{fig:appendix1}
\end{figure}

\begin{figure}[h]
    \centering
    
    \includegraphics[width=0.5\textwidth]{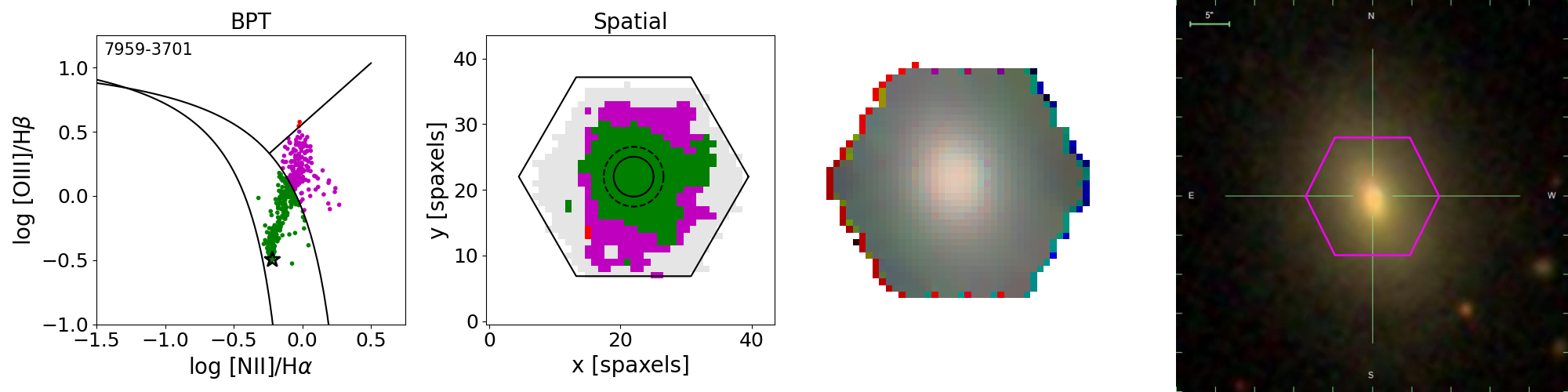}\hfill
    \includegraphics[width=0.5\textwidth]{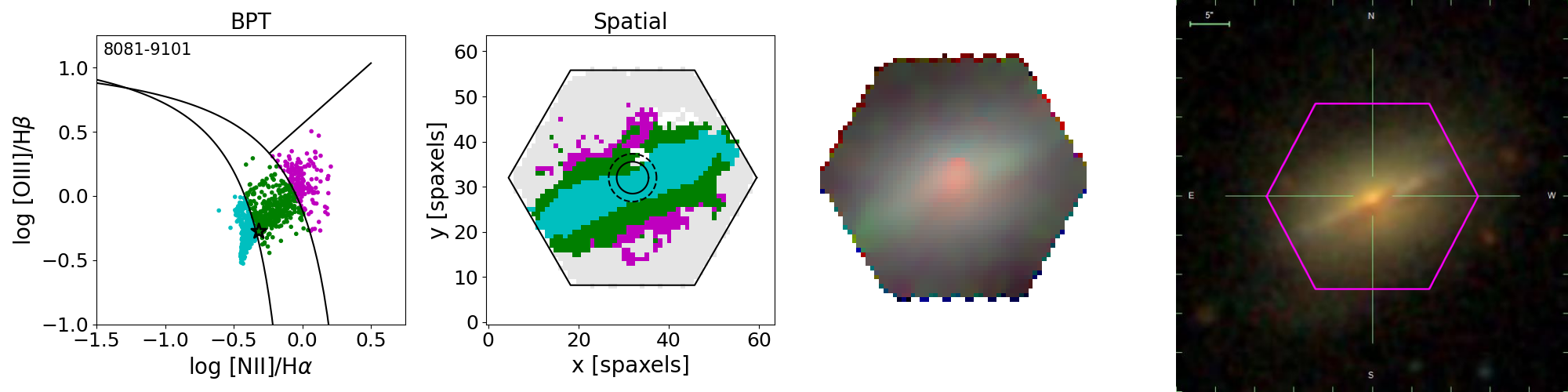}

    \includegraphics[width=0.5\textwidth]{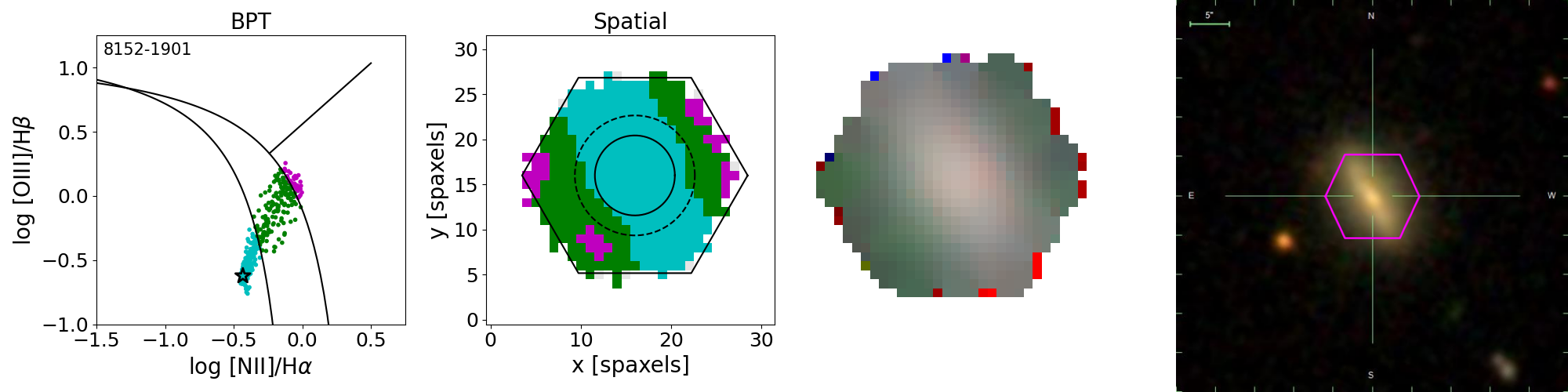}\hfill
    \includegraphics[width=0.5\textwidth]{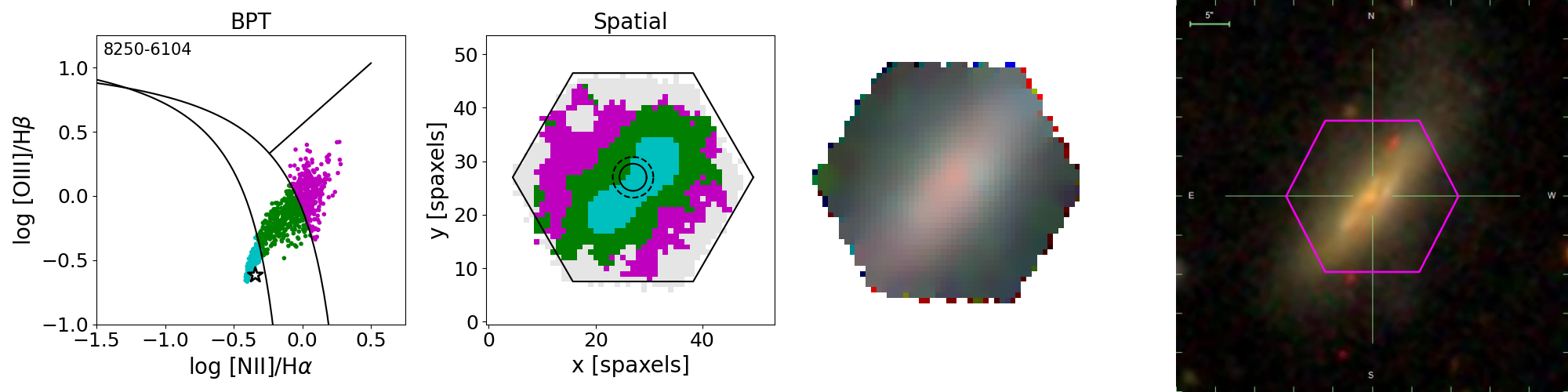}

    \includegraphics[width=0.5\textwidth]{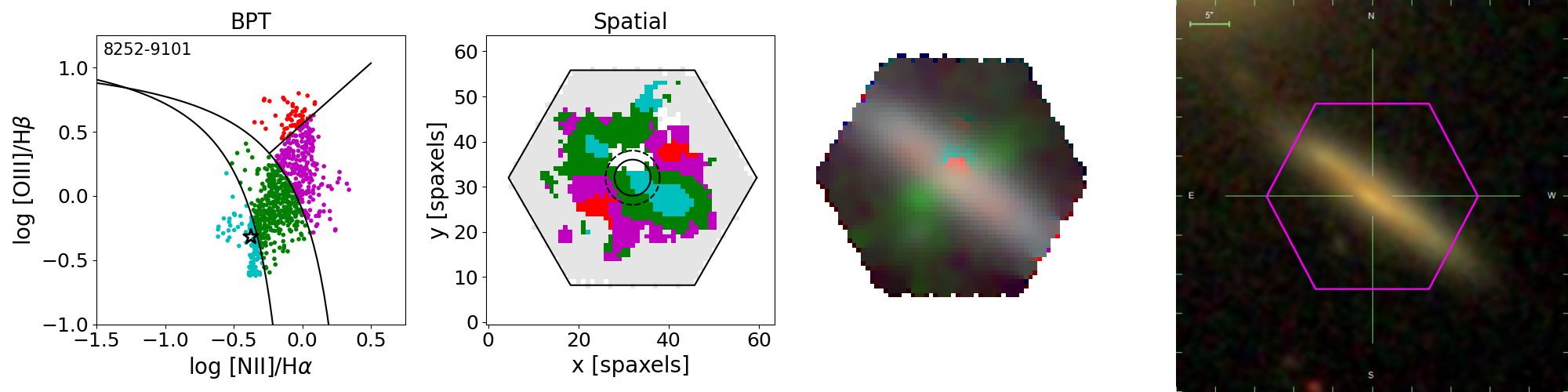}\hfill
    \includegraphics[width=0.5\textwidth]{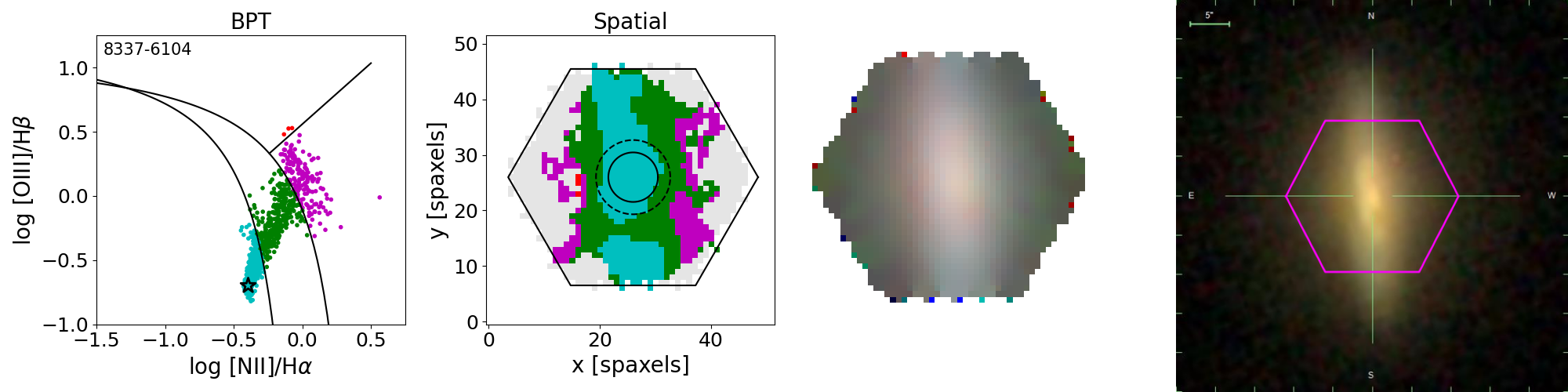}

    \includegraphics[width=0.5\textwidth]{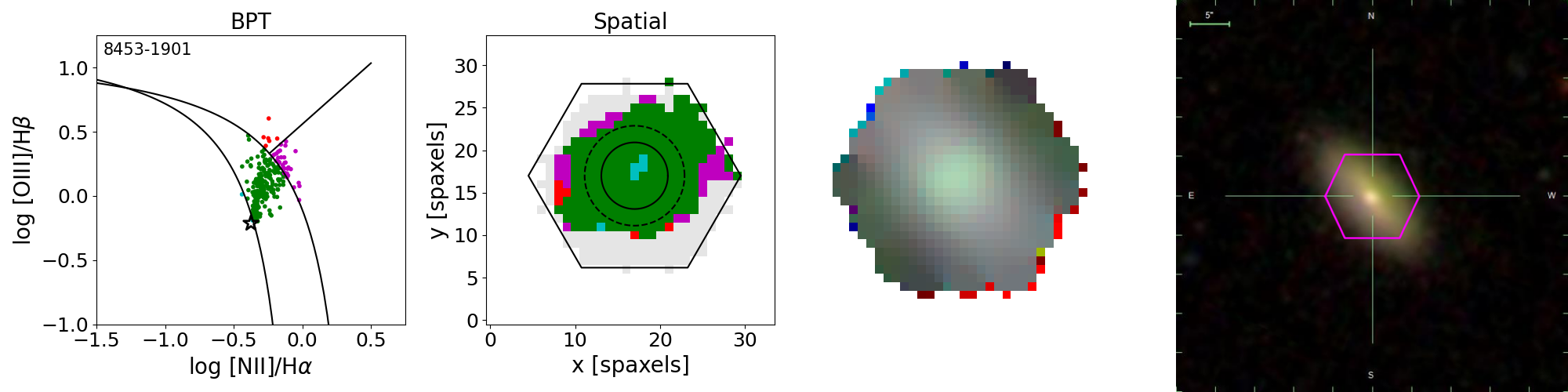}\hfill
    \includegraphics[width=0.5\textwidth]{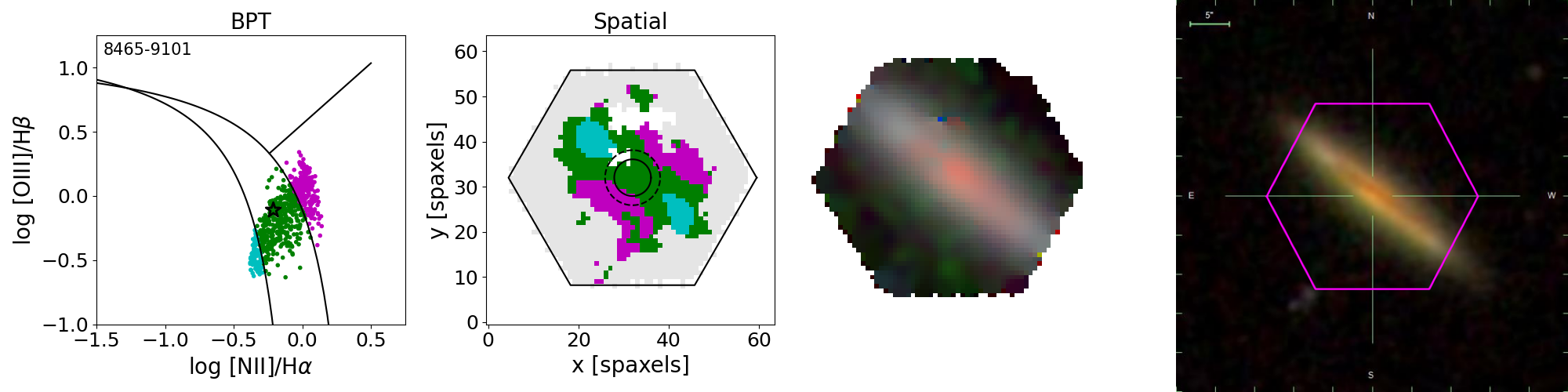}

    \includegraphics[width=0.5\textwidth]{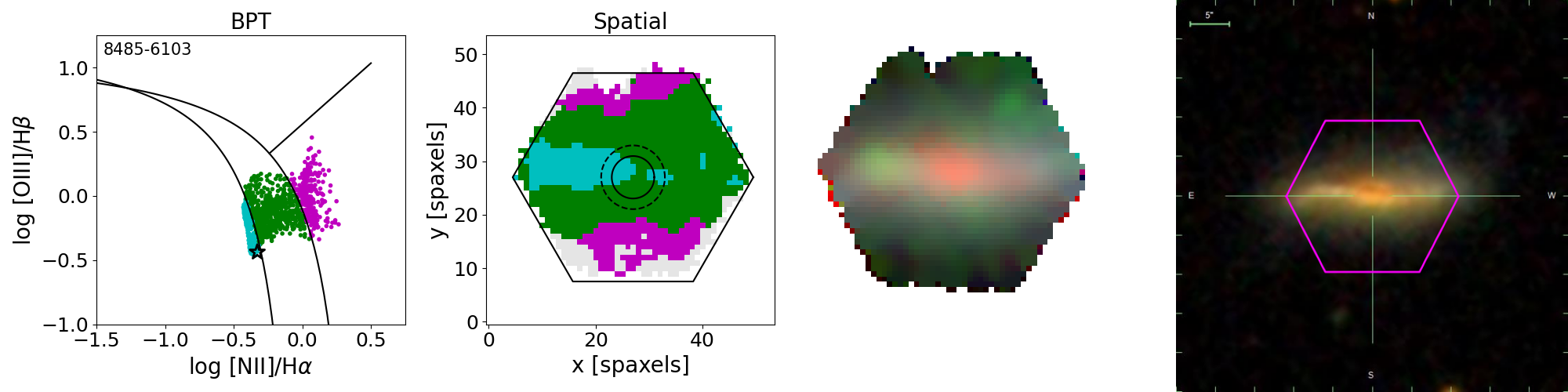}\hfill
    \includegraphics[width=0.5\textwidth]{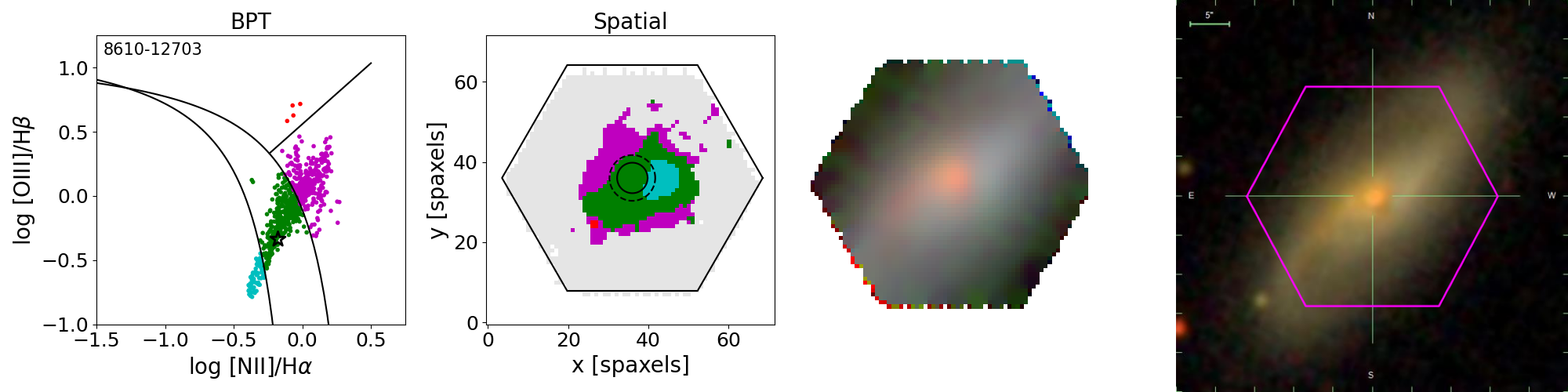}
    
    \includegraphics[width=0.5\textwidth]{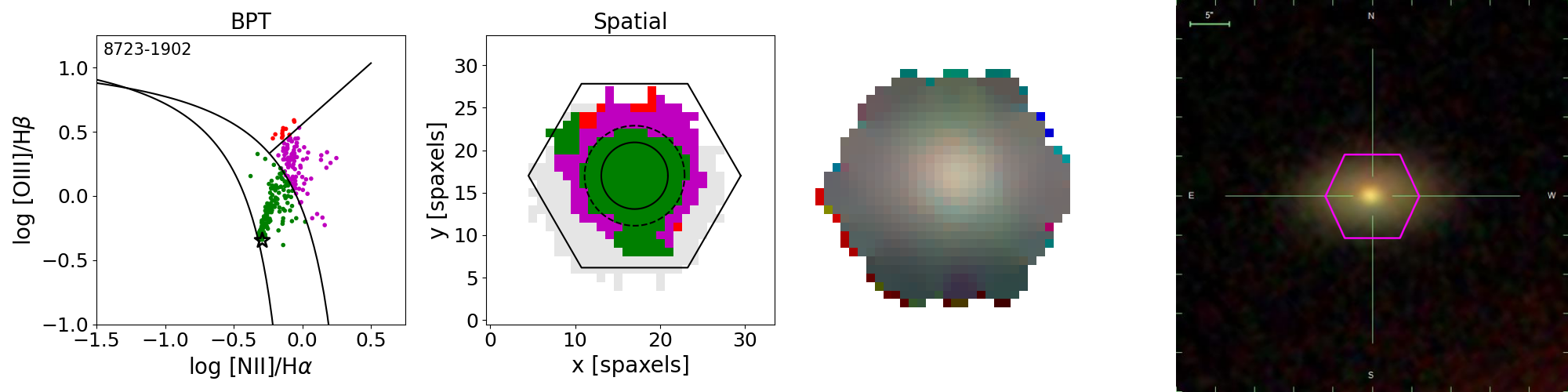}\hfill
    \includegraphics[width=0.5\textwidth]{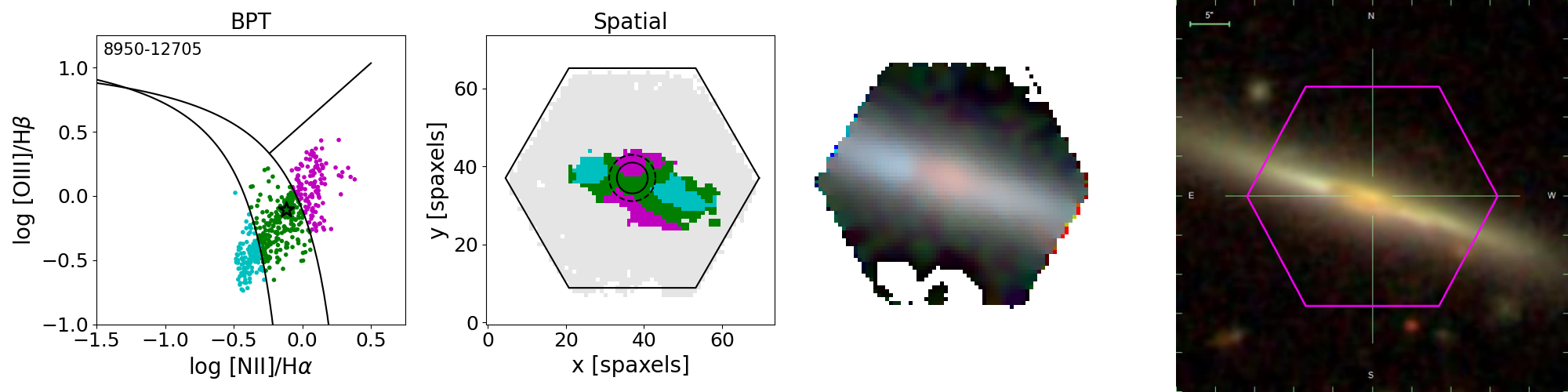}
    
    \includegraphics[width=0.5\textwidth]{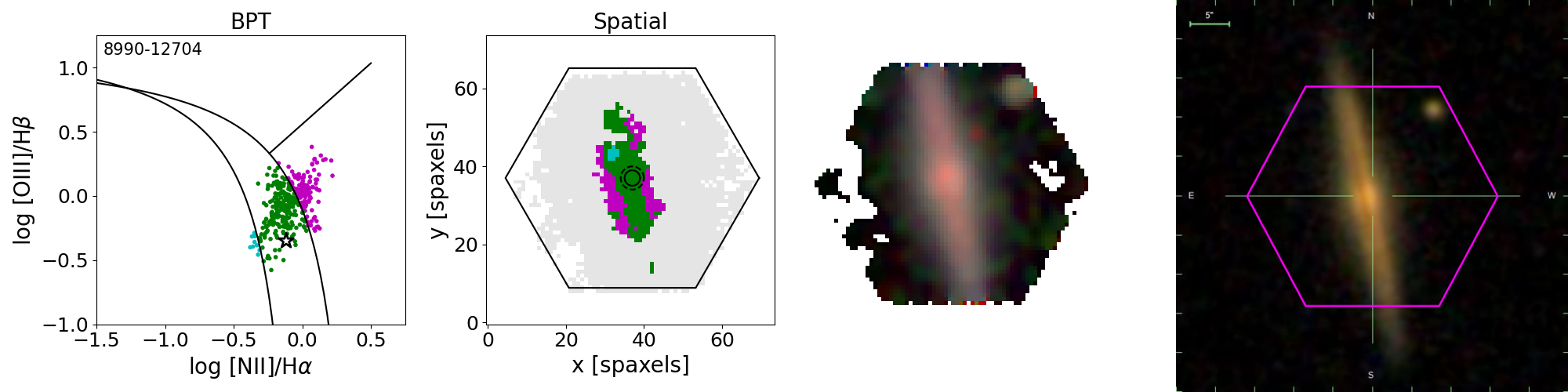}\hfill
    \includegraphics[width=0.5\textwidth]{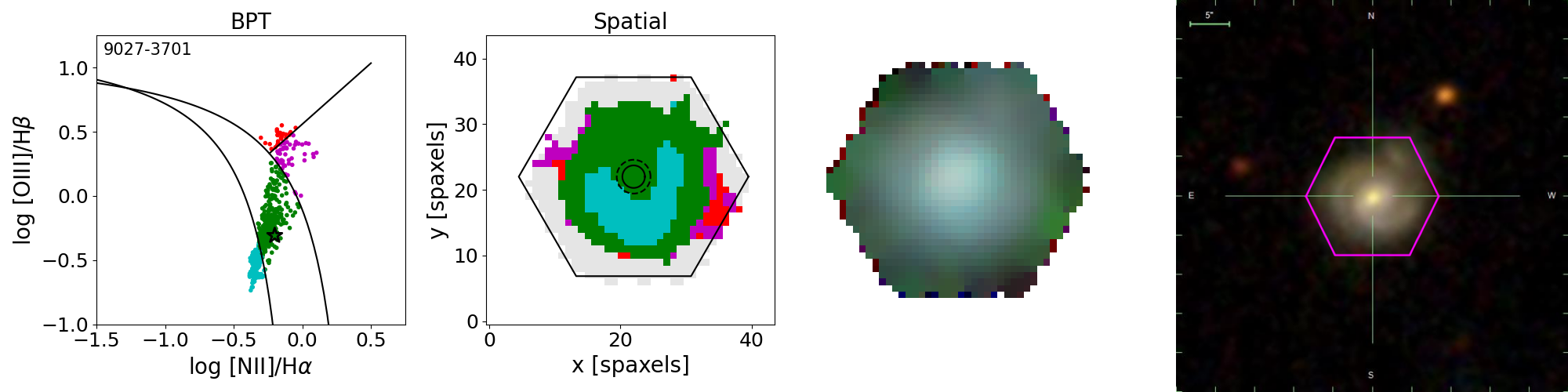}
    
    \includegraphics[width=0.5\textwidth]{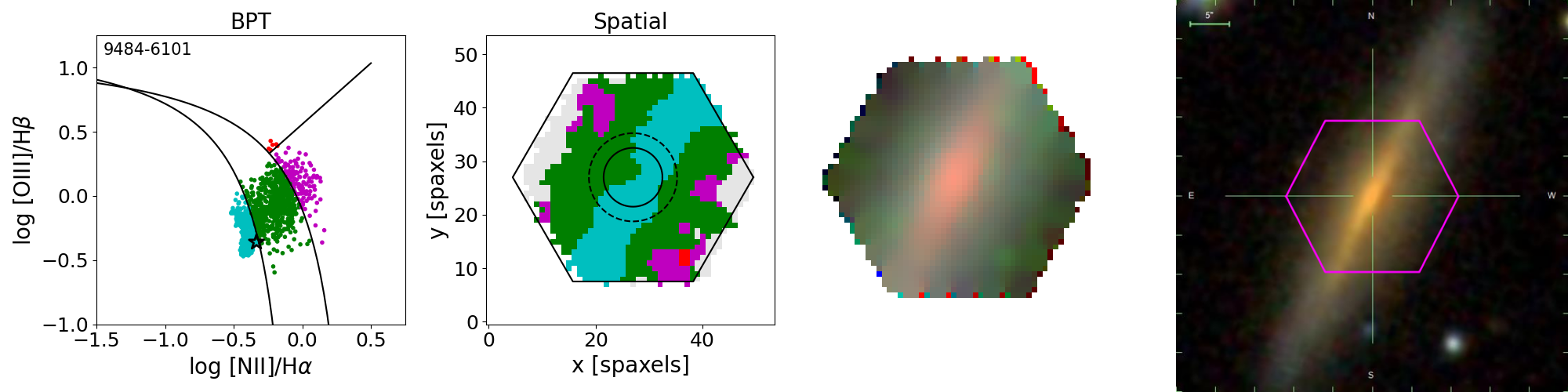}\hfill
    \includegraphics[width=0.5\textwidth]{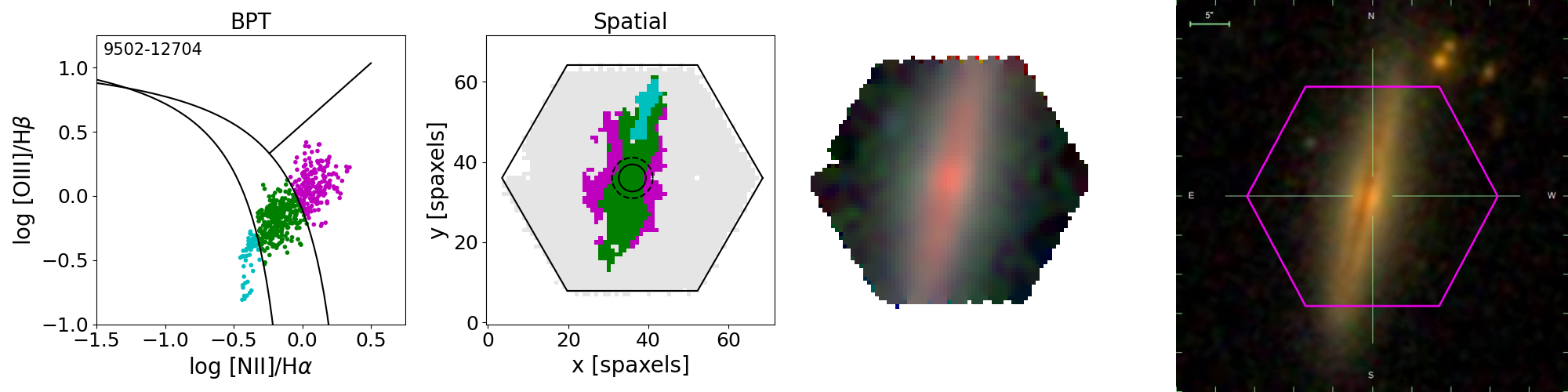}
    
    \includegraphics[width=0.5\textwidth]{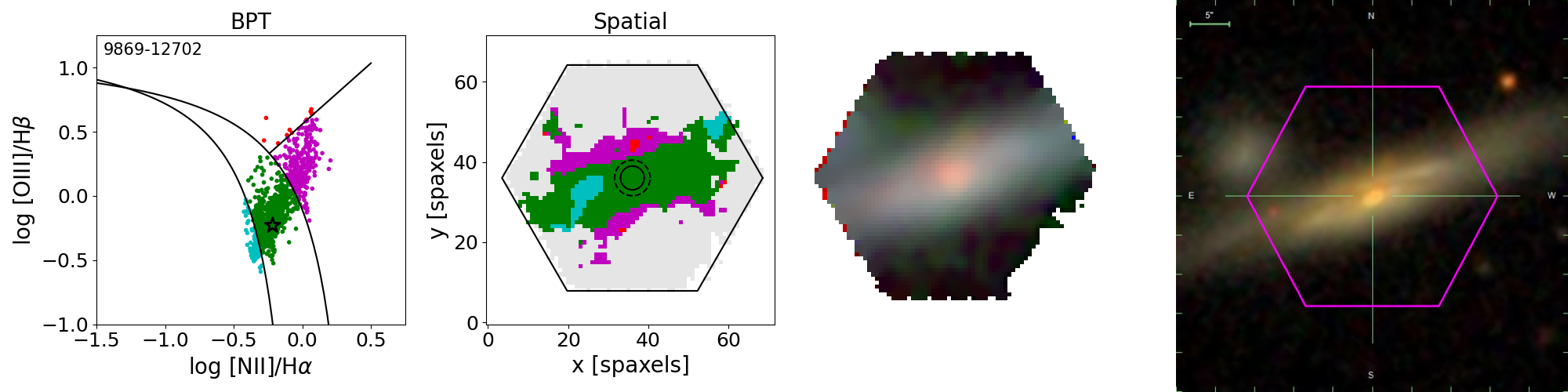}\hfill
    \includegraphics[width=0.5\textwidth]{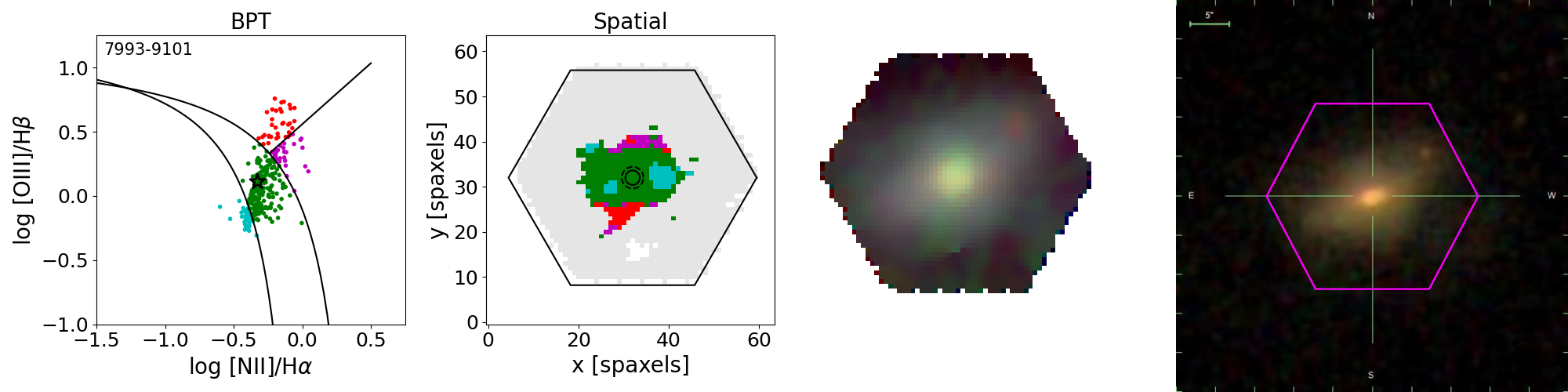}
    
    \caption{The same as Figure \ref{fig:appendix1}.}
    \label{fig:appendix2}
\end{figure}

\appendix
Figures \ref{fig:appendix1} and \ref{fig:appendix2} show diagnostic figures for the 34 candidate fading-AGN in our final sample. 

\begin{thebibliography}{}

\bibitem[Abdurro'uf et al.(2022)]{Abdurrouf22} Abdurro'uf, Accetta, K., Aerts, C., et al.\ 2022, \apjs, 259, 35. doi:10.3847/1538-4365/ac4414

\bibitem[Alb{\'a}n \& Wylezalek(2023)]{Alban23} Alb{\'a}n, M. \& Wylezalek, D.\ 2023, \aap, 674, A85. doi:10.1051/0004-6361/202245437

\bibitem[Alb{\'a}n et al.(2026)]{Alban26} Alb{\'a}n, M., Wylezalek, D., Kukreti, P., et al.\ 2026, \aap, 707, A149. doi:10.1051/0004-6361/202555882

\bibitem[Angl{\'e}s-Alc{\'a}zar et al.(2021)]{Alcazar21} Angl{\'e}s-Alc{\'a}zar, D., Quataert, E., Hopkins, P.~F., et al.\ 2021, \apj, 917, 2, 53. doi:10.3847/1538-4357/ac09e8

\bibitem[Alexander et al.(2025)]{Alexander25} Alexander, D.~M., Hickox, R.~C., Aird, J., et al.\ 2025, \nar, 101, 101733. doi:10.1016/j.newar.2025.101733

\bibitem[Amor{\'\i}n et al.(2015)]{Amorin15} Amor{\'\i}n, R., P{\'e}rez-Montero, E., Contini, T., et al.\ 2015, \aap, 578, A105. doi:10.1051/0004-6361/201322786

\bibitem[Antonucci(1993)]{Antonucci93} Antonucci, R.\ 1993, \araa, 31, 473. doi:10.1146/annurev.aa.31.090193.002353

\bibitem[Assef et al.(2010)]{Assef10} Assef, R.~J., Kochanek, C.~S., Brodwin, M., et al.\ 2010, \apj, 713, 2, 970. doi:10.1088/0004-637X/713/2/970

\bibitem[Assef et al.(2018)]{Assef18} Assef, R.~J., Stern, D., Noirot, G., et al.\ 2018, \apjs, 234, 23. doi:10.3847/1538-4365/aaa00a

\bibitem[Astropy Collaboration et al.(2022)]{astropy2022} Astropy Collaboration, Price-Whelan, A.~M., Lim, P.~L., et al.\ 2022, \apj, 935, 167. doi:10.3847/1538-4357/ac7c74

\bibitem[Baldwin et al.(1981)]{Baldwin81} Baldwin, J.~A., Phillips, M.~M., \& Terlevich, R.\ 1981, \pasp, 93, 5. doi:10.1086/130766

\bibitem[Barthelmy et al.(2005)]{Barthelmy05} Barthelmy, S.~D., Barbier, L.~M., Cummings, J.~R., et al.\ 2005, \ssr, 120, 3-4, 143. doi:10.1007/s11214-005-5096-3

\bibitem[Becker et al.(1995)]{Becker95} Becker, R.~H., White, R.~L., \& Helfand, D.~J.\ 1995, \apj, 450, 559. doi:10.1086/176166

\bibitem[Belfiore et al.(2019)]{Belfiore19} Belfiore, F., Westfall, K.~B., Schaefer, A., et al.\ 2019, \aj, 158, 160. doi:10.3847/1538-3881/ab3e4e

\bibitem[Benedetti et al.(2024)]{Benedetti24} Benedetti, J.~P.~V., Riffel, R., Ricci, T., et al.\ 2024, \mnras, 534, 1, 400. doi:10.1093/mnras/stae2077

\bibitem[Berg et al.(2021)]{Berg21} Berg, D.~A., Chisholm, J., Erb, D.~K., et al.\ 2021, \apj, 922, 2, 170. doi:10.3847/1538-4357/ac141b

\bibitem[Binette \& Robinson(1987)]{Binette87} Binette, L. \& Robinson, A.\ 1987, \aap, 177, 11

\bibitem[Binney \& Tremaine(2008)]{Binney08} Binney, J. \& Tremaine, S.\ 2008, Galactic Dynamics: Second Edition (Princeton, NJ: Princeton Univ. Press)

\bibitem[Blanton et al.(2017)]{Blanton17} Blanton, M.~R., Bershady, M.~A., Abolfathi, B., et al.\ 2017, \aj, 154, 28. doi:10.3847/1538-3881/aa7567

\bibitem[Bongiorno et al.(2010)]{Bongiorno10} Bongiorno, A., Mignoli, M., Zamorani, G., et al.\ 2010, \aap, 510, A56. doi:10.1051/0004-6361/200913229

%\bibitem[Bradley et al.(2024)]{Bradley24} Bradley, L., Sip{\H{o}}cz, B., Robitaille, T., et al.\ 2024, Zenodo

\bibitem[Brinchmann et al.(2008)]{Brinchmann08} Brinchmann, J., Kunth, D., \& Durret, F.\ 2008, \aap, 485, 3, 657. doi:10.1051/0004-6361:200809783

\bibitem[Buchner \& Bauer(2017)]{Buchner17} Buchner, J. \& Bauer, F.~E.\ 2017, \mnras, 465, 4, 4348. doi:10.1093/mnras/stw2955

\bibitem[Bundy et al.(2015)]{Bundy15} Bundy, K., Bershady, M.~A., Law, D.~R., et al.\ 2015, \apj, 798, 7. doi:10.1088/0004-637X/798/1/7

\bibitem[Byrne-Mamahit et al.(2025)]{Byrne25} Byrne-Mamahit, S., Ellison, S.~L., Patton, D.~R., et al.\ 2025, \mnras, 544, 2, 1673. doi:10.1093/mnras/staf1765

\bibitem[Calzetti et al.(2000)]{Calzetti00} Calzetti, D., Armus, L., Bohlin, R.~C., et al.\ 2000, \apj, 533, 682. doi:10.1086/308692

\bibitem[Capelo et al.(2015)]{Capelo15} Capelo, P.~R., Volonteri, M., Dotti, M., et al.\ 2015, \mnras, 447, 3, 2123. doi:10.1093/mnras/stu2500

\bibitem[Cappellari \& Copin(2003)]{Cappellari03} Cappellari, M. \& Copin, Y.\ 2003, \mnras, 342, 345. doi:10.1046/j.1365-8711.2003.06541.x

\bibitem[Cappellari \& Emsellem(2004)]{Cappellari04} Cappellari, M. \& Emsellem, E.\ 2004, \pasp, 116, 138. doi:10.1086/381875

\bibitem[Cappellari(2017)]{Cappellari17} Cappellari, M.\ 2017, \mnras, 466, 798. doi:10.1093/mnras/stw3020

\bibitem[Cappellari(2023)]{Cappellari23} Cappellari, M.\ 2023, \mnras, 526, 3, 3273. doi:10.1093/mnras/stad2597

\bibitem[Comerford \& Greene(2014)]{Comerford14} Comerford, J.~M. \& Greene, J.~E.\ 2014, \apj, 789, 2, 112. doi:10.1088/0004-637X/789/2/112

\bibitem[Comerford et al.(2017)]{Comerford17} Comerford, J.~M., Barrows, R.~S., M{\"u}ller-S{\'a}nchez, F., et al.\ 2017, \apj, 849, 102. doi:10.3847/1538-4357/aa8e4b

\bibitem[Comerford et al.(2020)]{Comerford20} Comerford, J.~M., Negus, J., M{\"u}ller-S{\'a}nchez, F., et al.\ 2020, \apj, 901, 159. doi:10.3847/1538-4357/abb2ae

\bibitem[Comerford et al.(2022)]{Comerford22} Comerford, J.~M., Negus, J., Barrows, R.~S., et al.\ 2022, \apj, 927, 1, 23. doi:10.3847/1538-4357/ac496a

\bibitem[Comerford et al.(2024)]{Comerford24} Comerford, J.~M., Nevin, R., Negus, J., et al.\ 2024, \apj, 963, 1, 53. doi:10.3847/1538-4357/ad1a15

\bibitem[Condon et al.(1998)]{Condon98} Condon, J.~J., Cotton, W.~D., Greisen, E.~W., et al.\ 1998, \aj, 115, 1693. doi:10.1086/300337

\bibitem[Corrick et al.(2020)]{Corrick20} Corrick, E.~C., Drysdale, R.~N., Hellstrom, J.~C., et al.\ 2020, Science, 369, 6506, 963. doi:10.1126/science.aay5538

\bibitem[Curti et al.(2022)]{Curti22} Curti, M., Hayden-Pawson, C., Maiolino, R., et al.\ 2022, \mnras, 512, 4136. doi:10.1093/mnras/stac544

\bibitem[D'Agostino et al.(2018)]{Dagostino18} D'Agostino, J.~J., Poetrodjojo, H., Ho, I.-T., et al.\ 2018, \mnras, 479, 4907. doi:10.1093/mnras/sty1676

\bibitem[D'Agostino et al.(2019)]{Dagostino19} D'Agostino, J.~J., Kewley, L.~J., Groves, B.~A., et al.\ 2019, \mnras, 485, L38. doi:10.1093/mnrasl/slz028

\bibitem[Dahmer-Hahn et al.(2019)]{Dahmer19} Dahmer-Hahn, L.~G., Riffel, R., Ricci, T.~V., et al.\ 2019, \mnras, 489, 4, 5653. doi:10.1093/mnras/stz2453

\bibitem[Davies et al.(2014a)]{Davies14a} Davies, R.~L., Rich, J.~A., Kewley, L.~J., et al.\ 2014a, \mnras, 439, 3835. doi:10.1093/mnras/stu234

\bibitem[Davies et al.(2014b)]{Davies14b} Davies, R.~L., Kewley, L.~J., Ho, I.-T., et al.\ 2014b, \mnras, 444, 3961. doi:10.1093/mnras/stu1740

\bibitem[Dawson et al.(2016)]{Dawson16} Dawson, K.~S., Kneib, J.-P., Percival, W.~J., et al.\ 2016, \aj, 151, 44. doi:10.3847/0004-6256/151/2/44

\bibitem[Dey et al.(2019)]{Dey19} Dey, A., Schlegel, D.~J., Lang, D., et al.\ 2019, \aj, 157, 168. doi:10.3847/1538-3881/ab089d

\bibitem[Di Matteo et al.(2005)]{DiMatteo05} Di Matteo, T., Springel, V., \& Hernquist, L.\ 2005, \nat, 433, 7026, 604. doi:10.1038/nature03335

\bibitem[Dopita \& Sutherland(1995)]{Dopita95} Dopita, M.~A. \& Sutherland, R.~S.\ 1995, \apj, 455, 468. doi:10.1086/176596

\bibitem[Dopita et al.(2006)]{Dopita06} Dopita, M.~A., Fischera, J., Sutherland, R.~S., et al.\ 2006, \apjs, 167, 2, 177. doi:10.1086/508261

\bibitem[Dopita et al.(2016)]{Dopita16} Dopita, M.~A., Kewley, L.~J., Sutherland, R.~S., et al.\ 2016, \apss, 361, 61. doi:10.1007/s10509-016-2657-8

\bibitem[Dors(2021)]{Dors21} Dors, O.~L.\ 2021, \mnras, 507, 466. doi:10.1093/mnras/stab2166

\bibitem[Drory et al.(2015)]{Drory15} Drory, N., MacDonald, N., Bershady, M.~A., et al.\ 2015, \aj, 149, 77. doi:10.1088/0004-6256/149/2/77

\bibitem[Esparza-Arredondo et al.(2020)]{Esparza20} Esparza-Arredondo, D., Osorio-Clavijo, N., Gonz{\'a}lez-Mart{\'\i}n, O., et al.\ 2020, \apj, 905, 1, 29. doi:10.3847/1538-4357/abc425

\bibitem[Evans et al.(2010)]{Evans10} Evans, I.~N., Primini, F.~A., Glotfelty, K.~J., et al.\ 2010, \apjs, 189, 37. doi:10.1088/0067-0049/189/1/37

\bibitem[Falc{\'o}n-Barroso et al.(2011)]{Falcon11} Falc{\'o}n-Barroso, J., S{\'a}nchez-Bl{\'a}zquez, P., Vazdekis, A., et al.\ 2011, \aap, 532, A95. doi:10.1051/0004-6361/201116842

\bibitem[Cid Fernandes et al.(2010)]{Fernandes10} Cid Fernandes, R., Stasi{\'n}ska, G., Schlickmann, M.~S., et al.\ 2010, \mnras, 403, 2, 1036. doi:10.1111/j.1365-2966.2009.16185.x

\bibitem[Cid Fernandes et al.(2011)]{Fernandes11} Cid Fernandes, R., Stasi{\'n}ska, G., Mateus, A., et al.\ 2011, \mnras, 413, 1687. doi:10.1111/j.1365-2966.2011.18244.x

%\bibitem[Finlez et al.(2025)]{Finlez25} Finlez, C., Treister, E., Bauer, F., et al.\ 2025, , arXiv:2507.01115. doi:10.48550/arXiv.2507.01115
\bibitem[Finlez et al.(2025)]{Finlez25} Finlez, C., Treister, E., Bauer, F., et al.\ 2025, \aap, 703, A63. doi:10.1051/0004-6361/202452404

\bibitem[Fischer et al.(2013)]{Fischer13} Fischer, T.~C., Crenshaw, D.~M., Kraemer, S.~B., et al.\ 2013, \apjs, 209, 1, 1. doi:10.1088/0067-0049/209/1/1

\bibitem[French et al.(2023)]{French23} French, K.~D., Earl, N., Novack, A.~B., et al.\ 2023, \apj, 950, 153. doi:10.3847/1538-4357/acd249

\bibitem[Fu et al.(2018)]{Fu18} Fu, H., Steffen, J.~L., Gross, A.~C., et al.\ 2018, \apj, 856, 2, 93. doi:10.3847/1538-4357/aab364

\bibitem[Gabor \& Bournaud(2013)]{Gabor13} Gabor, J.~M. \& Bournaud, F.\ 2013, \mnras, 434, 1, 606. doi:10.1093/mnras/stt1046

\bibitem[Gabor \& Bournaud(2014)]{Gabor14} Gabor, J.~M. \& Bournaud, F.\ 2014, \mnras, 441, 2, 1615. doi:10.1093/mnras/stu677

\bibitem[Gagne et al.(2014)]{Gagne14} Gagne, J.~P., Crenshaw, D.~M., Kraemer, S.~B., et al.\ 2014, \apj, 792, 1, 72. doi:10.1088/0004-637X/792/1/72

\bibitem[Garg et al.(2022)]{Garg22} Garg, P., Narayanan, D., Byler, N., et al.\ 2022, \apj, 926, 80. doi:10.3847/1538-4357/ac43b8

\bibitem[Gezari(2021)]{Gezari21} Gezari, S.\ 2021, \araa, 59, 21. doi:10.1146/annurev-astro-111720-030029

\bibitem[Gkini et al.(2021)]{Gkini21} Gkini, A., Plionis, M., Chira, M., et al.\ 2021, \aap, 650, A75. doi:10.1051/0004-6361/202140278

\bibitem[Goel et al.(2026)]{Goel26} Goel, A., Salim, S., Ellison, S.~L., et al.\ 2026, \apj, 1008, 2, 232. doi:10.3847/1538-4357/ae930d

\bibitem[Goulding et al.(2012)]{Goulding12} Goulding, A.~D., Alexander, D.~M., Bauer, F.~E., et al.\ 2012, \apj, 755, 1, 5. doi:10.1088/0004-637X/755/1/5

\bibitem[Groves et al.(2006)]{Groves06} Groves, B.~A., Heckman, T.~M., \& Kauffmann, G.\ 2006, \mnras, 371, 1559. doi:10.1111/j.1365-2966.2006.10812.x

\bibitem[Gunn et al.(2006)]{Gunn06} Gunn, J.~E., Siegmund, W.~A., Mannery, E.~J., et al.\ 2006, \aj, 131, 2332. doi:10.1086/500975

\bibitem[Harris et al.(2020)]{numpy2020}Harris, C.R., Millman, K.J., van der Walt, S.J. et al.\ 2020, \nat, 585, 357. doi:10.1038/s41586-020-2649-2

\bibitem[He et al.(2018)]{He18} He, Z., Sun, A.-L., Zakamska, N.~L., et al.\ 2018, \mnras, 478, 3, 3614. doi:10.1093/mnras/sty1322

\bibitem[Heckman(1980)]{Heckman80} Heckman, T.~M.\ 1980, \aap, 87, 152

\bibitem[Hickox et al.(2014)]{Hickox14} Hickox, R.~C., Mullaney, J.~R., Alexander, D.~M., et al.\ 2014, \apj, 782, 9. doi:10.1088/0004-637X/782/1/9

\bibitem[Hickox et al.(2017)]{Hickox17} Hickox, R.~C., Myers, A.~D., Greene, J.~E., et al.\ 2017, \apj, 849, 1, 53. doi:10.3847/1538-4357/aa8c77

\bibitem[Hickox \& Alexander(2018)]{Hickox18} Hickox, R.~C. \& Alexander, D.~M.\ 2018, \araa, 56, 625. doi:10.1146/annurev-astro-081817-051803

\bibitem[Ho et al.(1993)]{Ho93} Ho, L.~C., Shields, J.~C., \& Filippenko, A.~V.\ 1993, \apj, 410, 567. doi:10.1086/172775

\bibitem[Ho et al.(2014)]{Ho14} Ho, I.-T., Kewley, L.~J., Dopita, M.~A., et al.\ 2014, \mnras, 444, 3894. doi:10.1093/mnras/stu1653

\bibitem[Hopkins et al.(2008)]{Hopkins08} Hopkins, P.~F., Hernquist, L., Cox, T.~J., et al.\ 2008, \apjs, 175, 2, 356. doi:10.1086/524362

\bibitem[Hunter(2007)]{matplotlib} Hunter, J.~D.\ 2007, Computing in Science and Engineering, 9, 3, 90. doi:10.1109/MCSE.2007.55

\bibitem[Ichikawa \& Tazaki(2017)]{Ichikawa17} Ichikawa, K. \& Tazaki, R.\ 2017, \apj, 844, 1, 21. doi:10.3847/1538-4357/aa7891

\bibitem[Ishibashi \& Fabian(2015)]{Ishibashi15} Ishibashi, W. \& Fabian, A.~C.\ 2015, \mnras, 451, 1, 93. doi:10.1093/mnras/stv944

\bibitem[Ji et al.(2022)]{Ji22} Ji, Z., Giavalisco, M., Kirkpatrick, A., et al.\ 2022, \apj, 925, 1, 74. doi:10.3847/1538-4357/ac3837

\bibitem[Jogee(2006)]{Jogee06} Jogee, S.\ 2006, Physics of Active Galactic Nuclei at all Scales, 693, 143. doi:10.1007/3-540-34621-X\_6

\bibitem[Jogee et al.(2009)]{Jogee09} Jogee, S., Miller, S.~H., Penner, K., et al.\ 2009, \apj, 697, 2, 1971. doi:10.1088/0004-637X/697/2/1971

\bibitem[Jones et al.(2016)]{Jones16} Jones, M.~L., Hickox, R.~C., Black, C.~S., et al.\ 2016, \apj, 826, 12. doi:10.3847/0004-637X/826/1/12

\bibitem[Kauffmann et al.(2003)]{Kauffmann03} Kauffmann, G., Heckman, T.~M., Tremonti, C., et al.\ 2003, \mnras, 346, 1055. doi:10.1111/j.1365-2966.2003.07154.x

\bibitem[Kauffmann \& Heckman(2009)]{Kauffmann09} Kauffmann, G. \& Heckman, T.~M.\ 2009, \mnras, 397, 135. doi:10.1111/j.1365-2966.2009.14960.x

\bibitem[Keel et al.(2012)]{Keel12} Keel, W.~C., Chojnowski, S.~D., Bennert, V.~N., et al.\ 2012, \mnras, 420, 878. doi:10.1111/j.1365-2966.2011.20101.x

\bibitem[Keel et al.(2013)]{Keel13} Keel, W.~C., Manning, A.~M., Holwerda, B.~W., et al.\ 2013, \pasp, 125, 2. doi:10.1086/669233

\bibitem[Keel et al.(2017)]{Keel17} Keel, W.~C., Lintott, C.~J., Maksym, W.~P., et al.\ 2017, \apj, 835, 256. doi:10.3847/1538-4357/835/2/256

\bibitem[Keel et al.(2022)]{Keel22} Keel, W.~C., Moiseev, A., Kozlova, D.~V., et al.\ 2022, \mnras, 510, 4608. doi:10.1093/mnras/stab3656

%\bibitem[Keel et al.(2024a)]{Keel24a} Keel, W., Moiseev, A., Bennert, V.~N., et al.\ 2024, \aas, 243, 175.08

\bibitem[Keel et al.(2024)]{Keel24} Keel, W.~C., Moiseev, A., Uklein, R., et al.\ 2024, \mnras, 530, 1624. doi:10.1093/mnras/stae946

\bibitem[Kelly et al.(2009)]{Kelly09} Kelly, B.~C., Bechtold, J., \& Siemiginowska, A.\ 2009, \apj, 698, 1, 895. doi:10.1088/0004-637X/698/1/895

\bibitem[Kewley et al.(2001)]{Kewley01} Kewley, L.~J., Dopita, M.~A., Sutherland, R.~S., et al.\ 2001, \apj, 556, 121. doi:10.1086/321545

\bibitem[Kewley et al.(2006)]{Kewley06} Kewley, L.~J., Groves, B., Kauffmann, G., et al.\ 2006, \mnras, 372, 961. doi:10.1111/j.1365-2966.2006.10859.x

\bibitem[Kewley et al.(2013)]{Kewley13} Kewley, L.~J., Dopita, M.~A., Leitherer, C., et al.\ 2013, \apj, 774, 100. doi:10.1088/0004-637X/774/2/100

\bibitem[Kewley et al.(2019)]{Kewley19} Kewley, L.~J., Nicholls, D.~C., \& Sutherland, R.~S.\ 2019, \araa, 57, 511. doi:10.1146/annurev-astro-081817-051832

\bibitem[Knese et al.(2020)]{Knese20} Knese, E.~D., Keel, W.~C., Knese, G., et al.\ 2020, \mnras, 496, 1035. doi:10.1093/mnras/staa1510

\bibitem[Koulouridis et al.(2013)]{Koulouridis13} Koulouridis, E., Plionis, M., Ch{\'a}vez, R., et al.\ 2013, \aap, 554, A13. doi:10.1051/0004-6361/201220917

\bibitem[Kumari et al.(2021)]{Kumari21} Kumari, N., Maiolino, R., Trussler, J., et al.\ 2021, \aap, 656, A140. doi:10.1051/0004-6361/202140757

\bibitem[Lagos et al.(2011)]{Lagos11} Lagos, C.~D.~P., Padilla, N.~D., Strauss, M.~A., et al.\ 2011, \mnras, 414, 3, 2148. doi:10.1111/j.1365-2966.2011.18531.x

\bibitem[Lamastra et al.(2009)]{Lamastra09} Lamastra, A., Bianchi, S., Matt, G., et al.\ 2009, \aap, 504, 73. doi:10.1051/0004-6361/200912023

\bibitem[LaMassa et al.(2015)]{Lamassa15} LaMassa, S.~M., Cales, S., Moran, E.~C., et al.\ 2015, \apj, 800, 144. doi:10.1088/0004-637X/800/2/144

\bibitem[Law et al.(2015)]{Law15} Law, D.~R., Yan, R., Bershady, M.~A., et al.\ 2015, \aj, 150, 19. doi:10.1088/0004-6256/150/1/19

\bibitem[Law et al.(2016)]{Law16} Law, D.~R., Cherinka, B., Yan, R., et al.\ 2016, \aj, 152, 83. doi:10.3847/0004-6256/152/4/83

\bibitem[Law et al.(2021a)]{Law21a} Law, D.~R., Westfall, K.~B., Bershady, M.~A., et al.\ 2021a, \aj, 161, 52. doi:10.3847/1538-3881/abcaa2

\bibitem[Law et al.(2021b)]{Law21b} Law, D.~R., Ji, X., Belfiore, F., et al.\ 2021b, \apj, 915, 35. doi:10.3847/1538-4357/abfe0a

\bibitem[Lebowitz et al.(2025)]{Lebowitz25} Lebowitz, S., Hainline, K., Juneau, S., et al.\ 2025, \apj, 984, 1, 13. doi:10.3847/1538-4357/adc07c

\bibitem[Liang et al.(2020)]{Liang20} Liang, F.-H., Li, C., Li, N., et al.\ 2020, \apj, 896, 2, 121. doi:10.3847/1538-4357/ab9596

\bibitem[Lintott et al.(2008)]{Lintott08} Lintott, C.~J., Schawinski, K., Slosar, A., et al.\ 2008, \mnras, 389, 1179. doi:10.1111/j.1365-2966.2008.13689.x

\bibitem[Lintott et al.(2009)]{Lintott09} Lintott, C.~J., Schawinski, K., Keel, W., et al.\ 2009, \mnras, 399, 129. doi:10.1111/j.1365-2966.2009.15299.x

\bibitem[Lugo-Aranda et al.(2024)]{Lugo24} Lugo-Aranda, A.~Z., S{\'a}nchez, S.~F., Barrera-Ballesteros, J.~K., et al.\ 2024, \mnras, 528, 4, 6099. doi:10.1093/mnras/stae345

\bibitem[Luridiana et al.(2015)]{Luridiana15} Luridiana, V., Morisset, C., \& Shaw, R.~A.\ 2015, \aap, 573, A42. doi:10.1051/0004-6361/201323152

\bibitem[Maiolino \& Mannucci(2019)]{Maiolino19} Maiolino, R. \& Mannucci, F.\ 2019, \aapr, 27, 3. doi:10.1007/s00159-018-0112-2

\bibitem[Maksym et al.(2017)]{Maksym17} Maksym, W.~P., Fabbiano, G., Elvis, M., et al.\ 2017, \apj, 844, 1, 69. doi:10.3847/1538-4357/aa78a4

\bibitem[Matt et al.(2003)]{Matt03} Matt, G., Guainazzi, M., \& Maiolino, R.\ 2003, \mnras, 342, 422. doi:10.1046/j.1365-8711.2003.06539.x

\bibitem[Matthews Acu{\~n}a et al.(2025)]{MatthewsAcuna25} Matthews Acu{\~n}a, O.~S., Tremonti, C.~A., Westfall, K.~B., et al.\ 2025, arXiv:2512.18076. doi:10.48550/arXiv.2512.18076

\bibitem[Menezes et al.(2016)]{Menezes16} Menezes, R.~B., Steiner, J.~E., \& da Silva, P.\ 2016, \apj, 817, 150. doi:10.3847/0004-637X/817/2/150

\bibitem[Merloni et al.(2024)]{Merloni24} Merloni, A., Lamer, G., Liu, T., et al.\ 2024, \aap, 682, A34. doi:10.1051/0004-6361/202347165

\bibitem[Mezcua \& Dom{\'\i}nguez S{\'a}nchez(2020)]{Mezcua20} Mezcua, M. \& Dom{\'\i}nguez S{\'a}nchez, H.\ 2020, \apjl, 898, 2, L30. doi:10.3847/2041-8213/aba199

\bibitem[Mezcua \& Dom{\'\i}nguez S{\'a}nchez(2024)]{Mezcua24} Mezcua, M. \& Dom{\'\i}nguez S{\'a}nchez, H.\ 2024, \mnras, 528, 3, 5252. doi:10.1093/mnras/stae292

\bibitem[Mo et al.(2024)]{Mo24} Mo, H., Chen, Y.-M., Zhang, Z.-Y., et al.\ 2024, \mnras, 529, 4500. doi:10.1093/mnras/stae753

\bibitem[Moiseev \& Smirnova(2023)]{Moiseev23} Moiseev, A.~V. \& Smirnova, A.~A.\ 2023, Galaxies, 11, 118. doi:10.3390/galaxies11060118

\bibitem[Moustakas et al.(2010)]{Moustakas10} Moustakas, J., Kennicutt, R.~C., Tremonti, C.~A., et al.\ 2010, \apjs, 190, 233. doi:10.1088/0067-0049/190/2/233

\bibitem[Moy \& Rocca-Volmerange(2002)]{Moy02} Moy, E. \& Rocca-Volmerange, B.\ 2002, \aap, 383, 46. doi:10.1051/0004-6361:20011727

\bibitem[Negus et al.(2024)]{Negus24} Negus, J., Comerford, J.~M., \& S{\'a}nchez, F.~M.\ 2024, \apj, 971, 1, 92. doi:10.3847/1538-4357/ad4c68

\bibitem[Nevin et al.(2023)]{Nevin23} Nevin, R., Blecha, L., Comerford, J., et al.\ 2023, \mnras, 522, 1, 1. doi:10.1093/mnras/stad911

\bibitem[Novak et al.(2011)]{Novak11} Novak, G.~S., Ostriker, J.~P., \& Ciotti, L.\ 2011, \apj, 737, 1, 26. doi:10.1088/0004-637X/737/1/26

\bibitem[Oh et al.(2015)]{Oh15} Oh, K., Yi, S.~K., Schawinski, K., et al.\ 2015, \apjs, 219, 1. doi:10.1088/0067-0049/219/1/1

\bibitem[Oh et al.(2018)]{Oh18} Oh, K., Koss, M., Markwardt, C.~B., et al.\ 2018, \apjs, 235, 4. doi:10.3847/1538-4365/aaa7fd

\bibitem[Osterbrock \& Ferland(2006)]{Osterbrock06} Osterbrock, D.~E. \& Ferland, G.~J.\ 2006, . 

\bibitem[Pai et al.(2024)]{Pai24} Pai, A., Blanton, M.~R., \& Moustakas, J.\ 2024, \apj, 977, 1, 102. doi:10.3847/1538-4357/ad89b8

\bibitem[Padovani et al.(2017)]{Padovani17} Padovani, P., Alexander, D.~M., Assef, R.~J., et al.\ 2017, \aapr, 25, 2. doi:10.1007/s00159-017-0102-9

\bibitem[Pier et al.(2003)]{Pier03} Pier, J.~R., Munn, J.~A., Hindsley, R.~B., et al.\ 2003, \aj, 125, 3, 1559. doi:10.1086/346138

\bibitem[Peterson(1993)]{Peterson93} Peterson, B.~M.\ 1993, \pasp, 105, 247. doi:10.1086/133140

\bibitem[Peterson et al.(1998)]{Peterson98} Peterson, B.~M., Wanders, I., Bertram, R., et al.\ 1998, \apj, 501, 1, 82. doi:10.1086/305813

\bibitem[Poetrodjojo et al.(2018)]{Poetrodjojo18} Poetrodjojo, H., Groves, B., Kewley, L.~J., et al.\ 2018, \mnras, 479, 5235. doi:10.1093/mnras/sty1782

\bibitem[Reback et al.(2021)]{pandas2021} Reback, J., Jbrockmendel, McKinney, W., et al.\ 2021, Zenodo

\bibitem[Ricci \& Trakhtenbrot(2023)]{Ricci23} Ricci, C. \& Trakhtenbrot, B.\ 2023, Nature Astronomy, 7, 1282. doi:10.1038/s41550-023-02108-4

\bibitem[Rich et al.(2010)]{Rich10} Rich, J.~A., Dopita, M.~A., Kewley, L.~J., et al.\ 2010, \apj, 721, 505. doi:10.1088/0004-637X/721/1/505

\bibitem[Rich et al.(2011)]{Rich11} Rich, J.~A., Kewley, L.~J., \& Dopita, M.~A.\ 2011, \apj, 734, 87. doi:10.1088/0004-637X/734/2/87

\bibitem[Riffel et al.(2013)]{Riffel13} Riffel, R., Rodr{\'\i}guez-Ardila, A., Aleman, I., et al.\ 2013, \mnras, 430, 3, 2002. doi:10.1093/mnras/stt026

\bibitem[Riffel et al.(2021)]{Riffel21} Riffel, R.~A., Storchi-Bergmann, T., Riffel, R., et al.\ 2021, \mnras, 504, 3, 3265. doi:10.1093/mnras/stab998

\bibitem[Riffel et al.(2023)]{Riffel23} Riffel, R., Mallmann, N.~D., Rembold, S.~B., et al.\ 2023, \mnras, 524, 4, 5640. doi:10.1093/mnras/stad2234

\bibitem[Riffel et al.(2024)]{Riffel24} Riffel, R., Dahmer-Hahn, L.~G., Vazdekis, A., et al.\ 2024, \mnras, 531, 1, 554. doi:10.1093/mnras/stae1192

\bibitem[Rosa-Gonz{\'a}lez et al.(2007)]{Rosa07} Rosa-Gonz{\'a}lez, D., Schmitt, H.~R., Terlevich, E., et al.\ 2007, \apj, 654, 226. doi:10.1086/509105

\bibitem[Salim et al.(2018)]{Salim18} Salim, S., Boquien, M., \& Lee, J.~C.\ 2018, \apj, 859, 11. doi:10.3847/1538-4357/aabf3c

\bibitem[S{\'a}nchez(2020)]{Sanchez20} S{\'a}nchez, S.~F.\ 2020, \araa, 58, 99. doi:10.1146/annurev-astro-012120-013326

\bibitem[Sargent \& Searle(1970)]{Sargent70} Sargent, W.~L.~W. \& Searle, L.\ 1970, \apjl, 162, L155. doi:10.1086/180644

\bibitem[Schawinski et al.(2010)]{Schawinski10} Schawinski, K., Evans, D.~A., Virani, S., et al.\ 2010, \apjl, 724, 1, L30. doi:10.1088/2041-8205/724/1/L30

\bibitem[Schawinski et al.(2015)]{Schawinski15} Schawinski, K., Koss, M., Berney, S., et al.\ 2015, \mnras, 451, 3, 2517. doi:10.1093/mnras/stv1136

\bibitem[Schirmer et al.(2013)]{Schirmer13} Schirmer, M., Diaz, R., Holhjem, K., et al.\ 2013, \apj, 763, 1, 60. doi:10.1088/0004-637X/763/1/60

\bibitem[Schweizer et al.(2013)]{Schweizer13} Schweizer, F., Seitzer, P., Kelson, D.~D., et al.\ 2013, \apj, 773, 2, 148. doi:10.1088/0004-637X/773/2/148

\bibitem[Seyfert(1943)]{Seyfert43} Seyfert, C.~K.\ 1943, \apj, 97, 28. doi:10.1086/144488

\bibitem[Shappee et al.(2014)]{Shappee14} Shappee, B.~J., Prieto, J.~L., Grupe, D., et al.\ 2014, \apj, 788, 48. doi:10.1088/0004-637X/788/1/48

\bibitem[Sijacki et al.(2015)]{Sijacki15} Sijacki, D., Vogelsberger, M., Genel, S., et al.\ 2015, \mnras, 452, 1, 575. doi:10.1093/mnras/stv1340

\bibitem[Skipper \& Browne(2018)]{Skipper18} Skipper, C.~J. \& Browne, I.~W.~A.\ 2018, \mnras, 475, 4, 5179. doi:10.1093/mnras/sty114

\bibitem[Smee et al.(2013)]{Smee13} Smee, S.~A., Gunn, J.~E., Uomoto, A., et al.\ 2013, \aj, 146, 32. doi:10.1088/0004-6256/146/2/32

\bibitem[Springel et al.(2005)]{Springel05} Springel, V., Di Matteo, T., \& Hernquist, L.\ 2005, \mnras, 361, 3, 776. doi:10.1111/j.1365-2966.2005.09238.x

\bibitem[Steffen et al.(2023)]{Steffen23} Steffen, J.~L., Fu, H., Brownstein, J.~R., et al.\ 2023, \apj, 942, 2, 107. doi:10.3847/1538-4357/aca768

\bibitem[Stern et al.(2012)]{Stern12} Stern, D., Assef, R.~J., Benford, D.~J., et al.\ 2012, \apj, 753, 1, 30. doi:10.1088/0004-637X/753/1/30

\bibitem[Storchi-Bergmann \& Schnorr-M{\"u}ller(2019)]{Storchi19} Storchi-Bergmann, T. \& Schnorr-M{\"u}ller, A.\ 2019, Nature Astronomy, 3, 48. doi:10.1038/s41550-018-0611-0

\bibitem[Sun et al.(2017)]{Sun17} Sun, A.-L., Greene, J.~E., \& Zakamska, N.~L.\ 2017, \apj, 835, 222. doi:10.3847/1538-4357/835/2/222

\bibitem[Taylor(2005)]{Taylor2005} Taylor, M.~B.\ 2005, Astronomical Data Analysis Software and Systems XIV, 347, 29

\bibitem[Thomas et al.(2018)]{Thomas18} Thomas, A.~D., Kewley, L.~J., Dopita, M.~A., et al.\ 2018, \apjl, 861, 1, L2. doi:10.3847/2041-8213/aacce7

\bibitem[Trump et al.(2015)]{Trump15} Trump, J.~R., Sun, M., Zeimann, G.~R., et al.\ 2015, \apj, 811, 1, 26. doi:10.1088/0004-637X/811/1/26

\bibitem[Ulrich et al.(1997)]{Ulrich97} Ulrich, M.-H., Maraschi, L., \& Urry, C.~M.\ 1997, \araa, 35, 445. doi:10.1146/annurev.astro.35.1.445

\bibitem[Vaughan et al.(2003)]{Vaughan03} Vaughan, S., Edelson, R., Warwick, R.~S., et al.\ 2003, \mnras, 345, 1271. doi:10.1046/j.1365-2966.2003.07042.x

\bibitem[V{\'a}zquez-Mata et al.(2022)]{Vazquez22} V{\'a}zquez-Mata, J.~A., Hern{\'a}ndez-Toledo, H.~M., Avila-Reese, V., et al.\ 2022, \mnras, 512, 2, 2222. doi:10.1093/mnras/stac635

\bibitem[Virtanen et al.(2020)]{scipy} Virtanen, P., Gommers, R., Oliphant, T.~E., et al.\ 2020, Nature Medicine, 17, 261. doi:10.1038/s41592-019-0686-2

\bibitem[Wake et al.(2017)]{Wake17} Wake, D.~A., Bundy, K., Diamond-Stanic, A.~M., et al.\ 2017, \aj, 154, 3, 86. doi:10.3847/1538-3881/aa7ecc

\bibitem[Watkins et al.(2018)]{Watkins18} Watkins, A.~E., Mihos, J.~C., Bershady, M., et al.\ 2018, \apjl, 858, 2, L16. doi:10.3847/2041-8213/aabba1

\bibitem[Westfall et al.(2019)]{Westfall19} Westfall, K.~B., Cappellari, M., Bershady, M.~A., et al.\ 2019, \aj, 158, 231. doi:10.3847/1538-3881/ab44a2

\bibitem[Wild et al.(2010)]{Wild10} Wild, V., Heckman, T., \& Charlot, S.\ 2010, \mnras, 405, 2, 933. doi:10.1111/j.1365-2966.2010.16536.x

\bibitem[Wright et al.(2010)]{Wright10} Wright, E.~L., Eisenhardt, P.~R.~M., Mainzer, A.~K., et al.\ 2010, \aj, 140, 6, 1868. doi:10.1088/0004-6256/140/6/1868

\bibitem[Wylezalek et al.(2018)]{Wylezalek18} Wylezalek, D., Zakamska, N.~L., Greene, J.~E., et al.\ 2018, \mnras, 474, 2, 1499. doi:10.1093/mnras/stx2784

\bibitem[Wylezalek et al.(2020)]{Wylezalek20} Wylezalek, D., Flores, A.~M., Zakamska, N.~L., et al.\ 2020, \mnras, 492, 4, 4680. doi:10.1093/mnras/staa062

%\bibitem[Xiong et al.(2025)]{Xiong25} Xiong, Y., Jiang, N., Pan, Z., et al.\ 2025, , arXiv:2503.19722. doi:10.48550/arXiv.2503.19722
\bibitem[Xiong et al.(2025)]{Xiong25} Xiong, Y., Jiang, N., Pan, Z., et al.\ 2025, \apj, 989, 1, 49. doi:10.3847/1538-4357/ade2d7

\bibitem[Xu \& Wang(2023)]{Xu23} Xu, X. \& Wang, J.\ 2023, \apj, 943, 28. doi:10.3847/1538-4357/acac82

\bibitem[Yan et al.(2016)]{Yan16} Yan, R., Tremonti, C., Bershady, M.~A., et al.\ 2016, \aj, 151, 8. doi:10.3847/0004-6256/151/1/8

\bibitem[Yang et al.(2019)]{Yang19} Yang, C., Ge, J.-Q., \& Lu, Y.-J.\ 2019, Research in Astronomy and Astrophysics, 19, 12, 177. doi:10.1088/1674-4527/19/12/177

\bibitem[Zhang et al.(2017)]{Zhang17} Zhang, K., Yan, R., Bundy, K., et al.\ 2017, \mnras, 466, 3, 3217. doi:10.1093/mnras/stw3308

%\bibitem[Zhu et al.(2025)]{Zhu25} Zhu, P., Kewley, L.~J., Sutherland, R.~S., et al.\ 2025, , arXiv:2506.09962. doi:10.48550/arXiv.2506.09962
\bibitem[Zhu et al.(2025)]{Zhu25} Zhu, P., Kewley, L.~J., Sutherland, R.~S., et al.\ 2025, \apj, 988, 2, 261. doi:10.3847/1538-4357/ade78f

\end{thebibliography}
\end{document}